\documentclass[11pt]{article}
\usepackage{dsfont}

\usepackage{amsmath,amssymb,graphicx}
\usepackage{diagbox}
\usepackage{hyperref}
\usepackage[nosort]{cite}

\usepackage{tikz}
\usetikzlibrary{calc} \usetikzlibrary{patterns} \usetikzlibrary{decorations.pathreplacing} \usetikzlibrary{decorations.markings} \usetikzlibrary{decorations.pathmorphing} \usetikzlibrary{positioning}

\usepackage{mathtools}

\usepackage{color}
\definecolor{darkred}{rgb}{0.65,0.15,0}
\definecolor{newgreen}{rgb}{0.2,0.62,0.14}
\hypersetup{pdfborder={0 0 0},colorlinks=true,urlcolor=blue,citecolor=blue,linkcolor=darkred,linktocpage=true}

\DeclareFontFamily{U}{mathx}{}
\DeclareFontShape{U}{mathx}{m}{n}{ <-> mathx10 }{}
\DeclareSymbolFont{mathx}{U}{mathx}{m}{n}
\DeclareFontSubstitution{U}{mathx}{m}{n}

\DeclareMathAccent{\widecheck}{0}{mathx}{"71}

\numberwithin{equation}{section}

\def\spa#1.#2{\left\langle#1\,#2\right\rangle}
\def\spb#1.#2{\left[#1\,#2\right]}

\newcommand{\bsvBR}[3]{\beta^{\rm sv} \! \left[\begin{smallmatrix}#1\\#2\end{smallmatrix};#3\right]}
\newcommand{\bsvBRno}[2]{\beta^{\rm sv}\! \left[\begin{smallmatrix}#1\\#2\end{smallmatrix}\right]}
\newcommand{\beqv}[3]{\beta^{\rm eqv} \! \left[\begin{smallmatrix}#1\\#2\end{smallmatrix};#3\right]}
\newcommand{\beqvno}[2]{\beta^{\rm eqv}\! \left[\begin{smallmatrix}#1\\#2\end{smallmatrix}\right]}
\newcommand{\bplus}[3]{\beta_+ \! \left[\begin{smallmatrix}#1\\#2\end{smallmatrix};#3\right]}
\newcommand{\bplusno}[2]{\beta_+ \! \left[\begin{smallmatrix}#1\\#2\end{smallmatrix}\right]}
\newcommand{\bminus}[3]{\beta_- \! \left[\begin{smallmatrix}#1\\#2\end{smallmatrix};#3\right]}
\newcommand{\bminusno}[2]{\beta_-\! \left[\begin{smallmatrix}#1\\#2\end{smallmatrix}\right]}
\newcommand{\bphiminus}[3]{\beta^\phi_- \! \left[\begin{smallmatrix}#1\\#2\end{smallmatrix};#3\right]}
\newcommand{\bphiminusno}[2]{\beta^\phi_-\! \left[\begin{smallmatrix}#1\\#2\end{smallmatrix}\right]}
\newcommand{\bsvCUSP}[3]{\beta^{\rm sv}_{\Delta} \! \left[\begin{smallmatrix}#1\\#2\end{smallmatrix};#3\right]}
\newcommand{\bsvCUSPno}[2]{\beta^{\rm sv}_{\Delta} \! \left[\begin{smallmatrix}#1\\#2\end{smallmatrix}\right]}
\newcommand{\alphaBR}[3]{\alpha\! \left[\begin{smallmatrix}#1\\#2\end{smallmatrix};#3\right]}

\newcommand{\alphaeasy}[3]{\alpha_{{\rm easy}}\! \left[\begin{smallmatrix}#1\\#2\end{smallmatrix};#3\right]}
\newcommand{\alphahard}[3]{\alpha_{{\rm hard}}\! \left[\begin{smallmatrix}#1\\#2\end{smallmatrix};#3\right]}
\newcommand{\rhopart}[3]{\sigma\! \left[\begin{smallmatrix}#1\\#2\end{smallmatrix};#3\right]}
\newcommand{\kappaBR}[3]{\kappa\! \left[\begin{smallmatrix}#1\\#2\end{smallmatrix};#3\right]}
\newcommand{\kappaBRno}[2]{\kappa\! \left[\begin{smallmatrix}#1\\#2\end{smallmatrix}\right]}
\newcommand{\cplus}[2]{{\cal C}^+\! \left[\begin{smallmatrix}#1\\#2\end{smallmatrix}\right] \! (\tau)}
\newcommand{\cplusno}[2]{{\cal C}^+\! \left[\begin{smallmatrix}#1\\#2\end{smallmatrix}\right]}
\newcommand{\ompm}[3]{\omega_{\pm}\! \left[\begin{smallmatrix}#1\\#2\end{smallmatrix};#3\right]}
\newcommand{\omplus}[3]{\omega_{+}\! \left[\begin{smallmatrix}#1\\#2\end{smallmatrix};#3\right]}
\newcommand{\omminus}[3]{\omega_{-}\! \left[\begin{smallmatrix}#1\\#2\end{smallmatrix};#3\right]}
\newcommand{\coeffs}[2]{\epsilon \! \left[\begin{smallmatrix}#1\\#2\end{smallmatrix}\right]}
\newcommand{\ccsv}[2]{c^{\rm sv}\! \left[\begin{smallmatrix}#1\\#2\end{smallmatrix}\right]}
\newcommand{\ccsvpure}{c^{\rm sv}}
\newcommand{\ddsv}[3]{d^{\rm sv}\! \left[\begin{smallmatrix}#1\\#2\end{smallmatrix};#3\right]}
\newcommand{\ddsvno}[2]{d^{\rm sv}\! \left[\begin{smallmatrix}#1\\#2\end{smallmatrix}\right]}
\newcommand{\ddsvpure}{d^{\rm sv}}
\newcommand{\bbsv}[3]{b^{\rm sv}\! \left[\begin{smallmatrix}#1\\#2\end{smallmatrix};#3\right]}
\newcommand{\bbsvno}[2]{b^{\rm sv}\! \left[\begin{smallmatrix}#1\\#2\end{smallmatrix}\right]}
\newcommand{\bbsvpure}{b^{\rm sv}}
\newcommand{\nicedel}{\delta}
\newcommand{\isom}{\rho}
\newcommand{\newuscdelta}[5]{\underline{\Delta}_{#1}\! \left[\begin{smallmatrix}#2\\#3 \\ #4\end{smallmatrix};#5\right]}

\newcommand{\ezero}[3]{\overline{ {\cal E}_0 \! \left[\begin{smallmatrix}#1\\#2\end{smallmatrix};#3\right]}}

\font\tenshuffle=shuffle10 \font\sevenshuffle=shuffle7 \font\fiveshuffle=shuffle7 at 5pt
\def\shuffle{{%
\def\Dshuffle{\mathbin{\hbox{\tenshuffle\char'001}}}%
\def\Sshuffle{\mathbin{\hbox{\sevenshuffle\char'001}}}%
\def\SSshuffle{\mathbin{\hbox{\fiveshuffle\char'001}}}%
\mathchoice{\Dshuffle}{\Dshuffle}{\Sshuffle}{\SSshuffle}}}

\def\beq{\begin{equation}}
\def\eeq{\end{equation}}
\let\Re\relax
\let\Im\relax
\DeclareMathOperator{\Re}{Re}
\DeclareMathOperator{\Im}{Im}

\newcommand{\eq}{\begin{equation}}
\newcommand{\eqe}{\end{equation}}
\newcommand{\eqa}{\begin{eqnarray}}
\newcommand{\eqae}{\end{eqnarray}}

\newcommand{\bea}{\begin{eqnarray}}
\newcommand{\eea}{\end{eqnarray}}
\newcommand{\dd}{\mathrm{d}}

\newcommand{\ZZ}{\mathbb Z}

\newcommand{\ratc}{A}

\newbox\charbox
\newbox\slabox
\def\s#1{{      
        \setbox\charbox=\hbox{$#1$}
        \setbox\slabox=\hbox{$/$}
        \dimen\charbox=\ht\slabox
        \advance\dimen\charbox by -\dp\slabox
        \advance\dimen\charbox by -\ht\charbox
        \advance\dimen\charbox by \dp\charbox
        \divide\dimen\charbox by 2
        \raise-\dimen\charbox\hbox to \wd\charbox{\hss/\hss}
        \llap{$#1$}
}}

\begin{document}

 {\flushright  
 UUITP-37/22\\}

\begin{center}

{\bf {\LARGE \sc Modular graph forms from\\[2mm] equivariant iterated Eisenstein integrals}}

\vspace{6mm}
\normalsize
{\large  Daniele Dorigoni$^1$,
Mehregan Doroudiani$^2$,
Joshua Drewitt$^3$,
Martijn Hidding$^4$, \\
Axel Kleinschmidt$^{2,5}$,
Nils Matthes$^6$,
Oliver Schlotterer$^{4}$ 
and Bram Verbeek$^{4}$}

\vspace{10mm}
${}^1${\it Centre for Particle Theory \& Department of Mathematical Sciences\\
Durham University, Lower Mountjoy, Stockton Road, Durham DH1 3LE, UK}
\vskip 1 em
${}^2${\it Max-Planck-Institut f\"{u}r Gravitationsphysik (Albert-Einstein-Institut)\\
Am M\"{u}hlenberg 1, 14476 Potsdam, Germany}
\vskip 1 em
${}^3${\it University of Nottingham, University Park, Nottingham NG7 2RD, UK}
\vskip 1 em
${}^4${\it Department of Physics and Astronomy, Uppsala University, 75108 Uppsala, Sweden}
\vskip 1 em
${}^5${\it International Solvay Institutes\\
ULB-Campus Plaine CP231, 1050 Brussels, Belgium}
\vskip 1 em
${}^6${\it Department of Mathematical Sciences, University of Copenhagen\\ Universitetsparken 5,
2100 Copenhagen {\O}, Denmark}

\vspace{10mm}

\hrule

\vspace{5mm}

\begin{tabular}{p{14cm}}
The low-energy expansion of closed-string scattering amplitudes at genus one introduces infinite families of non-holomorphic modular forms called {\it modular graph forms}. Their differential and number-theoretic properties motivated Brown's alternative construction of non-holomorphic modular forms in the recent mathematics literature from so-called {\it equivariant iterated Eisenstein integrals}. In this work, we provide the first validations beyond depth one  of Brown's conjecture that equivariant iterated Eisenstein integrals contain modular graph forms. Apart from a variety of examples at depth two and three, we spell out the systematics of the dictionary and make certain elements of Brown's construction fully explicit to all orders.
\end{tabular}

\vspace{6mm}
\hrule

\end{center}

\thispagestyle{empty}

\newpage
\setcounter{page}{1}

\setcounter{tocdepth}{2}
\tableofcontents

\section{Introduction}
\label{sec:1}

Scattering amplitudes in string theories have turned out to be rewarding laboratories to
encounter deep mathematical structures in a physics context. Already at tree level, the multiple
zeta values (MZVs) in the low-energy expansion of string amplitudes 
furnish elegant physics applications of motivic MZVs \cite{Schlotterer:2012ny}, the Drinfeld
associator \cite{Terasoma, Drummond:2013vz, Broedel:2013aza, Kaderli:2019dny} and 
single-valued MZVs \cite{Schlotterer:2012ny, Stieberger:2013wea, Stieberger:2014hba, Schlotterer:2018abc, Vanhove:2018elu, Brown:2019wna}. At one loop,
the common themes of string amplitudes, number theory, and algebraic geometry are centered 
around elliptic polylogarithms \cite{BrownLev, Enriquez:Emzv, Broedel:2014vla, Broedel:2017jdo} 
and non-holomorphic modular forms.
 The latter arise from integrating over closed-string insertions on a torus
\cite{Green:1999pv, Green:2008uj, DHoker:2015gmr, Gerken:2018jrq} 
and were dubbed {\it modular graph forms} (MGFs) 
\cite{DHoker:2015wxz, DHoker:2016mwo}. We refer to \cite{Gerken:review}
for an overview of MGFs as of fall 2020 and to 
\cite{Berkovits:2022ivl, Dorigoni:2022iem, DHoker:2022dxx} for 
recent discussions in a broader context.

The remarkable properties of MGFs attracted considerable 
attention among mathematicians, for instance, their intricate network of algebraic
and differential relations \cite{DHoker:2015gmr, DHoker:2015sve, DHoker:2016mwo, 
DHoker:2016quv, Basu:2016kli, Gerken:2018zcy}
 or the appearance of (conjecturally single-valued) MZVs 
in their Fourier expansion \cite{Zerbini:2015rss, DHoker:2015wxz, 
DHoker:2017zhq, Panzertalk, DHoker:2019xef, Zagier:2019eus, Vanhove:2020qtt}. In particular, 
the advent of MGFs inspired Brown's construction of non-holomorphic modular forms
from iterated integrals of holomorphic Eisenstein series and their complex conjugates
\cite{Brown:mmv, Brown:2017qwo, Brown:2017qwo2}.
More specifically, Brown's infinite families of non-holomorphic modular forms arise
as expansion coefficients of certain generating series dubbed {\it equivariant iterated 
Eisenstein integrals} (EIEIs) and are conjectured to contain MGFs.

Brown's EIEIs are built from two implicitly defined
ingredients \cite{Brown:mmv, Brown:2017qwo2}: (i) a generating series $b^{\rm sv}$ of single-valued 
MZVs and (ii) a change of alphabet $\phi^{\rm sv}$
for the bookkeeping variables of antiholomorphic iterated Eisenstein integrals akin
to the construction of single-valued polylogarithms at genus zero in \cite{Brown:2004ugm}.
This close relation to the theory of single-valued polylogarithms is just one reason Brown's non-holomorphic modular forms are of great interest across several communities. Other reasons include their potential application to solving arithmetic problems involving periods and their link to the study of universal mixed motives. Further recent evidence for their arithmetic significance was also provided by~\cite{Diamantis:2019}, where classical and important number-theoretic objects, such as period polynomials, were associated with the space of such non-holomorphic modular forms. Despite this interest, the explicit form of Brown's EIEIs beyond depth one is essentially uncharted territory, apart from the simplest contributions of (i) to non-holomorphic Eisenstein series.

An explicit example of how MGFs at depth two relate to Brown's non-holomorphic modular forms was given first in~\cite{Brown:2017qwo} and then further investigated in~\cite{Drewitt:2021}, where the depth-three case was also briefly discussed. However, the equations determining (i) and (ii) have not yet been solved to the orders that probe the generic properties of Brown's non-holomorphic modular forms or their connection to MGFs at depth $\geq 2$. 

An alternative way of reducing MGFs to iterated Eisenstein integrals and their complex
conjugates was initiated by certain generating series of closed-string genus-one integrals
which contain all convergent MGFs in their low-energy expansion \cite{Gerken:2019cxz}. 
The explicitly-known reality properties
and first-order differential equations in $\tau$ of these closed-string integrals imply that
MGFs can be uniquely expressed in terms of real-analytic iterated Eisenstein integrals 
$\beta^{\rm sv}$ \cite{Gerken:2020yii}. These representations of MGFs expose both the 
entirety of their relations over rational combinations of MZVs and their expansion around the cusp.
The dictionary between MGFs and $\beta^{\rm sv}$ at depth two is known from
\cite{Dorigoni:2021jfr, Dorigoni:2021ngn}, and we will report on generalizations to depth three in future work \cite{depth3paper}.

In this work, we relate the organization of iterated Eisenstein integrals via $\beta^{\rm sv}$
to Brown's construction of non-holomorphic modular forms and confirm his conjecture
that EIEIs contain MGFs in a variety of cases. Moreover,
we present an all-order proposal for the explicit form of the aforementioned change of alphabet $\phi^{\rm sv}$
in terms of commutator relations between certain derivations $\{\epsilon_{2k+2},z_{2k+1}\}$ with $k \in \mathbb{N}$ which are well known in the mathematics literature. These derivations originated in pioneering work of Ihara \cite{Ihara:1990}, where deep connections to Grothendieck's theory of motives (and particularly to Deligne's motivic fundamental group of the projective line minus three points \cite{Deligne:1987}) were found. The link between quadratic relations among these derivations and modular forms was first described by Ihara--Takao 
(see the summary in \cite{IharaTakao:alt}), then studied in detail by Tsunogai \cite{Tsunogai}, Goncharov \cite{Gonchtalk}, Gangl--Kaneko--Zagier \cite{GKZ:2006}, Schneps \cite{Schneps:2006}, Pollack \cite{Pollack}, Baumard--Schneps \cite{BaumardSchneps:2015}, Hain--Matsumoto \cite{hain_matsumoto_2020} and Brown \cite{Brown:Anatomy, Brown:depth3}.

Our results close a notorious gap between the recent physics and mathematics literature
and may pave the way for deducing properties of closed-string amplitudes from powerful 
theorems in algebraic geometry and number theory.

\subsection*{Outline}

This work is organized as follows: We start by reviewing the basics of MGFs and
their iterated-integral building blocks from the string-theory literature in section \ref{sec:2}.
Section \ref{sec.3} then introduces new results on these string-theory motivated
building blocks with a focus on the construction of non-holomorphic modular forms.
This description of MGFs is compared with Brown's EIEIs in section \ref{sec:4}: We first make
a connection with Brown's generating series in terms of Tsunogai's derivations
\cite{Brown:2017qwo2} in subsections \ref{sec:4.1} and \ref{sec:4.2}. The role of Brown's 
equivariant double integrals including depth-one integrals of holomorphic cusp forms
\cite{Brown:mmv,Brown:2017qwo} is then discussed in section \ref{sec:4.3}.

\section{Basics}
\label{sec:2}

We start by reviewing selected aspects of MGFs and setting up the notation
to connect with Brown's work in later sections.

\subsection{Modular graph forms}
\label{sec:2.1}

The original definition of MGFs as integrals over marked points on a torus can be applied to any labeled directed graph~\cite{DHoker:2015wxz,DHoker:2016mwo}. In lieu of the full definition, we restrict ourselves to simple instances that suffice to illustrate our construction. For the case of a dihedral graph, the definition of MGFs reduces to the following nested
sums over discrete torus momenta $p_1,\ldots,p_R$~\cite{DHoker:2016mwo} 
\beq
\cplus{a_1   &\ldots &a_R}{b_1  &\ldots &b_R} = 
\bigg( \prod_{j=1}^R \frac{ (\Im \tau)^{a_j} }{ \pi^{b_j}} \bigg) 
\sum_{p_1,\ldots,p_R \in \Lambda'} \frac{\delta(p_1{+}\ldots{+}p_R) }{p_1^{a_1}\bar p_1^{b_1}
\ldots p_R^{a_R}\bar p_R^{b_R}}\, .
\label{MGFtoBR01}
\eeq
MGFs depend non-holomorphically on the modular parameter $\tau \in \mathbb C$ of a
torus with $\Im \tau>0$. The sums over lattice momenta
\beq
p_j  \in \Lambda' \, , \ \ \ \ \Lambda'  = (\ZZ \tau {+}\ZZ) \setminus \{0\}
\label{MGFtoBR02}
\eeq
converge absolutely if their exponents $a_j ,b_j \in \mathbb Z$ obey 
$a_i{+}b_i{+}a_{j}{+}b_j \geq 3$ for all $1{\leq} i{<}j{\leq} R$. The conventions for the
normalization factor $ (\Im \tau)^{a_j}   \pi^{-b_j}$ in (\ref{MGFtoBR01}) ensure that MGFs transform
with purely antiholomorphic modular weight $(0,\sum_{j=1}^R(b_j{-}a_j))$
under the modular group $SL(2,\mathbb Z)$,
\beq
\cplusno{a_1   &\ldots &a_R}{b_1  &\ldots &b_R}\! \big( \tfrac{ a \tau{+} b }{c\tau{+}d} \big)
= \bigg( \prod_{j=1}^R (c \bar \tau{+}d)^{b_j-a_j} \bigg) \cplus{a_1   &\ldots &a_R}{b_1  &\ldots &b_R}
\, , \ \ \ \
\big(\begin{smallmatrix} a&b \\ c &d \end{smallmatrix}\big) \in SL(2,\mathbb Z)\, .
\eeq
Expressions similar to (\ref{MGFtoBR01}) for  more general
topologies are for instance discussed in \cite{DHoker:2016mwo, Gerken:2020aju}.

The simplest non-vanishing
examples of MGFs are non-holomorphic Eisenstein series
\beq
{\rm E}_k(\tau) = \cplus{k   &0}{k  &0}
= \bigg( \frac{ \Im \tau}{\pi}\bigg)^k \sum_{p \in \Lambda'} \frac{1}{|p |^{2k}}
\, , \ \ \ \ k \geq 2
\label{MGFtoBR03}
\eeq
and their $\tau,\bar \tau$-derivatives. Infinitely many instances of (\ref{MGFtoBR01})
with $R\geq 3$ lattice momenta obey non-trivial relations 
such as \cite{DHoker:2015gmr, DHoker:2015sve}
\begin{align}
 \cplus{1 &1  &1}{1  &1 &1}
&= {\rm E}_3(\tau) + \zeta_3 \, ,
\label{MGFtoBR04} \\
 \cplus{1 &1  &1 &1}{1  &1 &1 &1}
 &=
 24  \cplus{2 &1  &1}{2 &1 &1} - 18  {\rm E}_4(\tau) + 3 {\rm E}_2(\tau)^2\, ,
\notag
\end{align}
which are mysterious from the lattice-sum representations of MGFs
but are exposed by the iterated-integral representations below.
A datamine of relations can be found within the {\sc Mathematica} package \cite{Gerken:2020aju}.
Relations and expansions of MGFs around the cusp $\tau \rightarrow i\infty$ 
introduce (conjecturally single-valued \cite{Zerbini:2015rss, DHoker:2015gmr}) MZVs
\cite{DHoker:2016quv, DHoker:2017zhq, Panzertalk, 
DHoker:2019xef, Zagier:2019eus, Vanhove:2020qtt}
\beq
\zeta_{n_1,n_2,\ldots,n_r}= \! \! \!  \sum_{0<k_1<k_2<\ldots < k_r} \frac{1}{k_1^{n_1}  k_2^{n_2} 
\ldots k_r^{n_r} } \, , \ \ \ \ n_r\geq 2
\label{MGFtoBR05}
\eeq
of weight $n_1{+}n_2{+}\ldots {+}n_r$ and depth $r$.

\subsection{Real-analytic iterated Eisenstein integrals}
\label{sec:2.2}

Non-holomorphic Eisenstein series (\ref{MGFtoBR03}) can
be written as
\beq
{\rm E}_k(\tau) = \frac{(2k{-}1)!}{ (k{-}1)!^2} \bigg\{
{-}\bsvBR{k-1}{2k}{\tau} 
+ \frac{2 \zeta_{2k-1} }{(2k{-}1) (4y)^{k-1} }\bigg\}\,,
\label{MGFtoBR06}
\eeq
involving the real-analytic depth-one integral \cite{Gerken:2020yii, Dorigoni:2021jfr}
\begin{align}
\bsvBR{j}{k}{\tau} &= \frac{1 }{ 2\pi i  }\bigg\{
\int_{\tau}^{i\infty}  \dd \tau_1  \bigg(\frac{\tau{-}\tau_1}{4y}\bigg)^{k-2-j}
 (\bar \tau{-}\tau_1)^j {\rm G}_k(\tau_1) \notag \\
&\quad - \! \int_{\bar \tau}^{-i\infty} \! \dd \bar \tau_1 \, 
 \bigg(\frac{\tau{-} \bar \tau_1}{4y}\bigg)^{k-2-j} 
 (\bar \tau{-}\bar \tau_1)^j \overline{{\rm G}_k(\tau_1) } \bigg\}\,,
\label{MGFtoBR07}
\end{align}
with $y= \pi \Im\tau$, holomorphic Eisenstein series,
\beq
{\rm G}_k(\tau) = (\Im \tau)^{-k}\, \cplus{k   &0}{0  &0}\, , \ \ \ \ k\geq 4
\label{MGFtoBR08}
\eeq
and tangential-basepoint regularization of the endpoint divergence
at $\tau_1 \rightarrow i\infty$ \cite{Brown:mmv}. Earlier discussions of
iterated-integral representations of non-holomorphic Eisenstein series
can for instance be found in \cite{Ganglzagier, DHoker:2015gmr}.
We emphasize that the integrals in (\ref{MGFtoBR07}) and similar iterated
integrals below are homotopy invariant: In spite of the appearance of  
$\tau$ and $\bar \tau$ in both lines, the integration variable $\tau_1$ ($\bar \tau_1$)
only appears holomorphically (antiholomorphically) along with $\dd \tau_1$ ($\dd  \bar \tau_1$).

Generic MGFs (\ref{MGFtoBR01}) can be uniquely\footnote{Uniqueness follows
from the linear-independence results of \cite{Nilsnewarticle} on
holomorphic iterated Eisenstein integrals.} represented via 
higher-depth generalizations of the real-analytic Eisenstein integral
(\ref{MGFtoBR07}) which are constructed from kernels
\begin{align}
\omplus{j}{k}{\tau,\tau_1} &= \frac{  \dd \tau_1 }{2\pi i }  \bigg(\frac{\tau{-}\tau_1}{4y}\bigg)^{k-2-j}
 (\bar \tau{-}\tau_1)^j {\rm G}_k(\tau_1)\, ,
 \label{MGFtoBR09} \\
 \omminus{j}{k}{\tau,\tau_1} &={-} \frac{  \dd \bar \tau_1 }{2\pi i }  \bigg(\frac{\tau{-}\bar \tau_1}{4y}\bigg)^{k-2-j}
 (\bar \tau{-}\bar \tau_1)^j \overline{{\rm G}_k(\tau_1)}\, ,
 \notag
\end{align}
with $k\geq 4$ even and $0\leq j\leq k{-}2$.
In terms of these kernels, the depth-one expression~\eqref{MGFtoBR07} simply reads
\begin{align}
\label{MGFtoBR10a}
\bsvBR{j}{k}{\tau} = \int_{\tau}^{i \infty} \omplus{j}{k}{\tau,\tau_1} +  \int_{\bar\tau}^{-i \infty} \omminus{j}{k}{\tau,\tau_1} \, .
\end{align}
Starting from the depth-two instance \cite{Gerken:2020yii, Dorigoni:2021jfr}
\begin{align}
&\bsvBR{j_1 &j_2}{k_1 &k_2}{\tau}  =
\int\limits_{\tau}^{i\infty}  \omplus{j_2}{k_2}{\tau,\tau_2} \int\limits_{\tau_2}^{i\infty} \omplus{j_1}{k_1}{\tau,\tau_1}  
 +  \int\limits_{ \tau}^{i\infty}  \omplus{j_2}{k_2}{\tau,\tau_2} 
\int\limits_{\bar \tau}^{-i\infty}   \omminus{j_1}{k_1}{\tau,\tau_1}
\label{MGFtoBR10} \\
 &\ \ + \int\limits_{\bar \tau}^{-i\infty}   \omminus{j_1}{k_1}{\tau,\tau_1} \int\limits_{\bar{\tau}_1}^{-i\infty}  \omminus{j_2}{k_2}{\tau,\tau_2} 
 + \sum_{p_1=0}^{k_1-2-j_1}
\sum_{p_2=0}^{k_2-2-j_2} \frac{ 
{ k_1{-}2 {-}j_1 \choose p_1} { k_2{-}2 {-}j_2 \choose p_2}
}{(4y)^{p_1+p_2}} 
\overline{\alphaBR{ j_1{+}p_1 &j_2{+}p_2 }{ k_1 &k_2 }{\tau} }\, ,
\notag
\end{align}
the real-analytic $\beta^{\rm sv}$ involve antiholomorphic
building blocks $\overline{\alphaBR{ \ldots }{ \ldots }{\tau} }$
featuring MZVs in each term which resemble the admixtures of MZVs
to single-valued polylogarithms at genus zero in \cite{Brown:2004ugm}.
Just like the $\beta^{\rm sv}$ at arbitrary depth, the
$\overline{\alphaBR{ \ldots }{ \ldots }{\tau} }$
are invariant under the modular $T$-transformation
$\tau \rightarrow \tau{+}1$, see (\ref{closedform}) below for an all-order formula at depth two. 
Note the reversal of the ordering of labels for the $\omega_-$ kernels 
in the definition~\eqref{MGFtoBR10}.

The kernels (\ref{MGFtoBR09}) lead to specific linear combinations of
Brown's iterated Eisenstein integrals over kernels $\tau_1^j {\rm G}_k(\tau_1)$
with $k\geq 4$ and $0\leq j \leq k{-}2$ \cite{Brown:mmv}.
Their accompanying polynomials in $\tau$ and $\bar \tau$ can be understood
from their generating function
\begin{align}
&\sum_{j=0}^{k-2} \ompm{j}{k}{\tau,\tau_1}(X{-}\tau Y)^{j}(X-\bar \tau Y)^{k-j-2} {k{-}2\choose j} \frac{1}{(-4y)^j}
\notag \\
&= \left\{ \begin{array}{rl}  
\frac{\dd  \tau_1}{(2\pi i )^{k-1}} (X{-} \tau_1Y )^{k-2} \, {\rm G}_k(\tau_1)
&: \ \omega_+\, , \\
\\
-\frac{\dd \bar \tau_1}{(2\pi i)^{k-1}} (X{-}\bar \tau_1Y )^{k-2} \,  \overline{ {\rm G}_k(\tau_1)} 
&: \ \omega_-\, ,
\end{array} \right.
\label{omegavsXY}
\end{align}
which translate into the kernels $(X{-} \tau_1Y )^{k-2} \, {\rm G}_k(\tau_1)$
of Brown's EIEIs \cite{Brown:2017qwo}:
Once the commutative bookkeeping variables $X,Y$ are taken to transform
as a vector under $ \big(\begin{smallmatrix} a&b \\ c &d \end{smallmatrix}\big) \in SL(2,\mathbb Z)$ 
according to $(X,Y) \rightarrow (aX{+}bY,cX{+}dY)$, both cases of (\ref{omegavsXY}) are
modular invariant.

\subsection{Higher-depth generalization}
\label{sec:2.3}

With a capital-letter notation $P$ for words in the 
composite letters $\begin{smallmatrix} j \\ k \end{smallmatrix}$ 
of the kernels (\ref{MGFtoBR09}), we can
write the higher-depth generalization of~\eqref{MGFtoBR10a} and  (\ref{MGFtoBR10}) as
\begin{align}
\beta^{\rm sv}[P;\tau] = \sum_{P=XYZ} 
\overline{ \kappa[X ;\tau] } \beta_-[Y^t;\tau]   \beta_+[Z;\tau] \, ,
\label{MGFtoBR12}
\end{align}
where $Y^t$ is obtained from $Y$ by reversing the order of its composite letters
(e.g.\ $( \begin{smallmatrix} j_1 & j_2 \\ k_1 &k_2 \end{smallmatrix} )^t=
(\begin{smallmatrix}  j_2 & j_1 \\ k_2 &k_1 \end{smallmatrix}$)), 
and the sum over deconcatenations of $P$ into $XYZ$ includes empty words $X,Y,Z$
with $\overline{ \kappa[\emptyset ;\tau] } = \beta_-[\emptyset ;\tau]  = \beta_+[\emptyset ;\tau] =1$.
We use a $\beta_{\pm}$-notation to separate the contributions from
holomorphic and antiholomorphic Eisenstein series
\begin{align}
 \bplus{j_1 &j_2& \ldots &j_\ell}{k_1 &k_2 &\ldots &k_\ell}{\tau} &= \int_\tau^{i\infty} 
 \omplus{j_\ell}{k_\ell}{\tau,\tau_\ell}  
  \ldots  \int_{\tau_3}^{i\infty}\omplus{j_2}{k_2}{\tau,\tau_2}\int_{\tau_2}^{i\infty} \omplus{j_1}{k_1}{\tau,\tau_1}\,,
\label{MGFtoBR13}
\\
\bminus{j_1 &j_2 &\ldots &j_\ell}{k_1 &k_2 &\ldots &k_\ell}{\tau} &= \int  _{\bar \tau}^{-i\infty} 
 \omminus{j_\ell}{k_\ell}{\tau,\tau_\ell}  
  \ldots \int_{{\bar{\tau}}_3}^{-i\infty} \omminus{j_2}{k_2}{\tau,\tau_2}\int_{{\bar{\tau}}_2}^{-i\infty}\omminus{j_1}{k_1}{\tau,\tau_1}\, .
  \notag
\end{align}
Moreover, the composition of antiholomorphic $\overline{\alphaBR{ \ldots }{ \ldots }{\tau} }$ 
in the last line of (\ref{MGFtoBR10}) generalizes to multiple sums over $p_i$,
\begin{align}
\overline{ \kappaBR{\ldots &j_i &\ldots}{\ldots &k_i &\ldots}{\tau} }
= \sum_{p_i=0}^{k_i-2-j_i} \frac{ 
{ k_i{-}2 {-}j_i \choose p_i} 
}{(4y)^{p_i}} 
\overline{\alphaBR{ \ldots &j_i{+}p_i &\ldots}{ \ldots &k_i &\ldots }{\tau} }\, ,
\label{MGFtoBR14}
\end{align}
where one can infer the vanishing of their depth-one instances 
$\overline{\kappaBR{ j }{ k}{\tau}} =  \overline{\alphaBR{ j }{ k}{\tau}}= 0$ 
from (\ref{MGFtoBR07}). Antiholomorphicity of the $\overline{\alphaBR{ \ldots }{ \ldots }{\tau} }$
and the composition rule (\ref{MGFtoBR12}) imply a simple form of the holomorphic differential 
equations \cite{Gerken:2020yii}
\begin{align}
2\pi i(\tau{-}\bar \tau)^2 \partial_\tau \bsvBR{j_1 &\ldots &j_\ell}{k_1 &\ldots &k_\ell}{\tau}
&= \sum_{i=1}^\ell (k_i{-}j_i{-}2)
\bsvBR{j_1 &\ldots &j_i{+}1 &\ldots &j_\ell}{k_1 &\ldots
&k_i &\ldots &k_\ell}{\tau}
\label{bsvdiff} \\
&\quad - \delta_{j_\ell,k_\ell-2} (\tau{-}\bar \tau)^{k_\ell} {\rm G}_{k_\ell}(\tau) 
\bsvBR{j_1 &\ldots &j_{\ell-1}}{k_1 &\ldots &k_{\ell-1}}{\tau}\, ,
\notag
\end{align}
while $\partial_{\bar \tau}$-derivatives are in general more complicated and sensitive
to the expressions for $\overline{\alphaBR{ \ldots }{ \ldots }{\tau} }$.

Note that all the $\overline{ \kappa[X ;\tau] } $ 
and the combinations $ \sum_{Q=YZ} \beta_-[Y^t;\tau]   \beta_+[Z;\tau] $ 
in (\ref{MGFtoBR12}) at fixed $Q$ are separately invariant under $T: \ \tau \rightarrow \tau{+}1$.
Modular $S$-transformations $\tau \rightarrow -\frac{1}{\tau}$ in turn mix
$\beta^{\rm sv}$ of different depths \cite{Gerken:2020yii, Dorigoni:2021ngn} with rational 
functions of $\tau,\bar \tau$, MZVs and more general multiple modular values 
\cite{Brown:mmv, Brown2019} in their coefficients.
In other words, individual $\beta^{\rm sv}$ do not have good modular properties, unlike MGFs, which are specific linear combinations of $\beta^{\rm sv}$ of different depths with $y$-dependent coefficients.
One of the main aims of this paper is to give a more direct characterization of these linear combinations (see section~\ref{sec.3}) and to relate them to Brown's construction
in section \ref{sec:4}.

\subsection{Generating series of modular graph forms}
\label{sec:2.4}

The entirety of convergent MGFs which do not simplify under holomorphic subgraph
reduction \cite{DHoker:2016mwo, Gerken:2018zcy} are embedded
into generating series of closed-string integrals over $n\geq 2$ marked points
on a torus \cite{Gerken:2019cxz}. Their KZB-type differential equations in $\tau$
have been solved through a generating series in $\beta^{\rm sv}$ \cite{Gerken:2020yii}
\begin{align}
Y^\tau_{\vec\eta} = 
\sum_P 
R_{\vec{\eta}}(
\epsilon[P] ) \beta^{\rm sv}[P;\tau] \exp\bigg( {-}\frac{ R_{\vec{\eta}}(\epsilon_0) }{4y} \bigg) \widehat Y^{i\infty}_{\vec{\eta}} \, ,
\label{MGFtoBR21}
\end{align}
where we recall $y=\pi \Im \tau$, and the sum over $P$ comprises 
all words $P = \begin{smallmatrix} j_1 &\ldots &j_\ell \\ 
k_1 &\ldots &k_\ell \end{smallmatrix}$ of length $\ell \geq 0$ with $k_i\geq 4$ even and
$0\leq j_i \leq k_i{-}2$. The coefficients
\begin{align}
&\coeffs{  j_1 &j_2 &\ldots &j_\ell }{k_1 &k_2 &\ldots &k_\ell }=
\bigg( \prod_{i=1}^\ell \frac{ (-1)^{j_i}(k_i{-}1) }{(k_i{-}j_i{-}2)!} \bigg)
\epsilon^{(k_\ell-2-j_\ell)}_{k_\ell} \cdots \epsilon^{(k_2-2-j_2)}_{k_2}\epsilon^{(k_1-2-j_1)}_{k_1}
\label{MGFtoBR22} 
\end{align}
with the shorthand
\beq
\epsilon_k^{(j)} = {\rm ad}_{\epsilon_0}^j(\epsilon_k)
 \label{MGFtoBR43}
\eeq
and the exponential in (\ref{MGFtoBR21}) involve certain $(n{-}1)!\times (n{-}1)!$
matrix valued operators $R_{\vec{\eta}}(\epsilon_{k\in 2\mathbb N_0})$ which are conjectured
\cite{Gerken:2019cxz, Gerken:2020yii} to furnish matrix representations
$R_{\vec{\eta}}(\cdot)$ of Tsunogai's
derivation algebra $\{ \epsilon_{ k \in 2\mathbb N_0}\}$ \cite{Tsunogai}
with $R_{\vec{\eta}}(\epsilon_{2})=0$. The notation
$R_{\vec{\eta}}(\epsilon[P] ) $ in (\ref{MGFtoBR21}) instructs us to replace all the $\epsilon_k$ in 
(\ref{MGFtoBR22}) by $R_{\vec{\eta}}(\epsilon_k)$. The conjecture is supported
by a huge number of checks that the relations of the derivation algebra \cite{Tsunogai, LNT, Pollack}
such as
\begin{align}
 0 &= \epsilon^{(k-1)}_k 
 \, , \ \ \ \ \ \ k\geq 4 \ {\rm even} \, , \notag \\
0 &= [\epsilon_4,\epsilon_{10}] - 3 [\epsilon_6,\epsilon_{8}]\, ,
 \label{MGFtoBR23}
\\
0&=
- 462 \big[\epsilon_4,[\epsilon_4,\epsilon_8] \big]
- 1725 \big[\epsilon_6,[\epsilon_6,\epsilon_4] \big]
- 280 [\epsilon_8^{\phantom{(}},\epsilon_8^{(1)}]
\notag \\
&\quad 
+ 125 [\epsilon^{\phantom{(}}_6,\epsilon_{10}^{(1)}]
+ 250 [\epsilon_{10}^{\phantom{(}},\epsilon_6^{(1)}]
-80[\epsilon_{12}^{\phantom{(}},\epsilon_4^{(1)}]
- 16 [\epsilon_4^{\phantom{(}},\epsilon_{12}^{(1)}]
\notag
\end{align}
are preserved in passing to the matrix-valued operators $\epsilon_k \rightarrow 
R_{\vec{\eta}}(\epsilon_k)$.

Finally, the quantity $\widehat Y^{i\infty}_{\vec{\eta}}$ in (\ref{MGFtoBR21}) 
accounts for the $\tau \rightarrow i\infty$ degenerations of the genus-one
integrals \cite{Gerken:2019cxz, Gerken:2020yii}. Its matrix entries are $\tau$-independent
Laurent series in the bookkeeping variables $s_{ij},\eta_j$ of the reference
that the operators $R_{\vec{\eta}}(\epsilon_k)$ act on, with (conjecturally single-valued)
MZVs in their coefficients (the explicit form at $n=2$ can be generated
from (4.2) of \cite{Gerken:2020yii}).

\subsection{Examples}
\label{sec:2.5}

The contributions of $
\exp\big( {-}\frac{ R_{\vec{\eta}}(\epsilon_0) }{4y} \big) \widehat Y^{i\infty}_{\vec{\eta}}$ 
to the generating series (\ref{MGFtoBR21}) ensure that $\beta^{\rm sv}$ of
different depths are combined into modular forms. At depth one,
the formula is~\cite{Gerken:2020yii}
\begin{align}
\label{MGFtoBR25}
\cplus{a &0}{b &0} = - \frac{(2i)^{b-a} (a{+}b{-}1)! }{(a{-}1)!(b{-}1)!} \left(\bsvBR{a-1}{a+b}{\tau}- \frac{2 \zeta_{a+b-1} }{(a{+}b{-}1) (4y)^{b-1}}\right)\, ,
\end{align}
where $a,b\geq 1$, and $a{+}b\geq 4$ is an even integer.

Higher-depth instances of $\beta^{\rm sv}$ in (\ref{MGFtoBR10})
and (\ref{MGFtoBR12}) occur in MGFs with three and more columns, such 
as \cite{Gerken:2020yii}  
\begin{align}
\cplus{2 &1 &1}{2 &1 &1} &=  
-126 \bsvBR{3}{8}{\tau} - 18 \bsvBR{2& 0}{4& 4}{\tau} + 
 12  \zeta_3 \bsvBR{0}{4}{\tau}  + \frac{5 \zeta_5}{ 12 y} 
 - \frac{ \zeta_3^2}{4 y^2}  + \frac{ 9 \zeta_7}{16 y^3} \, ,
\notag \\
 \cplus{3 &2 &1}{1 &2 &1} &=  
\frac{ 279}{2} \bsvBR{5}{10}{\tau} + 30 \bsvBR{3& 1}{6& 4}{\tau} + 
 \frac{15}{2} \bsvBR{4& 0}{6& 4}{\tau} \label{MGFtoBR26} \\
 &\quad   - 3 \zeta_5 \bsvBR{0}{4}{\tau} 
  - \frac{ 3 \zeta_5}{y} \bsvBR{1}{4}{\tau}  
  - \frac{ 7 \zeta_7}{48 y} 
  + \frac{ 5 \zeta_3 \zeta_5}{16 y^2} 
   - \frac{ 31 \zeta_9}{64 y^3}
\notag
\end{align} \normalsize
as well as \cite{Gerken:2020yii} 
\begin{align}
&2i \Im \cplus{0 &1 &2 &2}{1 &1 &0 &3} =
60 \bsvBR{0& 3}{4& 6}{\tau} - 270 \bsvBR{1& 2}{4& 6}{\tau}   - 
 60 \bsvBR{1& 2}{6& 4}{\tau} + 390 \bsvBR{2& 1}{4& 6}{\tau}  \notag \\
&\quad + 
 270 \bsvBR{2& 1}{6& 4}{\tau}  - 390 \bsvBR{3& 0}{6& 4}{\tau} - 
 3 \zeta_3 \bsvBR{1}{4}{\tau} 
 + \frac{39  \zeta_5}{y} \bsvBR{0}{4}{\tau} 
   - \frac{27  \zeta_5}{4 y^2} \bsvBR{1}{4}{\tau}  \label{MGFtoBR27} \\
&\quad
  + \frac{3 \zeta_5}{ 8 y^3} \bsvBR{2}{4}{\tau} 
  - 260 \zeta_3 \bsvBR{1}{6}{\tau} 
  +  \frac{ 45 \zeta_3}{y} \bsvBR{2}{6}{\tau}  
 - \frac{5 \zeta_3}{ 2 y^2} \bsvBR{3}{6}{\tau}  - \frac{13 \zeta_5}{120}    .
\notag
\end{align} 
As a common theme of (\ref{MGFtoBR25}) to (\ref{MGFtoBR27}),
$\beta^{\rm sv} \big[ \begin{smallmatrix} j_1 &\ldots &j_\ell \\
k_1 &\ldots &k_\ell \end{smallmatrix};\tau \big]$ 
are completed to modular forms of weight $(0,\sum_{i=1}^\ell (k_i{-}2{-}2j_i))$ by 
adding lower-depth terms with $\mathbb Q[y^{-1}]$-linear combinations 
of MZVs in their coefficients. Note that only non-positive powers of $y$
can arise from the expansion of the exponential in (\ref{MGFtoBR21}).

The antiholomorphic $\overline{\alphaBR{ \ldots }{ \ldots }{\tau} }$ 
in (\ref{MGFtoBR10}) and at higher depth
are determined by the reality properties
\beq
\overline{ \cplus{a_1 &\ldots &a_R}{b_1 &\ldots &b_R} } =
\bigg( \prod_{r=1}^R y^{ a_r-b_r}\bigg) \cplus{b_1 &\ldots &b_R}{a_1 &\ldots &a_R} \, ,
 \label{MGFtoBR28}
 \eeq
once a $\beta^{\rm sv}$ representation of all basis
MGFs at given weight $\sum_{r=1}^R (a_r{+}b_r)$ is
available. Examples and general formulae at depth two
will be given in the next section (also see appendix \ref{app.A}).

\section{Modular forms from $\beta^{\rm sv}$}
\label{sec.3}

In this section, we describe the structure of modular forms
constructed from $\beta^{\rm sv}$ in preparation for the matching with
Brown's EIEIs. These modular versions will be denoted by $\beta^{\rm eqv}$ by slight abuse 
of notation since they are not equivariant but simply standard non-holomorphic modular forms. 
Since the  $\beta^{\rm sv}$ and their $y$-dependent coefficients
are invariant under $T$-transformations, we only need to ensure 
that the $S$-modular transformation law is correct.

The answer at depth one is straightforward to obtain since 
the modular $S$-transformation of (\ref{MGFtoBR07}) involves a depth-zero
term which cancels from the combination
\begin{align}
\beqv{j}{k}{\tau} = \bsvBR{j}{k}{\tau}- \frac{2 \zeta_{k-1} }{(k{-}1) (4y)^{k-j-2}} \, .
\label{MGFtoBR24}
\end{align}
This illustrates the fact that the $\beta^{\rm sv}$ have to be 
completed by lower-depth terms with non-positive powers of $y$ and (single-valued) 
MZVs.

\subsection{Construction and properties of $\beta^{\rm eqv}$}
\label{sec:3.1}

Modular versions of $\beta^{\rm sv}[P;\tau]$ as seen in (\ref{MGFtoBR25}) to (\ref{MGFtoBR27})
can be parameterized by constants $\ccsvpure[P]$ associated with the same type of words $P$
in composite letters $\begin{smallmatrix} j \\ k \end{smallmatrix}$ with 
$k\geq 4$ even and $0\leq j \leq k{-}2$,
\begin{align}
\beqv{j}{k}{\tau} &= \bsvBR{j}{k}{\tau} + \sum_{p=0}^{k-j-2} \frac{ {k-j-2 \choose p} }{(4y)^p} \ccsv{j+p}{k} \, , \notag \\
\beqv{j_1 &j_2}{k_1 &k_2}{\tau} &=   \bsvBR{j_1 & j_2}{k_1 & k_2}{\tau} + \sum_{p_1=0}^{k_1-j_1-2} \frac{ {k_1-j_1-2 \choose p_1} }{(4y)^{p_1}}  \ccsv{j_1+p_1 }{ k_1 } \bsvBR{ j_2}{k_2}{\tau}  \label{MGFtoBR31} \\
&\quad +  
\sum_{p_1=0}^{k_1-j_1-2} \sum_{p_2=0}^{k_2-j_2-2} 
\frac{ {k_1-j_1-2 \choose p_1} {k_2-j_2-2 \choose p_2} }{(4y)^{p_1+p_2}}
\ccsv{j_1+p_1 & j_2+p_2 }{ k_1 & k_2}
+  \bsvCUSP{j_1 & j_2}{k_1 & k_2}{\tau}\, ,
\notag
 \end{align}
and more generally
\begin{align}
\beta^{\rm eqv}[P;\tau] &=\beta_\Delta^{\rm sv}[P;\tau] + \sum_{P=XY} \ddsvpure[X;\tau] 
 \beta^{\rm sv}[Y;\tau] \, ,\label{MGFtoBR32}
\end{align}
where the constants $c^{\rm sv}$ have been reassembled into polynomials in $y^{-1}$,
\begin{align}
\label{MGFtoBR32b}
\ddsv{\ldots &j_i &\ldots}{\ldots &k_i &\ldots}{\tau} &= \sum_{p_i=0}^{k_i-2-j_i} \frac{ 
{ k_i{-}2 {-}j_i \choose p_i} }{(4y)^{p_i}}  \ccsv{ \ldots &j_i{+}p_i &\ldots}{ \ldots &k_i &\ldots }\, ,
\end{align}
and we again have one summation of this type per column.
The $\beta_{\Delta}^{\rm sv}$ appearing in~\eqref{MGFtoBR31} are iterated integrals with at least one
holomorphic cusp form $\Delta_{2s}(\tau)$ among its kernels. They do not
contribute to MGFs \cite{DHoker:2016mwo, Gerken:2018zcy, Gerken:2019cxz}, 
vanish at depth one, i.e.\ $ \bsvCUSP{j}{k}{\tau}=0$, 
and will be discussed in section~\ref{sec:cusp}. Up to these
cusp-form contributions, the differential equations
\begin{align}
2\pi i(\tau{-}\bar \tau)^2 \partial_\tau \beqv{j_1 &\ldots &j_\ell}{k_1 &\ldots &k_\ell}{\tau}
&= \sum_{i=1}^\ell (k_i{-}j_i{-}2)
\beqv{j_1 &\ldots &j_i{+}1 &\ldots &j_\ell}{k_1 &\ldots
&k_i &\ldots &k_\ell}{\tau}
\label{MGFtoBR33} \\
&\quad - \delta_{j_\ell,k_\ell-2} (\tau{-}\bar \tau)^{k_\ell} {\rm G}_{k_\ell}(\tau) 
\beqv{j_1 &\ldots &j_{\ell-1}}{k_1 &\ldots &k_{\ell-1}}{\tau}
{\rm mod} \ \beta_{\Delta}^{\rm sv}
\notag
\end{align}
take the same form as those of the 
$\beta^{\rm sv}$ in (\ref{bsvdiff}) \cite{Gerken:2020yii}.

A few comments on the equations above are in order.
Firstly, we stress that the combination of $(4y)^{-1}$ with
constants $\ccsvpure [ \begin{smallmatrix} \ldots \\ \ldots \end{smallmatrix} ]$ in (\ref{MGFtoBR32}) 
or $\overline{\alphaBR{ \ldots }{ \ldots }{\tau}}$ in (\ref{MGFtoBR14})
both preserve the form of (\ref{MGFtoBR33}). 
Secondly, we notice that the construction~\eqref{MGFtoBR32} of $\beta^{\rm eqv}[P;\tau]$ is expressed in terms of $ \ddsvpure[X;\tau] 
 \beta^{\rm sv}[Y;\tau]$ with deconcatenations $P=XY$ of the word $P$, such 
 that $\beta^{\rm sv}[Y;\tau]$ is precisely labelled by the second subword $Y$. 
This fact ensures that only the right-most letter $\begin{smallmatrix} j_\ell \\ k_\ell \end{smallmatrix}$ 
of $P$ causes the appearance of holomorphic Eisenstein integration kernels in (\ref{MGFtoBR33}).

Once the constants $\ccsvpure$ and the $T$-invariant cusp-form contributions
$\beta_{\Delta}^{\rm sv}$ are tailored to the multiple modular
values in the $S$-transformations of (\ref{MGFtoBR13}), the
$\beta^{\rm eqv}$ in (\ref{MGFtoBR32}) transform as non-holomorphic modular forms
\beq
\beqv{j_1 &\ldots &j_\ell}{k_1 &\ldots &k_\ell}{\tfrac{ a \tau{+}b }{c\tau {+} d}}=
\bigg( \prod_{i=1}^\ell (c \bar \tau{+}d)^{k_i{-}2{-}2j_i} \bigg)
\beqv{j_1 &\ldots &j_\ell}{k_1 &\ldots &k_\ell}{\tau}
\, , \ \ \ \
\big(\begin{smallmatrix} a&b \\ c &d \end{smallmatrix}\big) \in SL(2,\mathbb Z)\, .
\label{MGFtoBR34}
\eeq
The constants $\ccsvpure \big[ \begin{smallmatrix} j_1 &\ldots&j_\ell \\ k_1 &\ldots &k_\ell \end{smallmatrix} \big]$ are (conjecturally single-valued) MZVs (\ref{MGFtoBR05}) of
weight $\sum_{i=1}^\ell (j_i{+}1)$. The earlier expression (\ref{MGFtoBR24}) 
for $\beqv{j}{k}{\tau}$ at depth one implies the closed formulae
\begin{align}
\ccsv{ j }{ k } &= - \frac{2 \zeta_{k-1}}{k{-}1}\, \delta_{j,k-2}
\, , \ \ \ \
\ddsv{j }{ k}{\tau} = -\frac{2 \zeta_{k-1}}{(k{-}1) (4y)^{k-2-j}} \, .
\label{MGFtoBR35} 
\end{align}
At depth two, all the $\ccsvpure [ \begin{smallmatrix} j_1 &j_2 \\ k_1 &k_2 \end{smallmatrix} ]$ with $k_1{+}k_2 \leq 28$ and $0\leq j_i\leq k_i{-}2$
are fixed by the $\beta^{\rm sv}$-representations of certain modular invariants
${\rm F}^{\pm(s)}_{m,k}$ and their $\tau,\bar \tau$-derivatives at $m{+}k \leq 14$ 
in the ancillary files of \cite{Dorigoni:2021jfr, Dorigoni:2021ngn}.
Double integrals of ${\rm G}_4(\tau_1){\rm G}_4(\tau_2)$ give rise to constants
\begin{align}
\ccsv{0 & 0}{4& 4} &= 0\, ,&\! \! \ccsv{0& 1}{4& 4} &= -\tfrac{\zeta_3}{2160}\, ,  &\! \! \ccsv{0& 2}{4& 4} &= 0\, ,\notag \\
 \ccsv{1& 0}{4& 4} &= \tfrac{\zeta_3}{2160}\, , &\! \! \ccsv{1& 1}{4& 4} &= 0\, ,&\! \!  \ccsv{1& 2}{4& 4} &= \tfrac{5 \zeta_5}{108} \, ,\label{bequv.5} \\ 
\ccsv{2& 0}{4& 4} &= 0 \, ,&\! \!  \ccsv{2& 1}{4& 4}&= -\tfrac{5 \zeta_5}{108}\, ,&\! \! \ccsv{2& 2}{4& 4} &= \tfrac{2}{9} \zeta_3^2 \, , \notag
\end{align}
and the non-vanishing instances of $\ccsv{j_1& j_2}{4& 6}$ are 
 \begin{align}
\ccsv{0& 1}{4& 6} &= \tfrac{\zeta_3}{907200} \, , &\ccsv{1& 0}{4& 6} &= -\tfrac{\zeta_3}{226800}\, ,
\notag \\
\ccsv{0& 3}{4& 6} &= -\tfrac{\zeta_5}{7200}\, , 
& \ccsv{1& 2}{4& 6} &= \tfrac{\zeta_5}{21600} \, ,
& \ccsv{2& 1}{4& 6} &= -\tfrac{\zeta_5}{21600} \, , \label{bequv.564} \\
\ccsv{0& 4}{4& 6} &= -\tfrac{\zeta_3^2}{315} \, ,
&\ccsv{1& 3}{4& 6} &= \tfrac{\zeta_3^2}{1260} \, ,
&\ccsv{2& 2}{4& 6} &= -\tfrac{\zeta_3^2}{1890} \, , \notag\\
 \ccsv{1& 4}{4& 6} &= \tfrac{7 \zeta_7}{360}\, ,
&\ccsv{2& 3}{4& 6} &= -\tfrac{7 \zeta_7}{720} \, ,
&\ccsv{2& 4}{4& 6} &= \tfrac{2 \zeta_3 \zeta_5}{15} \, . \notag
\end{align} \normalsize
More general examples $\ccsvpure [ \begin{smallmatrix} j_1 &j_2 \\ k_1 &k_2 \end{smallmatrix} ]$
are composed of odd Riemann zeta values and bilinears thereof.
A variety of $\ccsvpure [ \begin{smallmatrix} j_1 &j_2&j_3 \\ k_1 &k_2 &k_3 \end{smallmatrix} ]$ to
be extracted from depth-three analogues of ${\rm F}^{\pm(s)}_{m,k}$ in future 
work \cite{depth3paper} involve indecomposable single-valued
MZVs of depth three, such as
\begin{align}
\ccsv{2 & 2 &4}{ 4 & 4 &6} &=
-\frac{1}{450} \zeta^{\rm sv}_{3,5,3}
 - \frac{ 2}{45} \zeta_3^2 \zeta_5
  - \frac{ 221 }{21600}\zeta_{11}\, ,
\notag\\
  \ccsv{2 & 4 &4}{ 4 & 6 &6} &=
     \frac{1}{3750} \zeta_{5,3,5}^{\rm sv} 
     + \frac{2}{375} \zeta_3 \zeta_5^2 
     + \frac{1804427}{124380000}\zeta_{13}\, , 
 \label{bequv.9} \\
    \ccsv{2 & 2 &6}{ 4 & 4 &8}  &= -\frac{1}{1764}\zeta^{\rm sv}_{3,7,3} + \frac{1}{1470} \zeta^{\rm sv}_{5,3,5}  - \frac{2}{63} \zeta_3^2\zeta_7 -\frac{137359}{24378480} \zeta_{13}\, ,
    \notag
\end{align}
where \cite{Schnetz:2013hqa, Brown:2013gia}
\begin{align}
  \zeta^{\rm sv}_{3,5,3}&= 2 \zeta_{3,5,3} - 2 \zeta_3 \zeta_{3,5} - 10 \zeta_3^2 \zeta_5\,,
  \notag\\
  \zeta^{\rm sv}_{5,3,5}&= 2 \zeta_{5,3,5} - 22 \zeta_{3,5}\zeta_5 - 120 \zeta_5^2 \zeta_3 - 
 10 \zeta_5 \zeta_8\,,
\label{svmzvs}\\
  \zeta^{\rm sv}_{3,7,3}&= 2 \zeta_{3, 7, 3} - 2\zeta_{3,7}\zeta_3- 28 \zeta_3^2\zeta_7- 
 24 \zeta_{3,5} \zeta_5 - 144 \zeta_5^2 \zeta_3- 12 \zeta_5 \zeta_8\,.
 \notag
 \end{align}
A list of all $\ccsvpure [ \begin{smallmatrix} j_1 &j_2 \\ k_1 &k_2 \end{smallmatrix} ]$
at $k_1{+}k_2\leq 28$ and $\ccsvpure [ \begin{smallmatrix} j_1 &j_2&j_3 \\ 
k_1 &k_2 &k_3 \end{smallmatrix} ]$ with $k_1{+}k_2{+}k_3\leq 16$
can be found in the ancillary files of the arXiv submission,
and their instances with $j_i=k_i{-}2$ are revisited in the light of the
$f$-alphabet in section \ref{sec:4.1.4}.

As will be detailed in sections \ref{sec:4.1.2}
and \ref{sec:unique}, some of the $\ccsvpure [ \begin{smallmatrix} j_1 &j_2&j_3 \\ 
k_1 &k_2 &k_3 \end{smallmatrix} ]$ at $j_1{+}j_2{+}j_3= \frac{1}{2}(k_1{+}k_2{+}k_3)-3$
and $k_1{+}k_2{+}k_3\leq 16$ admit redefinitions that one may be able to fix from depth-four
computations. We will track the one-parameter freedom $\sim c_{446} \zeta_7$ of 
$\ccsvpure [ \begin{smallmatrix} j_1 &j_2&j_3 \\ 
4 &4 &6 \end{smallmatrix} ]$ and $\sim c_{466} \zeta_3 \zeta_5$ of 
$\ccsvpure [ \begin{smallmatrix} j_1 &j_2&j_3 \\ 
4 &6 &6 \end{smallmatrix} ]$ (with $c_{446},c_{466} \in \mathbb Q$) in the 
ancillary file that amounts to shifting
certain combinations of modular invariant $\beta^{\rm eqv}[ \begin{smallmatrix} j_1 &j_2&j_3 \\ 
4 &4 &6 \end{smallmatrix} ]$ and 
$\beta^{\rm eqv} [ \begin{smallmatrix} j_1 &j_2&j_3 \\ 
4 &6 &6 \end{smallmatrix} ]$ in an $SL(2,\mathbb R)$-singlet by constants.

All of $\beta^{\rm sv},\beta^{\rm eqv}$ and $c^{\rm sv}$ are expected to obey shuffle relations inherited from iterated integrals such as
\beq
\ccsv{j_1 }{k_1 } \ccsv{j_2}{k_2} =
\ccsv{j_1& j_2}{k_1& k_2}+ \ccsv{j_2& j_1}{k_2& k_1}
\label{shff.01}
\eeq
and more generally
\begin{align}
\beta^{\rm sv}[X;\tau]\beta^{\rm sv}[Y;\tau] &= \sum_{P \in X \shuffle Y} \beta^{\rm sv}[P;\tau] \, ,  \notag \\
\beta^{\rm eqv}[X;\tau]\beta^{\rm eqv}[Y;\tau] &=  \sum_{P \in X \shuffle Y}  \beta^{\rm eqv}[P;\tau]\, ,
\label{shff.02} \\
\ccsvpure[X]\ccsvpure[Y] &= \sum_{P \in X \shuffle Y}  \ccsvpure[P] \, ,
\notag
\end{align}
with $X \shuffle Y$ denoting the shuffle product of the words $X$ and $Y$.
These shuffle relations are consequences of our central claim in (\ref{MGFtoBR63}) below
that the modular forms $\beta^{\rm eqv}$ can be alternatively generated from
Brown's EIEIs.

\subsection{Modular graph forms and beyond in terms of $\beta^{\rm eqv}$}
\label{sec:3.0}

As will be illustrated from the examples in this section, the
non-holomorphic modular forms $\beta^{\rm eqv}$
compactly encode iterated-integral representations of MGFs and more general
modular forms. More specifically, each MGF
can be written as a $\mathbb Q[{\rm MZV}]$-linear (conjecturally
$\mathbb Q[\textrm{single-valued}\, {\rm MZV}]$-linear) combination of $\beta^{\rm eqv}$,
though there are $\beta^{\rm eqv}$ with contributions from holomorphic cusp 
forms which cannot be realized via MGFs.

The two-column cases (\ref{MGFtoBR25}) take the form
\beq
\cplusno{a &0}{b &0} = - \frac{(2i)^{b-a} (a{+}b{-}1)! }{(a{-}1)!(b{-}1)!}  \beqvno{a-1}{a+b}
\label{moreex.01}
\eeq
upon inserting the $\beta^{\rm eqv}$ at depth one into (\ref{MGFtoBR24}). 
The depth-two examples in (\ref{MGFtoBR26}) and (\ref{MGFtoBR27}) condense to
\begin{align}
\cplusno{2 &1 &1}{2 &1 &1} &=  
-126 \beqvno{3}{8}  - 18 \beqvno{2& 0}{4& 4} \, ,
\notag \\
 \cplusno{3 &2 &1}{1 &2 &1} &=  
\frac{ 279}{2} \beqvno{5}{10} + 30 \beqvno{3& 1}{6& 4}  + 
 \frac{15}{2} \beqvno{4& 0}{6& 4}  \, ,\label{MGFtoBR36}
 \\
2i \Im \cplusno{0 &1 &2 &2}{1 &1 &0 &3} &= 
 60 (\beqvno{0& 3}{4& 6} {-} \beqvno{1& 2}{6& 4} ) 
 - 270 ( \beqvno{1& 2}{4& 6} {-}  \beqvno{2& 1}{6& 4})
 \notag \\
 &\quad 
 + 390 ( \beqvno{2& 1}{4& 6} {-} \beqvno{3& 0}{6& 4}) -  3 \zeta_3 \beqvno{1}{4}\, ,
\notag
\end{align}
by virtue of the $\ccsvpure [ \begin{smallmatrix} j_1 &j_2 \\ 4 &6 \end{smallmatrix} ]$
in (\ref{bequv.564}) and the $\ccsvpure [ \begin{smallmatrix} j_1 &j_2 \\ 6 &4 \end{smallmatrix} ]$
following from the shuffle relations (\ref{shff.01}). 

As a convenient organization of 
modular double integrals, certain modular invariant solutions  ${\rm F}^{\pm(s)}_{m,k}$ 
to inhomogeneous Laplace equations have been studied in \cite{Dorigoni:2021jfr, Dorigoni:2021ngn},
with $2\leq m\leq k$ referring to the ${\rm E}_m ,{\rm E}_k$ in the inhomogeneous terms. The 
real ones ${\rm F}^{+(s)}_{m,k}$ are characterized by Laplace eigenvalues $s(s{-}1)$ with 
$s \in \{ k{-}m{+}2,\, k{-}m{+}4,
\ldots,k{+}m{-}4,\, k{+}m{-}2\}$ and contain the three-column MGFs $ \cplusno{a &b &c}{a &b &c}$ 
whose Laplace equations can be found in \cite{DHoker:2015gmr}. The imaginary ones
${\rm F}^{-(s)}_{m,k}$ with $s \in \{ k{-}m{+}1,\, k{-}m{+}3,
\ldots,k{+}m{-}3,\, k{+}m{-}1\}$ contain cuspidal MGFs with three or more columns
such as $\Im \cplusno{0 &1 &2 &2}{1 &1 &0 &3}$ in (\ref{MGFtoBR36}).

The collection of $\beta^{\rm eqv} [ \begin{smallmatrix} j_1 &j_2 \\ 2m &2k \end{smallmatrix} ]$
and $\beta^{\rm eqv} [ \begin{smallmatrix} j_2 &j_1 \\ 2k &2m \end{smallmatrix} ]$
with $0\leq j_1 \leq 2m{-}2$ and $0\leq j_2 \leq 2k{-}2$ can be expressed in
terms of ${\rm F}^{\pm(s)}_{m,k}$, bilinears in ${\rm E}_m ,{\rm E}_k$, products
$\zeta_{2m-1} {\rm E}_s$ as well as respective modular derivatives 
$\nabla_\tau = 2i (\Im \tau)^2 \partial_\tau$
and $\overline{\nabla}_\tau = -2i (\Im \tau)^2 \partial_{\bar \tau}$. Conversely, 
we have compact identities
\begin{align}
{\rm F}^{+(2)}_{2,2}  &= 18 \beqvno{2& 0}{4& 4} \, ,\notag \\
{\rm F}^{-(2)}_{2,3}  &= -90 \beqvno{1& 2}{4& 6} + 90 \beqvno{2& 1}{4& 6} + 
90 \beqvno{2& 1}{6& 4} - 90 \beqvno{3& 0}{6& 4} - 
\frac{5}{7}\zeta_3 \beqvno{1}{4} \, , \label{moreex.02} \\
{\rm F}^{-(6)}_{2,5}  &= -1890 \beqvno{1& 4}{4& 10} - 1512 \beqvno{2& 3}{4& 10} + 
1890 \beqvno{4& 1}{10& 4} + 1512 \beqvno{5& 0}{10& 4} \, ,\notag 
\end{align}
and the analogous $\beta^{\rm eqv}$-representations of any ${\rm F}^{\pm(s)}_{m,k}$
with $m{+}k \leq 12$ can be found in an ancillary file.
Modular forms $\beta^{\rm eqv} [ \begin{smallmatrix} j_1 &j_2 \\ k_1 &k_2 \end{smallmatrix} ]$
of non-zero weight $(0,k_1{+}k_2{-}2j_1{-}2j_2{-}4)$ occur in derivatives such as
\begin{align}
\pi \nabla_\tau {\rm F}^{+(3)}_{2,3}  &=
-\frac{45}{2} \beqvno{2& 2}{4& 6} - 15 \beqvno{3& 1}{6& 4}
- \frac{15}{2} \beqvno{4& 0}{6& 4} \, , \label{moreex.03} \\
\frac{ \pi \overline{\nabla}_\tau  {\rm F}^{+(3)}_{2,3} }{y^2} &=
-240 \beqvno{1& 1}{4& 6} - 120 \beqvno{2& 0}{4& 6} - 360 \beqvno{2& 0}{6& 4}\, ,
\notag
\end{align}
which follow from the $\beta^{\rm eqv}$-representations of
${\rm F}^{\pm(s)}_{m,k}$ via (\ref{MGFtoBR33}) and their 
Laplace equations.

As detailed in \cite{Dorigoni:2021ngn}, some of the ${\rm F}^{\pm(s)}_{m,k}$
with $m{+}k \geq 7$ and $s \geq 6$ contain iterated integrals of holomorphic
cusp forms and therefore go beyond MGFs. These cases determine
the cusp-form contributions $\beta^{\rm sv}_{\Delta}$ in (\ref{MGFtoBR31}) and 
will be discussed in section \ref{sec:cusp} below, see in 
particular (\ref{moreex.00}) below for a simple example.

We conclude this section with depth-three examples of MGFs,
\begin{align}
 \cplusno{2 &2 &1 &1}{2 &2 &1 &1} &=
 216 \beqvno{2& 1& 0}{4& 4& 4} + 840 \beqvno{1& 3}{4& 8} - 
252 \beqvno{2& 2}{4& 8} + 840 \beqvno{3& 1}{8& 4} 
- 252 \beqvno{4& 0}{8& 4}  \notag \\
&\quad  + 3600 \beqvno{2& 2}{6& 6} - 
3200 \beqvno{3& 1}{6& 6} + 
2100 \beqvno{4& 0}{6& 6} 
+  3212 \beqvno{5}{12}\, ,
\notag \\
  \cplusno{2 &1 &1 &1 &1 &1}{2 &1 &1 &1 &1 &1} &= 
\bigg( \frac{47}{72} + 1814400\, c_{446} \bigg)  \zeta_7 - 3 \zeta_5 \beqvno{1}{4} - 
 1260 \zeta_3 \beqvno{3}{8} - 
 180 \zeta_3 \beqvno{2&0}{4&4}  \notag \\
 &\hspace{-1.7cm}+ 
  360 (150 \beqvno{1& 2& 1}{4& 4& 6}+ 
    150 \beqvno{3& 0& 1}{6& 4& 4}
    - 90 \beqvno{1& 1& 2}{4& 4& 6}
     -    90 \beqvno{2& 1& 1}{6& 4& 4}  
     - 90 \beqvno{1& 2& 1}{4& 6& 4} \label{moreex.11} \\
    &\quad \quad  \hspace{-1.7cm}  + 
    150 \beqvno{1& 3& 0}{4& 6& 4} + 
    150 \beqvno{2& 1& 1}{4& 6& 4}+ 
    195 \beqvno{2& 0& 2}{4& 4& 6}+ 
    195 \beqvno{2& 2& 0}{6& 4& 4} + 15 \beqvno{2& 2& 0}{4& 6& 4}
      \notag\\
    &\quad \quad  \hspace{-1.7cm}
     -     330 \beqvno{2& 1& 1}{4& 4& 6}- 
    330 \beqvno{3& 1& 0}{6& 4& 4} + 
    480 \beqvno{2& 2& 0}{4& 4& 6}    + 
    480 \beqvno{4& 0& 0}{6& 4& 4} + 315 \beqvno{2& 3}{6& 8} \notag \\
 &\quad \quad  \hspace{-1.7cm}
     + 315 \beqvno{3& 2}{8& 6}  - 
    1190 \beqvno{3& 2}{6& 8}  - 1190 \beqvno{4& 1}{8& 6} 
    +   2800 \beqvno{4& 1}{6& 8} + 2800 \beqvno{5& 0}{8& 6} \notag \\
    &\quad \quad  \hspace{-1.7cm}  + 
    243 \beqvno{4& 1}{10& 4} + 243 \beqvno{1& 4}{4& 10} + 
    432 \beqvno{5& 0}{10& 4} +    432 \beqvno{2& 3}{4& 10}
    + 3640 \beqvno{6}{14})   
  \, ,
  \notag
\end{align}
where the former has been previously expressed in terms of iterated integrals
\cite{Broedel:2018izr, Gerken:2020yii}. Moreover, the six-column MGF in (\ref{moreex.11}) 
is sometimes denoted by $D_{5,1,1}$ in the literature and provided the first example of 
an irreducible single-valued MZV beyond depth one
in the expansion of MGFs around the cusp \cite{Zerbini:2015rss}. In our setting,
the depth-three MZV $\zeta_{3,5,3}$ enters $ \cplusno{2 &1 &1 &1 &1 &1}{2 &1 &1 &1 &1 &1}$ 
via $c^{\rm sv}[\begin{smallmatrix} 2 &2 &4 \\4 &4 &6 \end{smallmatrix}]$ in (\ref{bequv.9}),
and the role of rational free parameter $c_{446}$ is explained below (\ref{svmzvs}).

\subsection{From Tsunogai's derivation algebra to $\overline{\alpha}$}
\label{sec:3.2}

Apart from the constants $\ccsvpure$ and the cusp-form 
contributions $\beta^{\rm sv}_{\Delta}$ in the modular forms (\ref{MGFtoBR32}),
we still need to supply the antiholomorphic $T$-invariants 
$\overline{\alphaBR{\ldots}{\ldots}{\tau}}$ (or their $\mathbb Q[y^{-1}]$-linear combinations
$\overline{\kappa[\begin{smallmatrix} \ldots \\ \ldots \end{smallmatrix};\tau]}$) entering the $\beta^{\rm sv}$ in (\ref{MGFtoBR12}).
A variety of depth-two cases has been determined from the reality
properties (\ref{MGFtoBR28}) of MGFs (or respective generating
functions \cite{Gerken:2020yii}) and those of the above solutions to inhomogeneous
Laplace equations, $\overline{ {\rm F}^{\pm(s)}_{m,k} } = \pm {\rm F}^{\pm(s)}_{m,k}$
\cite{Dorigoni:2021jfr, Dorigoni:2021ngn}: Any $\overline{\alpha[\begin{smallmatrix}
j_1 &j_2 \\ k_1 &k_2 \end{smallmatrix};\tau]}$ is a linear combination of
 \begin{align}
{\cal E}_0(k,0^p;\tau) &= \frac{(2\pi i)^{p+1-k} }{p!} \int_\tau^{i\infty} \! \! \dd \tau_1 
\, (\tau{-}\tau_1)^p \big[ {\rm G}_k(\tau_1)-2\zeta_k \big] \notag \\
&= -\frac{2}{(k{-}1)!} \sum_{m,n=1}^\infty \frac{q^{mn}}{m^{p-k+2} n^{p+1} }\, ,
\ \ \ \ q= e^{2\pi i\tau} \, ,
 \label{MGFtoBR41}
\end{align} 
with $\mathbb Q$-multiples of odd Riemann zeta values as coefficients. 
Modular $T$-invariance is attained through the subtraction of the zero 
mode ${\rm G}_k(\tau)= 2\zeta_k + { O}(q)$ in the integrand of (\ref{MGFtoBR41})
\cite{Broedel:2015hia} and manifest from the $q$-series representation in the second line.
The subtraction of the zero mode also makes the integral well-defined without the use 
of tangential-basepoint regularization. We note for future reference that the conjugates of these integrals can be rewritten as (with Bernoulli numbers ${\rm B}_k$)
\begin{align}
\overline{{\cal E}_0(k,0^p;\tau)}  =- \frac{{\rm B}_k  (-2\pi i \bar\tau)^{p+1}}{k! (p{+}1)!} +  \frac1{p!} \sum_{\ell=0}^{k-2-p} \frac{\binom{k-2-p}{\ell}}{(-4y)^\ell} \bminus{ p+\ell}{k}{\tau}\, ,
\label{forlater}
\end{align}
when using the integrals $\beta_-$ of the antiholomorphic kernels introduced in~\eqref{MGFtoBR13}.

The examples of the $\overline{\alphaBR{\ldots}{\ldots}{\tau}}$ in the ancillary files of 
\cite{Dorigoni:2021jfr, Dorigoni:2021ngn} can be lined up with the following
general formula at depth two,
\begin{align}
\overline{\alphaBR{ j_1 &j_2  }{ k_1 &k_2 }{\tau} } &= 
\overline{\alphaeasy{ j_1 &j_2  }{ k_1 &k_2 }{\tau} }  + 
\overline{\alphahard{ j_1 &j_2  }{ k_1 &k_2 }{\tau} }\, , \label{closedform} \\
\overline{\alphaeasy{ j_1 &j_2  }{ k_1 &k_2 }{\tau} }   &=
\frac{2 \zeta_{k_1-1} }{k_1{-}1} \delta_{j_1,k_1-2} j_2!  \overline{ {\cal E}_0(k_2,0^{j_2}) }  
- \frac{2 \zeta_{k_2-1} }{k_2{-}1} \delta_{j_2,k_2-2} j_1!  \overline{ {\cal E}_0(k_1,0^{j_1}) }\, .
\notag
\end{align}
The second part $\overline{\alpha_{\rm hard}}$ is the crucial hint to
anticipate the connection between MGFs or $\beta^{\rm sv}$ and Brown's
EIEIs: The $\overline{\alpha_{\rm hard}}$ are specified by a generating series
\begin{align}
&\sum_{k_1,k_2=4}^\infty \sum_{j_1=0}^{k_1-2} \sum_{j_2=0}^{k_2-2}
\frac{ (k_1{-}1)(k_2{-}1) (-1)^{j_1+j_2} }{(k_1{-}j_1{-}2)!(k_2{-}j_2{-}2)!}  
\overline{\alphahard{ j_1 &j_2  }{ k_1 &k_2 }{\tau} }  \epsilon_{k_2}^{(k_2-j_2-2)}
\epsilon_{k_1}^{(k_1-j_1-2)}  \notag\\
&\ \ =
\sum_{m=1}^\infty 2 \zeta_{2m+1} \sum_{k=4}^{\infty} \sum_{j=0}^{k-2}
\frac{ (k{-}1) (-1)^j }{(k{-}j{-}2)!}  \, j! \overline{ {\cal E}_0(k,0^j) }
[z_{2m+1}, \epsilon_k^{(k-j-2)}] \, \big|_{\rm depth\ 2}
 \label{MGFtoBR42} 
 \end{align}
akin to (\ref{MGFtoBR21}), where we use the shorthand (\ref{MGFtoBR43}), 
and an explicit closed form of $\overline{\alpha_{\rm hard}}$ derived from (\ref{MGFtoBR42})
can be found in appendix \ref{app.A}.
 
The brackets in the second line of (\ref{MGFtoBR42})
involve derivations $z_3,z_5,z_7,\ldots$ dual to odd zeta values 
subject to $[z_{2m+1},\epsilon_0]=0$ that normalize\footnote{The normalizer, $N_\mathfrak{g}(S)$, of a subset $S$ in a Lie algebra $\mathfrak{g}$ is defined by $N_\mathfrak{g}(S)=\{z\in \mathfrak{g}\,{\rm s.t.}\,[z,x]\in S,\,\forall \,x\in S\}$.} Tsunogai's derivation 
algebra: Any $[z_{2m+1},\epsilon_{k\neq 0}]$ in (\ref{MGFtoBR42}) is
expressible in terms of nested commutators of two and more $\epsilon_{k_i}^{(j_i)}$
through a procedure described in \cite{Pollack} which relies on expressing the derivations in terms of elements of a free Lie algebra with two generators.

The ``depth-two'' 
contributions from $[\epsilon_{k_1}^{(j_1)}, \epsilon_{k_2}^{(j_2)}]$
are known in closed form, see \cite[Sec.~25]{hain_matsumoto_2020},
\begin{align}
\!\!\![z_{2m-1},\epsilon_{2n+2}] \, \big|_{\rm depth\ 2} \!=\!  \frac{(2n{+}2)! \, {\rm B}_{2n+2m}}{(2n{+}2m{-}2)! (2n{+}2m)! \, {\rm B}_{2n+2}} \! \sum_{\ell=0}^{2m-2} \! \frac{(-1)^\ell}{\ell!}  (2n{+}\ell)!\,
[ \epsilon^{(\ell)}_{2m} , \epsilon_{2n+2m}^{(2m-2-\ell)} ]\,.
\label{MGFtoBR44}
\end{align}
The simplest examples of full $[z_{2m+1},\epsilon_{k\neq 0}]$-relations are
\begin{align}
[z_3,\epsilon_4] &= \frac{1}{504} \Big({-}  [ \epsilon_4^{\phantom{(}} , \epsilon_6^{(2)} ] 
+ 3 [ \epsilon_4^{(1)} , \epsilon_6^{(1)} ] 
- 6 [ \epsilon_4^{(2)} , \epsilon_6^{ \phantom{(}} ]   \Big)\, ,
\notag \\
[z_3,\epsilon_6] &= \frac{1}{1200}\Big({-} 
[ \epsilon_4^{\phantom{(}} , \epsilon_8^{(2)} ] 
+  5 [ \epsilon_4^{(1)} , \epsilon_8^{(1)} ]  
 - 15 [ \epsilon_4^{(2)} , \epsilon_8^{\phantom{(}} ]  
 +63 [ \epsilon^{\phantom{(}}_4, [ \epsilon^{(1)}_4 , \epsilon^{\phantom{(}}_4] ]   \Big)  
 \, ,
\notag \\
[z_5,\epsilon_4] &= 
\frac{1}{604800} 
\Big( [ \epsilon_6^{\phantom{(}} , \epsilon_8^{(4)} ]  - 
  3 [ \epsilon_6^{(1)} , \epsilon_8^{(3)} ]   + 6 [ \epsilon_6^{(2)} , \epsilon_8^{(2)} ]  
    -  10 [ \epsilon_6^{(3)} , \epsilon_8^{(1)} ] 
   + 15 [ \epsilon_6^{(4)} , \epsilon_8^{\phantom{(}}  ] 
   \label{brown.15}\\
     &\quad + 105 [\epsilon_4^{(1)},[\epsilon_4^{(1)},\epsilon_6^{(1)}]] - 1668 [\epsilon_4^{(1)},[\epsilon_4^{(2)},\epsilon_6^{\phantom{(}}]] - 729 [\epsilon_4^{(2)},[\epsilon_4^{\phantom{(}},\epsilon_6^{(1)}]]+ 1458 [\epsilon_4^{(2)},[\epsilon_4^{(1)},\epsilon_6^{\phantom{(}}]]\notag\\
     &\quad + 35 [\epsilon_4^{\phantom{(}},[\epsilon_4^{\phantom{(}},\epsilon_6^{(3)}]] - 313 [\epsilon_4^{\phantom{(}},[\epsilon_4^{(1)},\epsilon_6^{(2)}]]+ 834 [\epsilon_4^{\phantom{(}},[\epsilon_4^{(2)},\epsilon_6^{(1)}]]+ 208 [\epsilon_4^{(1)},[\epsilon_4^{\phantom{(}},\epsilon_6^{(2)}]]
     \Big)\, ,\notag
\end{align}
where the term $\sim [ \epsilon^{\phantom{(}}_4, [ \epsilon^{(1)}_4 ,
\epsilon^{\phantom{(}}_4]  ] $ in the middle equation is the first instance of the 
higher-depth terms that are suppressed in (\ref{MGFtoBR44}). Combining
(\ref{brown.15}) and (\ref{MGFtoBR42}) reproduces contributions such as 
\cite{Gerken:2020yii}
\begin{align}
\overline{\alphahard{ 0 &2  }{ 4 &6 }{\tau} } &= \frac{ \zeta_3}{630} \overline{ {\cal E}_0(4;\tau) }\, ,
\notag \\
\overline{\alphahard{ 2 &1  }{ 4 &6 }{\tau} } &= \frac{ \zeta_3}{210} \overline{ {\cal E}_0(4,0;\tau) }\, ,
 \label{MGFtoBR47} \\
\overline{\alphahard{ 0 &4  }{ 4 &6 }{\tau} } &= \frac{2 \zeta_3}{105} \overline{ {\cal E}_0(4,0,0;\tau) }\, ,
\notag
\end{align}
or \cite{Gerken:2020xfv, Dorigoni:2021jfr}
\begin{align}
\overline{\alphahard{ 1 &3  }{ 4 &8 }{\tau} } &=  - \frac{ \zeta_3}{140} \overline{ {\cal E}_0(6,0,0;\tau) }
\, , \ \ \ \
\overline{\alphahard{ 2 &3  }{ 6 &8 }{\tau} } = -\frac{ \zeta_5}{98000} \overline{ {\cal E}_0(4,0;\tau) }
\, . \label{MGFtoBR48} 
\end{align}
Note that (\ref{MGFtoBR44}) readily implies that $\overline{\alpha_{\rm hard}[\begin{smallmatrix}
j_1 &j_2 \\ k_1 &k_2 \end{smallmatrix};\tau]}$ vanishes
whenever $k_1=k_2$. Any contribution of $\zeta_{2a-1} \beqvno{j}{w} $ to
the $\beta^{\rm eqv}$-representation of ${\rm F}^{\pm(s)}_{m,k}$
(such as the term $- \frac{5}{7} \zeta_3 \beqvno{1}{4}$
in the expression (\ref{moreex.02}) for ${\rm F}^{-(2)}_{2,3}$)  
can be traced back to $\overline{\alpha_{\rm hard}}$ and only occurs
for $s=k{-}m{+}1$.

In principle, one could use the commutator relations among $\epsilon_k$ in
(\ref{MGFtoBR23}) to eliminate some of the terms $[ \epsilon_6^{(j)} , \epsilon_8^{(4-j)} ] $ in 
the expression (\ref{brown.15}) for $[z_5,\epsilon_4] $ in favor of
$[ \epsilon_4^{(j)} , \epsilon_{10}^{(4-j)} ] $. The coefficients
of $\epsilon_{k_1}^{(j_1)} \epsilon_{k_2}^{(j_2)}$ in the generating-series
identity (\ref{MGFtoBR42}) are understood to be compared after importing
the direct outcome of (\ref{MGFtoBR44}) for the commutators
$[z_{2m+1},\epsilon_k]$ and before inserting any relation 
among $[\epsilon_{k_1},\epsilon_{k_2}]$ \cite{Tsunogai, LNT, Pollack}.\footnote{It appears unnatural
to rewrite the expressions (\ref{MGFtoBR44}) for the commutators
$[z_{2m+1},\epsilon_k]$ via relations in Tsunogai's
derivation algebra: The relative factors $ [ \epsilon_6^{\phantom{(}} , \epsilon_8^{(4)} ]  - 
  3 [ \epsilon_6^{(1)} , \epsilon_8^{(3)} ]+\ldots$ in the expression (\ref{brown.15})
  for $[z_5,\epsilon_4]$ are not reproduced by the binomials in the corollary $0= {\rm ad}_{\epsilon_0}^{4}\big([\epsilon_4,\epsilon_{10}] - 3 [\epsilon_6,\epsilon_{8}] \big) = \sum_{j=0}^4 {4 \choose j}
  \big([\epsilon_4^{(j)},\epsilon_{10}^{(4-j)}] - 3 [\epsilon^{(j)}_6,\epsilon^{(4-j)}_{8}] \big)$ 
of (\ref{MGFtoBR23}). The representation of $[z_5,\epsilon_4]$ in (\ref{brown.15})
is singled out by having no contribution of $[\epsilon_4^{(j)},\epsilon_{10}^{(4-j)}] $,
but it is not possible to eliminate all the $[\epsilon^{(j)}_6,\epsilon^{(4-j)}_{8}]$ with $j=0,1,2,3,4$.}
With this convention, the $\overline{\alpha_{\rm hard}}$ are individually well-defined by
(\ref{MGFtoBR42}), see the closed formula in appendix \ref{app.A}, which would 
not be the case when employing relations in Tsunogai's derivation algebra from the beginning. 
Beyond depth two, where we do not present a closed form analogous to \eqref{MGFtoBR44}, 
this convention will no longer fix all possible shifts of 
$[z_{2m+1},\epsilon_{k\neq 0}]$-relations by relations in Tsunogai's algebra (or ${\rm ad}_{\epsilon_0}$-actions thereon). Hence, additional criteria must be imposed to land on 
a canonical form for depth $\geq 3$ contributions to the commutation relations.

\subsection{Projecting out cusp forms}
\label{sec:cusp}

Holomorphic cusp forms 
$\Delta_{2s}(\tau)$ do not arise in the differential equations of
MGFs \cite{DHoker:2016mwo, DHoker:2016quv, Gerken:2018zcy}
and their generating series \cite{Gerken:2019cxz}. Hence, iterated-integral
representations of MGFs can only involve those combinations of 
$\beta^{\rm eqv}$ in (\ref{MGFtoBR32}) where the cusp-form
contributions $\beta^{\rm sv}_{\Delta}$ cancel.
These cancellations can be conveniently implemented by
dressing with Tsunogai's derivations
\begin{align}
&J^{\rm eqv}\big( \{\epsilon_k\};\tau \big) =
\sum_P \epsilon[P] \beta^{\rm eqv}[P;\tau]
\notag \\ 
&\quad\quad = 1 + \sum_{k=4}^\infty \sum_{j=0}^{k-2} \frac{(-1)^j (k{-}1) }{(k{-}2{-}j)!}
\beqv{j }{ k}{\tau}  \epsilon_{k}^{(k-j-2)} 
\label{select.1} \\
&\quad\quad \quad +  \sum_{k_1=4}^\infty  \sum_{k_2=4}^\infty \sum_{j_1=0}^{k_1-2}   \sum_{j_2=0}^{k_2-2} \frac{(-1)^{j_1+j_2} (k_1{-}1) (k_2{-}1) }{(k_1{-}2{-}j_1)!(k_2{-}2{-}j_2)!} 
\beqv{  j_1 &j_2 }{ k_1 &k_2}{\tau}
 \epsilon_{k_2}^{(k_2-j_2-2)} \epsilon_{k_1}^{(k_1-j_1-2)}  + \ldots
\notag
\end{align}
as in the generating series (\ref{MGFtoBR21}) of closed-string integrals over the torus.
The words $\epsilon[P]$ in derivations $\epsilon_k$ are defined in 
(\ref{MGFtoBR22}) without committing to a matrix representation, and
the ellipsis in the last line of (\ref{select.1}) refers to 
$\beta^{\rm eqv}[\begin{smallmatrix}  j_1 &\ldots &j_\ell \\ k_1 
&\ldots &k_\ell \end{smallmatrix} ]$ of depth $\ell\geq 3$.

While all $\beta^{\rm sv}_\Delta[ \begin{smallmatrix} j \\ k \end{smallmatrix};\tau]$ at depth one
and $\beta^{\rm sv}_\Delta[ \begin{smallmatrix} j_1 &j_2 \\ k_1 &k_2 \end{smallmatrix};\tau]$
 in (\ref{MGFtoBR32}) at $k_1{+}k_2<14$ vanish, their simplest non-trivial instances occur in 
 the Laplace eigenfunctions $ {\rm F}^{\pm(s)}_{m,k}$ with $m{+}k\geq 7$ and $s\geq 6$
\cite{Dorigoni:2021jfr, Dorigoni:2021ngn} as well as their $\tau,\bar \tau$ derivatives.
The results in the references translate into
\beq
\bsvCUSP{j_1 & j_2}{k_1 & k_2}{\tau} = \sum_{\Delta_{2s} \ {\rm at} \atop{ 2s\leq k_1+k_2-2} }
\xi^{\Delta_{2s}}_{k_1,k_2} \ratc^{j_1,j_2}_{k_1,k_2,N} 
 \bsvBR{j_1 +j_2-N}{ \Delta_{2s}^{\pm} }{\tau}\, ,
   \label{bequv.12}
\eeq
where $N=\frac{1}{2}(k_1{+}k_2) {-} s {-} 1$, and the
rational numbers $\ratc^{j_1,j_2}_{k_1,k_2,N}$ are given by
\begin{align}
\ratc^{j_1,j_2}_{k_1,k_2,N} &= \frac{ 1 }{ k_1! k_2!}  \sum_{\ell=0}^N   (-1)^\ell {N \choose \ell} \,
\frac{j_1!  (k_1{-}2{-}j_1)!    j_2!  (k_2{-}2{-}j_2)! }{
 (j_1{-}N{+}\ell)!  (k_1{-}j_1{-}2{-}\ell)!  (k_2{-}2{-}j_2{-}N{+}\ell)! (j_2{-}\ell)! }  \, .
  \label{bequv.13}
\end{align}
Moreover, we introduce the following delta-variant of the $\beta^{\rm sv}$ 
in (\ref{MGFtoBR07}) with $j=0,1,\ldots,k{-}2$,
\begin{align}
\bsvBR{j}{\Delta^{\pm}_{k}}{\tau} &= ( 2\pi i )^{k-1} \bigg\{
\int_{\tau}^{i\infty}  \dd \tau_1  \bigg(\frac{\tau{-}\tau_1}{4y}\bigg)^{k-2-j}
 (\bar \tau{-}\tau_1)^j \,  \Delta_k(\tau_1)   \notag \\
&\quad \mp \int_{\bar \tau}^{-i\infty} \! \dd \bar \tau_1 \, 
 \bigg(\frac{\tau{-} \bar \tau_1}{4y}\bigg)^{k-2-j}  
 (\bar \tau{-}\bar \tau_1)^j  \, \overline{\Delta_k(\tau_1) }  \bigg\}\, .
\label{deltavar}
\end{align} 
The constants $\xi^{\Delta_{2s}}_{k_1,k_2}$ in (\ref{bequv.12})
do not depend on $j_1,j_2$ and are expected to 
be transcendental, for instance \cite{Dorigoni:2021ngn}
\begin{align}
\big(\xi^{\Delta_{12}}_{4,10}, \, \xi^{\Delta_{12}}_{6,8} \big)
&= \bigg({-} \frac{32}{45 }
, \,  \frac{ 96}{175 }
\bigg)  \frac{ \Lambda(\Delta_{12},12)}{  \Lambda(\Delta_{12}, 10)}\, ,
\notag \\
\big( \xi^{\Delta_{12}}_{4,12}   , \, \xi^{\Delta_{12}}_{6,10}    , \, \xi^{\Delta_{12}}_{8,8}    \big)   
&=\bigg({-}\frac{576}{3455}, \,
\frac{75}{691}, \,
{-}\frac{64}{691} \bigg)
  \frac{  \Lambda(\Delta_{12}, 13)}{ \Lambda(\Delta_{12}, 11)}
       \, ,
\label{xiexpl} \\
\big( \xi^{\Delta_{16}}_{4,14}  ,\,  \xi^{\Delta_{16}}_{6,12} , \, \xi^{\Delta_{16}}_{8,10} \big)
&=
\bigg(
{-} \frac{96}{ 35} ,\,
\frac{432}{385} ,\,
 {-} \frac{32}{ 49 }
\bigg)
 \frac{ \Lambda(\Delta_{16}, 16) }{ \Lambda(\Delta_{16}, 14) } \, ,
\notag \\
\big( \xi^{\Delta_{12}}_{4,14}  ,\,  \xi^{\Delta_{12}}_{6,12} , \, \xi^{\Delta_{12}}_{8,10} \big)
&=
\bigg(
{-} \frac{672}{  44915} ,\,
\frac{8}{975} ,\,
 {-} \frac{8}{ 1365 }
\bigg)
 \frac{ \Lambda(\Delta_{12}, 14) }{ \Lambda(\Delta_{12}, 10) } \, .
 \notag
\end{align}
As explained in the reference, the $\xi^{\Delta_{2s}}_{k_1,k_2}$
are ratios of $L$-values $ \Lambda(\Delta_{2s}, t)$ outside and 
inside the critical strip $t \in (0,2s)$ (with critical denominators 
$ \Lambda(\Delta_{2s}, 2s{-}2)$ and $ \Lambda(\Delta_{2s}, 2s{-}1)$
if $s{+}\frac{1}{2}(k_1{+}k_2)$ is odd and even, respectively). The complete
list of $\xi^{\Delta_{2s}}_{k_1,k_2}$ with $k_1{+}k_2\leq 24$ can be found 
in the ancillary files of the arXiv submission.

The simplest instance of integrals (\ref{deltavar}) over holomorphic cusp forms
occurs in the modular invariant ${\rm F}^{-(6)}_{2,5}(\tau)$, where the
$\beta^{\rm eqv}$-representation in (\ref{moreex.03})
gives rise to
\begin{align}
{\rm F}^{-(6)}_{2,5}(\tau) \, \big|_{\Delta} &=
1890 \bsvCUSP{4& 1 }{ 10& 4}{\tau} -1890 \bsvCUSP{1& 4 }{ 4&10}{\tau}
+ 1512 \bsvCUSP{5& 0 }{ 10& 4}{\tau}  - 1512 \bsvCUSP{2& 3 }{ 4& 10}{\tau} \notag \\
&=  \frac{ \Lambda(\Delta_{12}, 12) }{18000 \Lambda(\Delta_{12}, 10) } \bsvBR{5}{\Delta^{-}_{12}}{\tau} \, ,
\label{moreex.00}
\end{align}
and the notation $|_{\Delta}$ on the left-hand side instructs us to suppress 
all Eisenstein integrals and depth-zero terms. 
More generally, the sign in the superscript of ${\rm F}^{\pm(s)}_{m,k}$ 
matches that of the $\beta^{\rm sv} \big[\begin{smallmatrix} j \\ \Delta^{\pm}_{2s}
 \end{smallmatrix}  ;\tau \big] $ in the iterated-integral representations of
 ${\rm F}^{\pm(s)}_{m,k}$ and their modular derivatives. In particular, the 
$\beta^{\rm sv} \big[\begin{smallmatrix} j \\ \Delta^{\pm}_{2s} \end{smallmatrix}  ;\tau \big] $ 
with the middle value $j=s{-}1$ correspond to the solutions
${\rm H}^{\pm}_{\Delta_{2s}}$ of homogeneous Laplace equations\footnote{Larger 
(smaller) values of $j$ correspond to the $(j{-}s{+}1)^{\rm th}$ $\tau$-derivative
and $(s{-}1{-}j)^{\rm th}$ $\bar \tau$-derivative of ${\rm H}^{\pm}_{\Delta_{2s}}$, respectively,
where $ \nabla_\tau=2i (\Im \tau)^2 \partial_\tau$:
\begin{align*}
\bsvBR{s-1+m}{\Delta^{\pm}_{2s}}{\tau} &= -2(-4)^m (s{-}1{-}m)! (\pi \nabla_\tau)^m  {\rm H}^{\pm}_{\Delta_{2s}}(\tau)\,, \\
\bsvBR{s-1-m}{\Delta^{\pm}_{2s}}{\tau} &= \frac{ -2 (s{-}1{-}m)! (\pi \overline{\nabla}_\tau)^m  {\rm H}^{\pm}_{\Delta_{2s}}(\tau) }{ (-4)^m y^{2m} }\,.
\end{align*}}
\cite{Dorigoni:2021ngn},
\beq
\bsvBR{s-1}{\Delta^{\pm}_{2s}}{\tau} = -2(s{-}1)! \, {\rm H}^{\pm}_{\Delta_{2s}}(\tau)\, ,
\label{vanish.02}
\eeq
whose appearance in the ${\rm F}^{\pm(s)}_{m,k}$ is discussed in the reference.
Note that the shuffles (\ref{shff.02}) relate $\beta^{\rm sv}_{\Delta}
\big[\begin{smallmatrix} j_2 &j_1 \\ k_2 &k_1 \end{smallmatrix}  ;\tau \big]
= - \beta^{\rm sv}_{\Delta}
\big[\begin{smallmatrix} j_1 &j_2 \\ k_1 &k_2 \end{smallmatrix}  ;\tau \big]$
which together with the parity 
property $A_{k_2,k_1,N}^{j_2,j_1} = (-1)^N A_{k_1,k_2,N}^{j_1,j_2}$ of (\ref{bequv.13}) implies that 
\beq
\xi_{k_2,k_1}^{\Delta_{2s}} = (-1)^{s+\frac{1}{2}(k_1+k_2) } \xi_{k_1,k_2}^{\Delta_{2s}}\, .
\label{parityxi}
\eeq
Starting from depth three, the $\beta^{\rm sv}_{\Delta}$ will generically
involve both (\ref{deltavar}) and double integrals that combine kernels
${\rm G}_k$ and $\Delta_{2s}$ and will be investigated in \cite{depth3paper}.

All the $\beta^{\rm sv} \big[\begin{smallmatrix} j \\ \Delta^{\pm}_{k} \end{smallmatrix}  ;\tau \big] $ 
and their higher-depth generalizations cancel from the generating series (\ref{select.1})
since their coefficients vanish by the relations in the
derivation algebra \cite{Tsunogai, LNT, Pollack}:
The overall coefficient in \eqref{select.1} of $\frac{ \Lambda(\Delta_{12},12)}{  \Lambda(\Delta_{12}, 10)} 
\beta^{\rm sv} \big[\begin{smallmatrix} j \\ \Delta^{-}_{12} \end{smallmatrix}  ;\tau \big] $
is, for instance, proportional to
\beq
{\rm ad}_{\epsilon_0}^{10-j}\big( [\epsilon_4,\epsilon_{10} ] - 3 [\epsilon_6,\epsilon_{8}] \big) = 0
\, ,
\label{vanish.01}
\eeq
and more general coefficients of ${\rm H}^{\pm}_{\Delta_{2s}}$ in the series (\ref{select.1}) are
described in section~4 of \cite{Dorigoni:2021ngn}. By the depth-three
terms $[\epsilon_4,[\epsilon_4,\epsilon_8]]$ and
$[\epsilon_6,[\epsilon_6,\epsilon_4]]$ in the last relation of (\ref{MGFtoBR23}),
the cancellation of $ \frac{ \Lambda(\Delta_{12},13)}{  \Lambda(\Delta_{12}, 11)}
\beta^{\rm sv} \big[\begin{smallmatrix} j \\ \Delta^{+}_{12}\end{smallmatrix}  ;\tau \big]$
also hinges on contributions from $\beta^{\rm sv}_\Delta[ \begin{smallmatrix} j_1 &j_2 &j_3 \\ 4 &4 &8 \end{smallmatrix};\tau]$ and $\beta^{\rm sv}_\Delta[ \begin{smallmatrix} j_1 &j_2 &j_3 \\ 6 &6 &4 \end{smallmatrix};\tau]$, see \cite{depth3paper} for details.

Given the cancellation of cusp-form contributions, the differential equation
of the generating series (\ref{select.1}) is determined by the terms in the $\tau$-derivatives 
of $\beta^{\rm eqv}$ displayed in (\ref{MGFtoBR33}):
\begin{align}
-2\pi i &(\tau{-}\bar \tau)^2\partial_\tau J^{\rm eqv}\big( \{\epsilon_k\};\tau \big) 
= {\rm ad}_{\epsilon_0}  J^{\rm eqv}\big( \{\epsilon_k\};\tau \big)
 \label{MGFtoBR61} 
+ \sum_{m=4}^\infty
(m{-}1) (\tau{-}\bar \tau)^m {\rm G}_m(\tau) \epsilon_m 
J^{\rm eqv}\big( \{\epsilon_k\};\tau \big) \, .
\end{align}
Up to convention-dependent powers of $\tau{-}\bar \tau$ and
normalization factors of the holomorphic Eisenstein series, this
matches the holomorphic derivative of Brown's generating series $J^{\rm eqv}$
in section 8.2 of \cite{Brown:2017qwo2}. In comparison to the differential equation
of the generating series $Y^\tau_{\vec{\eta}}$ in section 2.4 of \cite{Gerken:2020yii},
left-action of the operator $R_{\vec{\eta}}(\epsilon_0)$ is replaced by the adjoint action
${\rm ad}_{\epsilon_0}$ in (\ref{MGFtoBR61}) since $\partial_\tau$ no longer acts on 
the exponential in (\ref{MGFtoBR21}).

\section{Connection with equivariant iterated Eisenstein integrals}
\label{sec:4}

In this section, we relate the modular forms $\beta^{\rm eqv}$ introduced in the preceding section to Brown's EIEIs by equating certain generating series.

\subsection{Matching $\beta^{\rm eqv}$ with Brown's construction}
\label{sec:4.1}

We begin by describing an alternative way to assemble the generating series
(\ref{select.1}) of the modular forms $\beta^{\rm eqv}$: Instead of combining the holomorphic and
anti-holomorphic iterated Eisenstein integrals $\beta_{\pm}$ into the $\beta^{\rm sv}$
as in (\ref{MGFtoBR12}), we organize them into separate generating series
\beq
J_{\pm}\big( \{ \epsilon_k\};\tau\big) = \sum_P \epsilon[P] \beta_{\pm}[P;\tau]\, ,
 \label{MGFtoBR62}
\eeq
see (\ref{MGFtoBR13}) for $\beta_{\pm}$ and (\ref{MGFtoBR22}) for the 
words $\epsilon[P]$ in $\epsilon_k$. The sums in (\ref{MGFtoBR62}) and below
over $P$ are again over all words $P=\begin{smallmatrix} j_1 &\ldots &j_\ell \\ 
k_1 &\ldots &k_\ell \end{smallmatrix}$ of length $\ell \geq 0$ with $k_i\geq 4$ even and
$0\leq j_i \leq k_i{-}2$.

The central result of this work is that the modular forms $\beta^{\rm eqv}$
constructed in (\ref{MGFtoBR32}) can be alternatively generated via
\beq
J^{\rm eqv}\big( \{\epsilon_k\};\tau \big) =
J_{+}\big( \{ \epsilon_k\};\tau\big)
B^{\rm sv}\big( \{ \epsilon_k\};\tau\big)
 \phi^{\rm sv} \big(\widetilde{ J_{-}}( \{ \epsilon_k\};\tau) \big)\, .
 \label{MGFtoBR63}
\eeq
The ingredients $B^{\rm sv}$ and $\phi^{\rm sv}$ follow Brown's construction in
\cite{Brown:2017qwo2}, see section \ref{sec:4.1.1} for the series $B^{\rm sv}$ in MZVs,
and the change of alphabet $\phi^{\rm sv}$ will be made fully 
explicit in section \ref{sec:4.1.2} below. The tilde of 
$\widetilde{J_{-}}( \{ \epsilon_k\};\tau)$ instructs us to reverse
the words $\epsilon_{k_1}^{(j_1)}   \ldots \epsilon_{k_\ell}^{(j_\ell)} 
\rightarrow \epsilon_{k_\ell}^{(j_\ell)}   \ldots \epsilon_{k_1}^{(j_1)} $
without additional $j_i$-dependent minus-signs that one may have expected from
the ${\rm ad}_{\epsilon_0}$-action. Note that the change of alphabet $\phi^{\rm sv}$ 
is performed {\it after} reversal of the word in 
$ \epsilon_{k_i}^{(j_i)} $, and it would lead to an incorrect expression
if the reversal is applied to the image of $\phi^{\rm sv}$.

The series $J^{\rm eqv}$ with
the modular transformation $\epsilon_k \rightarrow (c \bar \tau{+}d)^{k-2} \epsilon_k$
of its bookkeeping variables under $\big(\begin{smallmatrix} a&b \\ c &d 
\end{smallmatrix}\big) \in SL(2,\mathbb Z)$ \cite{Brown:2017qwo, Brown:2017qwo2}
is modular invariant and referred to as generating EIEIs. As emphasized before, by slight 
abuse of notation, we allude to equivariance in the superscript of the component 
integrals $\beta^{\rm eqv}$ in (\ref{select.1}) even though they transform as modular 
forms according to (\ref{MGFtoBR34}).

\subsubsection{The generating series $B^{\rm sv}$ in single-valued MZVs}
\label{sec:4.1.1}

The series $B^{\rm sv}$ of single-valued MZVs in (\ref{MGFtoBR63}) is claimed to be 
constructed from the constants $\ccsvpure$ in the conversion (\ref{MGFtoBR32}) from
$\beta^{\rm sv}$ to $\beta^{\rm eqv}$. However, the
rational dependence of the $\ddsvpure$ on $y=\pi \Im \tau$ in (\ref{MGFtoBR32b})
needs to be augmented by additional powers of $\bar \tau$ in the numerator
\begin{align}
\bbsv{\ldots &j_i &\ldots}{\ldots &k_i &\ldots}{\tau}
&= \sum_{p_i=0}^{k_i-2-j_i}  \sum_{\ell_i=0}^{j_i+p_i}  {k_i{-}2{-}j_i \choose p_i}
{j_i{+}p_i \choose \ell_i}  
\frac{ ({-}2\pi i \bar \tau)^{\ell_i} }{(4y)^{p_i}}
\ccsv{\ldots &j_i-\ell_i+p_i &\ldots}{\ldots &k_i &\ldots}\, ,
 \label{MGFtoBR64}
\end{align}
where we have one double-sum over $p_i,\ell_i$ per column.
The rational dependence on $\tau$ and $\bar \tau$ can be
understood from the transformation between the coefficient $\ccsvpure$
of $Y_i^{j_i} X_i^{k_i-j_i-2}$ and $\bbsvpure$ of 
$(X_i{-}\tau Y_i)^{j_i} (X_i{-}\bar \tau Y_i)^{k_i-j_i-2}$,
similar to the transformation of integration kernels in (\ref{omegavsXY}).

At depth one, the additional $\bar \tau$-dependence still cancels by
the simple form of $\ccsv{ j }{ k }$ in (\ref{MGFtoBR35}),
\beq
\bbsv{j }{ k}{\tau} = -\frac{2 \zeta_{k-1}}{(k{-}1) (4y)^{k-2-j}} \, ,
 \label{brown.07} 
\eeq
but already the simplest depth-two examples depend non-trivially
on $\bar \tau$ and therefore vary under the modular $T$-transformation,
\begin{align}
\bbsv{1 &0 }{ 4 &4}{\tau} &=  \frac{ \zeta_3}{2160}
- \frac{i \bar \tau \pi \zeta_3}{2160 y} 
 - \frac{ \bar \tau^2 \pi^2 \zeta_3}{8640 y^2} 
+ \frac{ \zeta_3^2}{288 y^3} 
- \frac{ 5 \zeta_5}{1728 y^2}\, ,
\notag \\
\bbsv{2 &0 }{ 4 &4}{\tau} &= -\frac{ i \bar \tau \pi \zeta_3 }{540}  - 
\frac{ \bar \tau^2 \pi^2 \zeta_3}{1080 y} + \frac{ \zeta_3^2}{72 y^2} - 
\frac{ 5 \zeta_5}{216 y}\, ,
 \label{brown.09}  \\
\bbsv{2 &1 }{ 4 &4}{\tau} &= \frac{ \bar \tau^2 \pi^2 \zeta_3}{540}  + \frac{ \zeta_3^2}{18 y} 
+ \frac{5 \zeta_5}{108}\, .
\notag 
\end{align}
The alternative expression (\ref{MGFtoBR63}) for $J^{\rm eqv}$
is then built from the following generating series of $\bbsvpure$:
\begin{align}
B^{\rm sv}\big( \{ \epsilon_k\};\tau\big) &= \sum_P \epsilon[P] \bbsvpure[P;\tau]\, .
 \label{MGFtoBR65}
\end{align}
%

\subsubsection{The change of alphabet}
\label{sec:4.1.2}

The change of alphabet $\epsilon_k\rightarrow \phi^{\rm sv}(\epsilon_k)$ in (\ref{MGFtoBR63})
acts on the series $\widetilde{J_{-}}$ 
of antiholomorphic iterated Eisenstein integrals in (\ref{MGFtoBR62})
and maps each derivation to an infinite series in single-valued MZVs and nested commutators 
of $\epsilon^{(j)}_k$ \cite{Brown:2017qwo2}. Theorem 7.2 in \cite{Brown:2017qwo2} implicitly  
determines $B^{\rm sv}$ and $\phi^{\rm sv}$ in terms of multiple modular values.
The map $\phi^{\rm sv}$ is an automorphism of the universal enveloping 
algebra of Tsunogai's derivations, and we find its explicit form on single derivations to be given~by
\beq
\phi^{\rm sv} (\epsilon_0) 
=  \epsilon_0\, , \ \ \ \ 
\phi^{\rm sv} (\epsilon_k) 
= \mathbb M^{\rm sv}  \epsilon_k (\mathbb M^{\rm sv})^{-1} \, , \ \ \ \
k\geq 4
\label{defphisv}
\eeq
with the following group-like element:
\begin{align}
\mathbb M^{\rm sv} &= \sum_{\ell=0}^\infty \sum_{i_1,i_2,\ldots,i_\ell  \atop{\in 2\mathbb N+1}}
z_{i_1} z_{i_2}\ldots z_{i_\ell}  \, \isom^{-1}\Big( {\rm sv}(f_{i_1} f_{i_2}\ldots f_{i_\ell}) \Big)
 \label{brown.16} \\
&= 1 + \! \sum_{i_1\in 2\mathbb N+1}   \!  z_{i_1} \, \rho^{-1} \big(  {\rm sv}(f_{i_1}) \big)
+ \!  \!  \sum_{i_1,i_2\in 2\mathbb N+1} \! \! 
z_{i_1} z_{i_2} \, \rho^{-1} \big(  {\rm sv}(f_{i_1} f_{i_2})  \big)
+ \ldots \, .
\notag
\end{align} 
The dependence on the derivations $z_3,z_5,\ldots$ discussed around (\ref{MGFtoBR42}) and (\ref{MGFtoBR44}) will not be displayed in the notation for $\mathbb M^{\rm sv}$. Moreover,
we use the $f$-alphabet description of (motivic) MZVs \cite{Brown:2011ik}\footnote{Both
the $f$-alphabet and the single-valued map are only well-defined in the context 
of motivic MZVs whose elaborate definition
can for instance be found in \cite{Goncharov:2005sla, Brown:2011mot}.} 
with one non-commutative generator $f_i$ for each $i\in 2\mathbb N{+}1$.
The isomorphism $\isom$ mapping MZVs to the $f$-alphabet
\begin{align}
\isom( \zeta_{i} ) &= f_i  \, , \ \ \ \  \isom( \zeta_i \zeta_j ) = f_i \shuffle f_j = f_i f_j + f_j f_i \, , \ \ \ \
i,j \in 2\mathbb N{+}1 \, ,
\notag \\
 \isom( \zeta_{3,5} ) &= - 5 f_3 f_5\, , \ \ \ \
\isom( \zeta_{3,5,3} ) = - 5 f_3 f_5 f_3 + \frac{299}{2} f_{11} \, ,
\label{conversiontof} \\
\isom( \zeta_{3,5,3}^{\rm sv} ) &= - 20 ( f_3 f_5 f_3  + f_5 f_3 f_3  ) + 299 f_{11} 
= {\rm sv} \bigg({-}5 f_3 f_5f_3 + \frac{299}{2} f_{11} \bigg)
\notag
\end{align}
is invertible and often denoted by $\phi$ instead of $\isom$ in the mathematics and
physics literature (see \cite{Brown:2011ik, Schlotterer:2012ny} for examples beyond depth one).

In (\ref{conversiontof}), we have given an example of the single-valued map in the $f$-alphabet. 
It takes the following simple form in the general case
\cite{Schnetz:2013hqa, Brown:2013gia}
\beq
{\rm sv}(f_2^N f_{i_1} f_{i_2}\ldots f_{i_\ell}) = \delta_{N,0}
\sum_{j=0}^\ell f_{i_j}  \ldots f_{i_2} f_{i_1} \shuffle 
f_{i_{j+1}} f_{i_{j+2}} \ldots f_{i_{\ell}}\, ,
\label{MGFtoBR66}
\eeq
such that the depth-one and depth-two contributions to the change of alphabet (\ref{defphisv}) 
reduce to odd Riemann zeta values by ${\rm sv}(f_i) = 2 f_i$ and ${\rm sv}(f_i f_j) = 2 f_i \shuffle f_j = 2(f_i f_j+f_j f_i)$,
\begin{align}
\phi^{\rm sv} (\epsilon_k) 
&= \epsilon_k +2 \sum_{i_1 \in 2\mathbb N+1} \zeta_{i_1} [z_{i_1},  \epsilon_k] 
 + 2 \sum_{i_1,i_2 \in 2\mathbb N+1} 
\zeta_{i_1} \zeta_{i_2} \big[z_{i_1},[z_{i_2},  \epsilon_k] \big]
+ \ldots\, .
\label{atlowdpt}
\end{align}
Also, at higher depth, each term boils down to nested brackets of $\epsilon_{k_i}^{(j_i)}$ since
the $z_3,z_5,\ldots$ normalize the derivation algebra, see e.g.\ (\ref{brown.15}). 
The all-depth expression for the adjoint action in (\ref{defphisv}) yields 
an infinite series of nested commutators
\begin{align}
\phi^{\rm sv} (\epsilon_k) 
&= \sum_{\ell=0}^\infty  \sum_{i_1,i_2,\ldots,i_\ell  \atop{\in 2\mathbb N+1}}
  [ z_{i_1},[ z_{i_2},\ldots ,[z_{i_{\ell-1}} ,[z_{i_\ell},\epsilon_k]] \ldots ]] 
  \,  \isom^{-1} \Big( {\rm sv}(f_{i_1} f_{i_2}\ldots f_{i_\ell}) \Big)\, ,
\label{brown.17}
\end{align}
which is implicit in Brown's work \cite{Brown:mmv, Brown:2017qwo2}. 
The action of $\phi^{\rm sv}$ on higher-depth 
expressions in the $\epsilon_k$ derivations can be deduced from (\ref{brown.17}) via
$\phi^{\rm sv} (\epsilon_{k_1}^{(j_1)}\epsilon_{k_2}^{(j_2)}\ldots) =
({\rm ad}_{\epsilon_0}^{j_1} \phi^{\rm sv} (\epsilon_{k_1}) )
({\rm ad}_{\epsilon_0}^{j_2} \phi^{\rm sv} (\epsilon_{k_2}))\ldots$
as can be seen from the conjugation action in~\eqref{defphisv}.
%
%
In the context of the series $J^{\rm eqv}$ in (\ref{select.1}), conjugation (\ref{defphisv}) 
of the individual $\epsilon_k$ can be converted to the overall adjoint action
\beq
\phi^{\rm sv}\big( \widetilde{J_{-}}(  \{ \epsilon_k\};\tau)\big) = \mathbb M^{\rm sv}  \widetilde{J_{-}}\big(   \{ \epsilon_k\};\tau\big) (\mathbb M^{\rm sv})^{-1}\, .
\label{beqv.new00}
\eeq
Based on the ideas of \cite{Pollack}, we have determined the 
commutators $[z_{3},\epsilon_{k\leq 14}]$ and $[z_{5},\epsilon_{k\leq 10}]$
and appended them in an ancillary file of the arXiv submission. 
In this way, one can extract the contributions of $\phi^{\rm sv}$
to all the $\beta^{\rm eqv} [ \begin{smallmatrix} j_1 &j_2&j_3 \\ 
k_1 &k_2 &k_3 \end{smallmatrix} ]$ with $k_1{+}k_2{+}k_3\leq 16$ from (\ref{MGFtoBR63}). 

Note that the analogue of $B^{\rm sv},\phi^{\rm sv}$ described in Theorem 7.2 
of \cite{Brown:2017qwo2}\footnote{The series $b^{\rm sv}$ in Brown's work
\cite{Brown:2017qwo2} corresponds to the inverse of $B^{\rm sv}$ defined
in (\ref{MGFtoBR65}).} is only well-defined up to $B^{\rm sv}\rightarrow B^{\rm sv} a^{-1}$
and $\phi^{\rm sv} ( \widetilde{J_{-}} ) \rightarrow a
\phi^{\rm sv} ( \widetilde{J_{-}} ) a^{-1}$ for some series $a$
in $\epsilon^{(j)}_k$ whose coefficients are $\mathbb Q$-linear combinations of single-valued 
MZVs. Such redefinitions amount to
right-multiplication of the series $J^{\rm eqv}$ in (\ref{MGFtoBR63}) by $a^{-1}$. One can 
view our realization (\ref{defphisv}) or (\ref{brown.17}) of $\phi^{\rm sv}$ as fixing a 
particular ``gauge-choice'' of the series $a$, and it would be interesting to compare it with
alternative choices. The one-parameter freedom of shifting 
certain $\ccsvpure [ \begin{smallmatrix} j_1 &j_2&j_3 \\ 
4 &4 &6 \end{smallmatrix} ]$ and $\ccsvpure [ \begin{smallmatrix} j_1 &j_2&j_3 \\ 
4 &6 &6 \end{smallmatrix} ]$ by $\zeta_7$ and $\zeta_3 \zeta_5$ -- see the discussion
below (\ref{svmzvs}) -- corresponds to
leading-depth contributions of schematic form $a = 1 + \zeta_7 [\epsilon^{(j_1)}_4,[\epsilon^{(j_2)}_4,\epsilon^{(j_3)}_6]]+ \zeta_3 \zeta_5 [\epsilon^{(j_1')}_6,[\epsilon^{(j_2')}_6,\epsilon^{(j_3')}_4]] +\ldots$ with $j_1{+}j_2{+}j_3=4$ and $j_1'{+}j_2'{+}j_3'=5$, respectively. This freedom should get
fixed once we impose all $\beta^{\rm eqv}$ at depth four to arise from
the change of alphabet $\phi^{\rm sv}$ in the form (\ref{brown.17}): It requires
at least one unit of depth from $\widetilde{J_{-}}$ to distinguish
$\phi^{\rm sv} ( \widetilde{J_{-}} ) $ from $a \phi^{\rm sv} ( \widetilde{J_{-}} ) a^{-1}$,
so the departure of $a \phi^{\rm sv} ( \widetilde{J_{-}} ) a^{-1}$
from (\ref{brown.17}) with the above depth-three contribution to $a$
is at least fourth order in $\epsilon_k^{(j)}$. The reason why the discussion
of $a$ becomes more pressing at depth three is discussed in section
\ref{sec:unique}.

Moreover, derivations $z_i$ at $i\geq 11$ are only well defined
up to nested commutators of $z_{i_1}z_{i_2}\ldots$ with $i_1{+}i_2{+}\ldots =i$,
which, for instance, leaves an ambiguity of adding $[z_3,[z_5,z_3]]$ to $z_{11}$.
This reflects the fact that the isomorphism $\isom$ to the $f$-alphabet is
non-canonical, e.g.\ the choice in (\ref{conversiontof}) sets the coefficient 
of $f_{11}$ in $\isom(\zeta_{3,3,5})$ to zero by convention. For a given
choice of setting up the $f$-alphabet, there is a preferred scheme of
fixing the ambiguity of $z_{i\geq 11}$, e.g.\ a specific representative of $z_{11}$ adapted
to having vanishing coefficient of $f_{11}$ in $\isom(\zeta_{3,3,5})$.\footnote{More generally,
the $\rho$-images of MZVs at weights $w\leq 16$ in \cite{Schlotterer:2012ny} are taken to 
have a vanishing coefficients of $f_{w}$ for each element of the bases in the
datamine \cite{Blumlein:2009cf} (where $f_{2k} = \frac{\zeta_{2k} }{(\zeta_2)^k}f_2^k$ 
for even weight). As a result, the $\rho$-images of the weight-thirteen $c^{\rm sv}$ in 
(\ref{bequv.9}) are computed from
\begin{align*}
\rho(\zeta_{5,3,5}) &= 
 {-} 60 f_3 f_5 f_5  - 5 f_5 f_3 f_5
 + \frac{24}{35}  f_2^4 f_5
 + \frac{1003 }{2} f_{13}
\, , \\
\rho(\zeta_{3,7,3}) &= 
 {-} 66 f_3 f_5 f_5   - 6 f_5 f_3 f_5 - 6 f_5 f_5 f_3 - 14 f_3 f_7 f_3
 + \frac{144 }{175} f_2^4 f_5
+ 716 f_{13}
\, ,
\end{align*}
which follow from the absence of $f_{13}$ in the basis elements $\rho(\zeta_{3,5,5})$ and $
\rho(\zeta_{3,3,7})$ \cite{Blumlein:2009cf, Schlotterer:2012ny}.} One then 
arrives at the same change of alphabet $\phi^{\rm sv}$ for any choice of $f$-alphabet
upon mapping back to (motivic) MZVs via $\isom^{-1}$ in (\ref{brown.17}). 

As will be detailed in future work, a conjugation formula similar to (\ref{defphisv})
applies to the letters in the generating series of antiholomorphic
genus-zero polylogarithms that enter Brown's construction of single-valued
polylogarithms in one variable \cite{Brown:2004ugm}.

\subsubsection{Extracting components}
\label{sec:4.1.3}

By extracting the coefficients of $\epsilon[P]$ in (\ref{MGFtoBR63}), one can read
off components
\begin{align}
\beta^{\rm eqv}[P;\tau] =
\beta_\Delta^{\rm sv}[P;\tau] +\! \!  \sum_{P=XYZ}  \! \! \beta^{\phi}_-[X^t;\tau]  \bbsvpure[Y;\tau]
\beta_+[Z;\tau]  \, ,
\label{beqv.new01}
\end{align}
where the cusp-form contributions $\beta_\Delta^{\rm sv}[P;\tau]$ are projected out
from $J^{\rm eqv}$ by the relations among the $\epsilon_k$, see section \ref{sec:cusp}. 
We employ the shorthand
\beq
 \beta^{\phi}_-[P^t;\tau] = \phi^{\rm sv} \big( \widetilde{J_{-}}(  \{ \epsilon_k\};\tau)\big) \, \big|_{  \epsilon[P] }
 \label{beqv.new02}
\eeq
for the iterated integrals of antiholomorphic Eisenstein series
deformed by the change of alphabet in (\ref{brown.17}). Since
any term in $[z_m,\epsilon_k]$ involves at least two letters
$\epsilon_{k_i}^{(j_i)}$ and no instance of $\epsilon_{k_1}^{(j_1)} \epsilon_{k_2}^{(j_2)}$
with $k_1=k_2$, we have
\beq
\bphiminus{j}{k}{\tau} = \bminus{j}{k}{\tau}\, , \ \ 
\bphiminus{j_1 &j_2}{k &k}{\tau} =  \bminus{j_1 &j_2}{k &k}{\tau} \, .
\label{beqv.new03}
\eeq
Hence, the simplest non-trivial corrections 
\beq
\nicedel_\phi\bminus{j_1  &\ldots &j_\ell}{k_1 &\ldots &k_\ell}{\tau}=
\bphiminus{j_1  &\ldots &j_\ell}{k_1 &\ldots &k_\ell}{\tau} 
-\bminus{j_1  &\ldots &j_\ell}{k_1 &\ldots &k_\ell}{\tau} 
\label{beqv.new04}
\eeq
via $\phi^{\rm sv}$ occur at depth two with $k_1\neq k_2$,
where the structure of (\ref{MGFtoBR44}) implies that 
$\nicedel_\phi \beta_-\big[ \begin{smallmatrix}j_1 &j_2 \\ k_1 &k_2\end{smallmatrix} \big]$
$= - \nicedel_\phi \beta_-\big[ \begin{smallmatrix}j_2 &j_1 \\ k_2 &k_1\end{smallmatrix} \big]$
are products of odd zeta values with $\beta_-\big[ \begin{smallmatrix}j \\ k\end{smallmatrix} \big]$ 
of depth one. As one can anticipate from the examples
\begin{align}
\nicedel_\phi\bminus{0 &2}{6&4}{\tau}&= \frac{  \zeta_3 }{105}  \bminus{0}{4}{\tau}\, ,  \notag \\
\nicedel_\phi\bminus{3 &2}{8&4}{\tau}&= \frac{  \zeta_3 }{420}  \bminus{3}{6}{\tau}\, , \label{exbphiminus2} \\
\nicedel_\phi\bminus{3 &2}{8&6}{\tau}&= - \frac{  \zeta_5 }{98000}  \bminus{1}{4}{\tau}\,,
\notag
\end{align}
as well as
\begin{align}
\nicedel_\phi\bminus{1 &2 &2}{4&4&4}{\tau}&= \frac{ 7 \zeta_3 }{360}  \bminus{4}{6}{\tau}\, ,
\notag \\
\nicedel_\phi\bminus{2 &0 &1}{6 &4& 4}{\tau} &=\frac{ 13 \zeta_5}{141750}
 \bminus{0}{4}{\tau}
 - \frac{ \zeta_3}{40}   \bminus{2}{8}{\tau}
 +  \frac{ \zeta_3}{630}   \bminus{0 & 1}{4 &4}{\tau}\, ,
\label{exbphiminus3} \\
\nicedel_\phi\bminus{3 &1 &0}{8 &4& 4}{\tau} &=
 -\frac{  \zeta_3^2 }{ 352800}  \bminus{0}{4}{\tau}
  - \frac{  \zeta_5 }{6048}  \bminus{1}{6}{\tau}
  + \frac{  33 \zeta_3 }{ 1750}  \bminus{3}{10}{\tau} - 
  \frac{ \zeta_3 }{280}   \bminus{2 &0}{6 &4}{\tau}\, ,
  \notag
\end{align}
the $\phi^{\rm sv}$-corrections (\ref{beqv.new04}) to $\beta_-$ of depth $\ell$ 
comprise iterated Eisenstein integrals of depth $\leq \ell{-}1$, and their coefficients
are single-valued MZVs with one to $\ell{-}1$ letters in the $f$-alphabet.
The simplest MGF that receives $\phi^{\rm sv}$-corrections is the non-holomorphic
cusp form $\Im \cplusno{0 &1 &2 &2}{1 &1 &0 &3}$ 
in (\ref{MGFtoBR36}). The real MGFs $\cplusno{a &b &c }{a &b &c}$ at $a{+}b{+}c\leq 6$
and $\cplusno{2 &2 &1 &1}{2 &2 &1 &1}$ with iterated-integral representations 
in \cite{Broedel:2018izr, Gerken:2020yii} and (\ref{moreex.11}) are unaffected by 
$\phi^{\rm sv}$, that is why cuspidal MGFs pioneered in \cite{DHoker:2019txf} and 
further investigated in \cite{Gerken:2020yii, Gerken:2020aju} are essential case
studies of the change of alphabet.

Note that the depth-two examples in (\ref{exbphiminus2}) can be lined up with
a closed formula
\begin{align}
\nicedel_\phi\bminusno{j_1&j_2}{k_1&k_2} &= \left\{ \begin{array}{rl}
\frac{2 (-1)^{j_1} \zeta_{k_1-1} (k_2{-}k_1{+}1)! (k_2{-}k_1{+}2)!
j_2! (k_2{-}j_2{-}2)! \, {\rm B}_{k_2} }{ (k_1{-}1) k_2! (k_2{-}1)!
(k_2{-}j_1{-}j_2{-}2)! (j_1{+}j_2{-}k_1{+}2)! \, {\rm B}_{k_2-k_1+2} }
 \bminusno{j_1+j_2-k_1+2}{k_2- k_1+2} 
&: \ k_1<k_2\, ,\vspace{0.2cm} \\
-\frac{2(-1)^{j_2} \zeta_{k_2-1} (k_1{-}k_2{+}1)!(k_1{-}k_2{+}2)!
j_1! (k_1{-}j_1{-}2)!\, {\rm B}_{k_1} }{(k_2{-}1) k_1!(k_1{-}1)!
(k_1{-}j_1{-}j_2{-}2)! (j_1{+}j_2{-}k_2{+}2)! \, {\rm B}_{k_1-k_2+2} }
  \bminusno{ j_1+j_2-k_2+2 }{ k_1-k_2+2 } 
&: \ k_1>k_2\, , \vspace{0.2cm}\\
0 \ \ \ \ \ \ \ \ \ \ \ \ \ \ \ \ \ \ \ \ \ \ \ \ \ \ \ 
 \ \ \ \ \ \ \ \ \ \ \ \ \ \ \ \ \, &: \ k_1=k_2\, ,
\end{array} \right.
\label{allphis}
\end{align}
which is equivalent to (\ref{beqv.new56}) below and the results of appendix \ref{app.A}.

Since $\phi^{\rm sv}$ does not contribute at
depth one, the simplest examples of the $\beta^{\rm eqv}$ in 
the new description (\ref{beqv.new01}) read
\begin{align}
 \beqvno{j_1}{k_1} &=  \bminusno{j_1}{ k_1} + \bbsvno{j_1}{k_1}  + \bplusno{j_1}{ k_1}\, ,
\notag \\
 \beqvno{j_1& j_2}{k_1& k_2} &=  \bsvCUSPno{j_1 & j_2}{k_1 & k_2}
 +  \bplusno{j_1& j_2}{k_1& k_2}
 + \bbsvno{j_1& j_2}{k_1& k_2}
+ \bminusno{j_2& j_1}{k_2& k_1}
+ \nicedel_\phi\bminusno{j_2 &j_1}{k_2&k_1} \label{beqv.new06} \\
&\quad
+  \bbsvno{j_1}{k_1} \bplusno{j_2}{ k_2} 
+ \bminusno{j_1}{ k_1} \bbsvno{j_2}{k_2} 
+  \bminusno{j_1}{ k_1} \bplusno{j_2}{ k_2}\, ,
\notag
\end{align}
see (\ref{MGFtoBR31}) for the earlier description in terms of $\beta^{\rm sv}$.
We have checked up to $k_1{+}k_2=24$ at depth two and $k_1{+}k_2{+}k_3=16$ 
at depth three that the expansions of (\ref{MGFtoBR63}) and
(\ref{select.1}) in terms of $\epsilon_{k_i}^{(j_i)}$ match 
after iteratively using the commutation relations $[z_m,\epsilon_k]$
in (\ref{brown.15}) and the ancillary files. 

The main conjecture of this work is
that the two constructions (\ref{MGFtoBR63}) and
(\ref{select.1}) of the generating series $J^{\rm eqv}$ agree to all orders
in $\epsilon_{k_i}^{(j_i)}$. This conjecture implies that all 
MGFs -- i.e.\ modular combinations
of $\beta^{\rm sv}$ -- are contained in the components of Brown's EIEIs 
generated by (\ref{MGFtoBR63}). Further corollaries of this conjecture include the
shuffle relations (\ref{shff.02}) of the building blocks $\beta^{\rm sv},\ccsvpure$
of MGFs and the exclusive appearance of single-valued MZVs in the expansion
of MGFs around the cusp as firstly proposed in \cite{Zerbini:2015rss, DHoker:2015gmr}.

\subsubsection{$\ccsvpure$ in the $f$-alphabet}
\label{sec:4.1.4}

The $f$-alphabet also reveals all-order properties of the single-valued MZVs in 
$\ccsvpure \big[ \begin{smallmatrix} j_1 &\ldots&j_\ell \\ k_1 &\ldots &k_\ell \end{smallmatrix} \big]$
 that determine the components (\ref{MGFtoBR64}) of the series $B^{\rm sv}$ in 
(\ref{MGFtoBR63}): The results at depth $\leq 3$ in section \ref{sec:3.1} and the
ancillary files suggest that their instances at $j_i = k_i{-}2$ obey a simple formula
\beq
\isom \big( \ccsv{k_1{-}2 &k_{2}{-}2 &\ldots &k_\ell{-}2 }{ k_1 & k_2 &\ldots &k_\ell} \big)
= \bigg( \prod_{i=1}^\ell \frac{1}{1{-}k_i}\bigg) \, {\rm sv}(f_{k_1-1} f_{k_2-1}
\ldots f_{k_\ell-1})  \ {\rm mod} \ {\rm lower} \ {\rm depth}\,,
\label{csvf.01}
\eeq
for their highest-depth terms, where words of length $<\ell$ in the $f$-alphabet
have been dropped. This is confirmed by (see (\ref{MGFtoBR35}) to
(\ref{bequv.564}))
\beq
\isom \big( \ccsv{k{-}2}{ k} \big)
= \frac{1}{1{-}k}\, {\rm sv}(f_{k-1})  \, , \ \ \ \
\isom\big(\ccsv{2 &2}{ 4 &4} \big) = \frac{1}{9} \,{\rm sv}(f_{3} f_3) \, , \ \ \ \
\isom\big(\ccsv{2 &4}{ 4 &6} \big) = \frac{1}{15} \,{\rm sv}(f_{3} f_5)
\label{csvf.02}
\eeq
as well as the irreducible depth-three MZVs in (\ref{bequv.9})
\begin{align}
\isom\big( \ccsv{2 &2 &4}{ 4 &4 &6} \big)&= - \frac{1}{45} \,{\rm sv}(f_{3} f_3 f_5) 
-\frac{14573}{43200} \,  f_{11}\, ,
\notag \\
\isom \big(\ccsv{2 &4 &4}{ 4 &6 &6}  \big)&= - \frac{1}{75} \, {\rm sv}(f_3 f_5 f_5)
+ \frac{35071931 }{124380000} \, f_{13}\, ,
\label{csvf.03}
\\
\isom \big(\ccsv{2 &2 &6}{ 4 &4 &8}  \big)&=  - \frac{1}{63}\,{\rm sv}(f_{3} f_3 f_7)
-\frac{365983 }{2708720} \, f_{13}\, .
\notag
\end{align}
This resonates with the results of Saad in Lemma 12.3 of \cite{Saad:2020mzv},
where motivic iterated Eisenstein integrals over ${\rm G}_{k_1}(\tau_1) 
{\rm G}_{k_2}(\tau_2) \ldots {\rm G}_{k_\ell}(\tau_\ell)$
are related to $\mathbb Q$-multiples of $f_{k_\ell-1}\ldots f_{k_2-1}f_{k_1-1}$
upon translating MZVs into the $f$-alphabet and discarding lower-depth terms. 
These iterated Eisenstein integrals arise as multiple modular values in the $S$-cocycle of 
$\beta^{\rm sv}\big[ \begin{smallmatrix} j_1 &\ldots&j_\ell \\ k_1 &\ldots &k_\ell \end{smallmatrix} \big]$
and need to be cancelled by the corresponding $\ccsvpure$ in the
modular completions $\beta^{\rm eqv}$. The single-valued map of the
$f_{i}$ encountered in (\ref{csvf.01}) can be traced back to the combination of 
holomorphic and antiholomorphic iterated Eisenstein integrals and the associated 
multiple modular values in the expression (\ref{MGFtoBR12}) for $\beta^{\rm sv}$.

It would be interesting if all-order formulae similar to (\ref{csvf.03}) 
could be found for $\ccsvpure$ with subleading entries $j_i < k_i{-}2$
which still feature irreducible single-valued MZVs beyond depth one.
The simplest depth-three examples of this type are
$c^{\rm sv} \big[ \begin{smallmatrix} j_1 &j_2 &j_3 \\ 4 &6 &6 \end{smallmatrix} \big]$
at $j_1 {+} j_2 {+} j_3 = 8$ (rather than $10$), where for instance
\begin{align}
c^{\rm sv} \big[ \begin{smallmatrix} 0 &4 &4 \\ 4 &6 &6 \end{smallmatrix} \big]
&=  -\frac{ \zeta_{3,5,3}^{\rm sv} }{31500} + \frac{ \zeta_3^2 \zeta_5 }{1575} 
- \frac{ 31 \zeta_{11}}{94500} \, , 
\label{anothercsv} \\
\isom \big( c^{\rm sv} \big[ \begin{smallmatrix} 0 &4 &4 \\ 4 &6 &6 \end{smallmatrix} \big] \big)
&= \frac{1}{3150} \, {\rm sv}(f_3 f_3 f_5 + f_3 f_5 f_3) - \frac{232 }{23625} \, f_{11}\,.
\notag
\end{align}

\subsection{Comparison with the construction via $\beta^{\rm sv}$}
\label{sec:4.2}

In order to compare the new generating function (\ref{MGFtoBR63})
of the modular forms $\beta^{\rm eqv}$ with their earlier construction in terms
of $\beta^{\rm sv}$, we also cast (\ref{MGFtoBR32}) and (\ref{MGFtoBR12}) 
into generating-function form,
\begin{align}
J^{\rm eqv}\big( \{\epsilon_k\};\tau \big) &=
J_{+}\big( \{ \epsilon_k\};\tau\big)
\widetilde{J_{-}}\big(  \{ \epsilon_k\};\tau\big)  
{\cal K}\big(  \{ \epsilon_k\};\tau\big)
D^{\rm sv}\big( \{ \epsilon_k\};\tau\big) \, .
\label{earlierconst}
\end{align}
The cancellation of cusp-form contributions is again incorporated through the relations
among the derivations $\epsilon_k$. The combinations $\ddsvpure$ and $\overline{\kappa}$
of constants $\ccsvpure$ and antiholomorphic $T$-invariants $\overline{\alpha}$
in (\ref{MGFtoBR32b}) and (\ref{MGFtoBR14}) are generated by
\begin{align}
D^{\rm sv}\big( \{ \epsilon_k\};\tau\big) &= \sum_P \epsilon[P] \ddsvpure[P;\tau]
\, , \ \ \ \
{\cal K}\big(  \{ \epsilon_k\};\tau\big)= \sum_P \epsilon[P] \overline{ \kappa[P;\tau]}\, .
 \label{beqv.new20}
\end{align}
In fact, all of $D^{\rm sv}$, ${\cal K}$ and the product $J_{+}\widetilde{J_{-}}$ 
in (\ref{earlierconst}) are individually $T$ invariant. This is different from
(\ref{MGFtoBR63}), where the series $B^{\rm sv}$ of single-valued MZVs
in (\ref{MGFtoBR65}) depends on both $\Re \tau$ and $\Im \tau$ with a non-trivial
$T$-variation.

The goal of this section is to compare the two presentations
(\ref{earlierconst}) and (\ref{MGFtoBR63}) of the generating series
of modular forms $\beta^{\rm eqv}$. In particular, we will
describe the antiholomorphic $T$-invariants $\overline{\alphaBR{ \ldots }{ \ldots }{\tau} }$
in the earlier construction of $\beta^{\rm sv}$ from the perspective of
Brown's work by equating (\ref{earlierconst}) with (\ref{MGFtoBR63})
\begin{align}
&\widetilde{J_{-}}\big(  \{ \epsilon_k\};\tau\big) 
{\cal K}\big(  \{ \epsilon_k\};\tau\big)
D^{\rm sv}\big( \{ \epsilon_k\};\tau\big) 
 = B^{\rm sv}\big( \{ \epsilon_k\};\tau\big)
 \phi^{\rm sv}\big( \widetilde{J_{-}}( \{ \epsilon_k\};\tau)\big)
 \label{beqv.new21}
\end{align}
and solving for the generating series ${\cal K}\big(  \{ \epsilon_k\};\tau\big)$
of the $\overline{\alphaBR{ \ldots }{ \ldots }{\tau} }$ in (\ref{MGFtoBR14})
\begin{align}
{\cal K}\big(  \{ \epsilon_k\};\tau\big)
&= \widetilde{J_{-}}\big(  \{ \epsilon_k\};\tau\big)^{-1}
B^{\rm sv}\big( \{ \epsilon_k\};\tau\big)
  \phi^{\rm sv}\big( \widetilde{J_{-}}( \{ \epsilon_k\};\tau)\big)
D^{\rm sv}\big( \{ \epsilon_k\};\tau\big)^{-1} \, .
 \label{beqv.new22}
\end{align}
The inversion of the series $ \widetilde{J_{-}}$ and $D^{\rm sv}$
can be readily implemented by reversing the entries $P$ of their
coefficients and inserting minus signs for the $|P|$ number of letters 
$\begin{smallmatrix} j \\ k \end{smallmatrix}$ in $P$, i.e.\ by employing
\beq
\bigg( \sum_P   \epsilon[P] C[P] \bigg)^{-1}
= \sum_P (-1)^{|P|} \epsilon[P] C[P^t]
 \label{beqv.new23} 
\eeq
for arbitrary coefficients $C[P]$ subject to shuffle relations
$C[X]C[Y] = \sum_{P \in X \shuffle Y} C[P]$. Hence, 
the components of (\ref{beqv.new22}) yield
\beq
\overline{ \kappa[P;\tau]} = \sum_{P=WXYZ}
(-1)^{|W|+|Z|}  \ddsvpure[W^t;\tau]
\beta^{\phi}_-[X^t;\tau]
 \bbsvpure[Y;\tau]
\beta_-[Z;\tau] \, ,
 \label{beqv.new24} 
\eeq
which determine the antiholomorphic $T$-invariants $\overline{\alphaBR{ \ldots }{ \ldots }{\tau} }$ 
through the inverse
\beq
\overline{ \alphaBR{\ldots &j_i &\ldots}{\ldots &k_i &\ldots}{\tau} }
= \sum_{p_i=0}^{k_i-2-j_i} \frac{ 
{ k_i{-}2 {-}j_i \choose p_i} 
}{(-4y)^{p_i}} 
\overline{\kappaBR{ \ldots &j_i{+}p_i &\ldots}{ \ldots &k_i &\ldots }{\tau} }
\label{invak}
\eeq
of the transformation (\ref{MGFtoBR14}) with one summation over $p_i$ per column.
The depth-one version of the deconcatenation formula (\ref{beqv.new24})
is not sensitive to the reversals of $W$ and $X$, and one can easily verify the vanishing of
\beq
\overline{\kappaBRno{j}{k}}  = 
\bphiminusno{j}{ k} - \bminusno{j}{ k}
+ \bbsvno{j}{k} -  \ddsvno{j}{k}  = 0
 \label{beqv.new25} 
\eeq
since both $\bbsvno{j}{k}-\ddsvno{j}{k} $ and $\bphiminusno{j}{ k}-\bminusno{j}{ k}$ cancel
in view of (\ref{MGFtoBR35}), (\ref{brown.07}) and (\ref{beqv.new03}). In the following,
we shall express $\overline{\kappa[\begin{smallmatrix} \ldots\\ \ldots \end{smallmatrix};\tau]}$ at depths two and three in terms of
the objects $\ddsvpure,\beta^{\phi}_-, \bbsvpure,\beta_-$ related to Brown's
construction. Together with the all-order results for the $\overline{\alphaBR{ \ldots }{ \ldots }{\tau} }$
in (\ref{closedform}) and appendix \ref{app.B}, the subsequent depth-two and depth-three
expressions were the central piece of evidence for the equivalence of the two constructions 
(\ref{MGFtoBR32}) and (\ref{beqv.new01}) of modular forms $\beta^{\rm eqv}$.

\subsubsection{Depth two}
\label{sec:4.2.1}

The expression (\ref{beqv.new24}) for $\overline{ \kappa[P;\tau]}$
at depth $|P|=2$ simplifies to
\begin{align}
\overline{\kappaBRno{j_1 &j_2}{k_1 &k_2} }&= 
\bminusno{j_1 &j_2}{ k_1 &k_2}  
\! - \! \bminusno{j_2}{ k_2} \bphiminusno{j_1 }{ k_1 }
\! + \! \bphiminusno{j_2 &j_1}{ k_2 &k_1}
+\bbsvno{j_1 &j_2}{k_1 &k_2}
\! - \! \bbsvno{j_2}{k_2} \ddsvno{j_1 }{k_1 }
\! +\! \ddsvno{j_2 &j_1}{k_2 &k_1} \notag \\
& \quad
+\bphiminusno{j_1 }{ k_1 } \bbsvno{j_2}{k_2}
- \bbsvno{j_1 }{k_1 }   \bminusno{j_2}{ k_2}
+ \ddsvno{j_1 }{k_1 } \Big(   \bminusno{j_2}{ k_2}  - \bphiminusno{j_2 }{ k_2 } \Big)
 \label{beqv.new26}  \\
&= \nicedel_\phi \bminusno{j_2 &j_1}{ k_2 &k_1}
+\bminusno{j_1 }{ k_1 } \ddsvno{j_2}{k_2}
- \ddsvno{j_1 }{k_1 }   \bminusno{j_2}{ k_2}
+ \bbsvno{j_1 &j_2}{k_1 &k_2}
- \ddsvno{j_1 &j_2}{k_1 &k_2}\, ,
\notag
\end{align}
with $\ddsvno{j }{k}$ given by (\ref{MGFtoBR35}). In passing to the last line,
we have used the agreement of $\beta_-^{\phi},\beta_-$ and $\bbsvpure,\ddsvpure$
at depth one as well as the shuffle property of $\beta_-$ and $\ddsvpure$.

The various terms in~\eqref{beqv.new26} have different interpretations and combine in such a way 
as to make this expression $T$-invariant even though this invariance is not manifest term by term. 
More precisely, the first three terms involving $\delta_\phi \beta_-$ and $\beta_-$ yield
antiholomorphic iterated Eisenstein integrals at depth one with odd zeta values in their coefficients
due to $\phi^{\rm sv}$ or the accompanying $\ddsvpure$. The difference 
$ \bbsvpure \big[ \begin{smallmatrix}j_1 &j_2 \\ k_1 &k_2\end{smallmatrix} \big]
- \ddsvpure \big[ \begin{smallmatrix}j_1 &j_2 \\ k_1 &k_2\end{smallmatrix} \big]$
at the end of (\ref{beqv.new26}) has depth zero from the viewpoint of Eisenstein 
integrals and introduces ratios $ \bar \tau^\ell / y^p$ as in (\ref{MGFtoBR64}) with at least one power 
$\ell \geq 1$ of $\bar \tau$. The relative factors and MZV coefficients of depth-one
and depth-zero contributions play out to recombine every term into the $T$-invariant 
integrals $\overline{{\cal E}_0(\ldots;\tau)}$ in (\ref{MGFtoBR41}), multiplied by
non-positive powers of $y$.

When extracting $ \overline{ \alpha \big[ \begin{smallmatrix}j_1 &j_2 \\ k_1 &k_2\end{smallmatrix} \big] }$ from (\ref{beqv.new26}) via (\ref{invak}), we can clearly identify the sources of the two
contributions $\overline{ \alpha_{\rm easy} }$ and $\overline{\alpha_{\rm hard}}$ in the all-order formula (\ref{closedform}):
\begin{itemize}
\item The simple expression for $\overline{\alpha_{\rm easy}}$ in (\ref{closedform}) can be traced back to
the second and third term $\beta_- \big[ \begin{smallmatrix} j_1 \\ k_1 \end{smallmatrix} \big]
\ddsvpure \big[ \begin{smallmatrix} j_2 \\ k_2\end{smallmatrix} \big]
-\ddsvpure \big[ \begin{smallmatrix} j_1 \\ k_1\end{smallmatrix} \big]
 \beta_- \big[ \begin{smallmatrix} j_2 \\ k_2 \end{smallmatrix} \big]$ in
the last line of (\ref{beqv.new26}).
\item The generating function (\ref{MGFtoBR42}) of $\overline{\alpha_{\rm hard}}$ stems from
the $\delta_\phi \beta_- \big[ \begin{smallmatrix}j_2 &j_1 \\ k_2 &k_1\end{smallmatrix} \big]$
in (\ref{beqv.new26}) which are already fixed by the contributions
$\sim \zeta_{i_1} [z_{i_1},\epsilon_k]$ to $\phi^{\rm sv}(\epsilon_k)$ 
in (\ref{atlowdpt}) and obey the closed formula (\ref{allphis}).
\end{itemize}
In both cases, the depth-zero terms 
$\bbsvpure \big[ \begin{smallmatrix}j_1 &j_2 \\ k_1 &k_2\end{smallmatrix} \big]
- \ddsvpure \big[ \begin{smallmatrix}j_1 &j_2 \\ k_1 &k_2\end{smallmatrix} \big]$ 
in (\ref{beqv.new26}) ensure that the antiholomorphic Eisenstein integrals 
in $\beta_- \big[ \begin{smallmatrix} j \\ k \end{smallmatrix} \big]$ and 
$\delta_\phi \beta_- \big[ \begin{smallmatrix}j_2 &j_1 \\ k_2 &k_1\end{smallmatrix} \big]$
conspire to reconstruct $\overline{{\cal E}_0}$.

Conversely, the generating function (\ref{MGFtoBR42}) of $\overline{\alpha_{\rm hard}}$
inferred from inspecting the variety of examples in \cite{Dorigoni:2021jfr} was
crucial in the early stages of this work to anticipate the significance of the derivations
$z_3,z_5,\ldots$ in the closed-form expression (\ref{brown.17}) for $\phi^{\rm sv}$.
While the matching of $\overline{\alpha_{\rm hard}}$ with
$\delta_\phi \beta_- \big[ \begin{smallmatrix}j_2 &j_1 \\ k_2 &k_1\end{smallmatrix} \big]$
guided the identification of $ \zeta_{i_1} [z_{i_1},\epsilon_k]$-contributions 
to $\phi^{\rm sv}(\epsilon_k)$, the second order $\zeta_{i_1} \zeta_{i_2}[z_{i_1}, [z_{i_2},\epsilon_k]]$
firstly became accessible from the depth-three analysis in the next section.

\subsubsection{Depth three}
\label{sec:4.2.2}

By suitably assembling the $\overline{ \kappa[P;\tau]}$
at depth $|P|=3$ from (\ref{beqv.new24}), we have
\begin{align}
\overline{\kappaBRno{j_1 &j_2 &j_3}{k_1 &k_2 &k_3} }&= 
\bminusno{j_2  &j_1 }{ k_2 &k_1} \ddsvno{ j_3  }{k_3 }
 +\bminusno{  j_2  &j_3  }{ k_2 &k_3 } \ddsvno{ j_1  }{ k_1 }
 - \bminusno{  j_1   }{ k_1 } \bminusno{ j_3  }{ k_3 }\ddsvno{ j_2  }{ k_2 } \notag \\
  &\quad - \bminusno{  j_2   }{ k_2 }
\ddsvno{ j_1 }{ k_1 }  \ddsvno{ j_3  }{ k_3 }
+ \Big(  \ddsvno{ j_1  }{ k_1 }  \ddsvno{ j_2 }{k_2 }
-  \bbsvno{ j_1 &j_2  }{ k_1 &k_2 } \Big)   
\bminusno{ j_3  }{ k_3 }
 \notag \\
 &\quad + \bminusno{  j_1}{ k_1}\bbsvno{ j_2 &j_3  }{ k_2 &k_3 }
 + \nicedel_\phi\bminusno{ j_2 &j_1  }{ k_2 &k_1  } \ddsvno{ j_3  }{ k_3 }
 -  \ddsvno{ j_1  }{ k_1 } \nicedel_\phi\bminusno{  j_3 &j_2  }{ k_3 &k_2 } \label{beqv.new31}\\
&\quad 
+\nicedel_\phi \bminusno{ j_3 &j_2 &j_1 }{ k_3 &k_2 &k_1 } \, \big|_{\rm depth \ 1}
+\nicedel_\phi \bminusno{ j_3 &j_2 &j_1 }{ k_3 &k_2 &k_1 } \, \big|_{\rm depth \ 2}
- \nicedel_\phi \bminusno{ j_2 &j_1 }{ k_2 &k_1 } \bminusno{ j_3  }{ k_3 }\notag \\
 &\quad + \bbsvno{ j_1 &j_2 &j_3  }{ k_1 &k_2 &k_3 } 
 -  \ddsvno{ j_1  }{ k_1 }  \bbsvno{ j_2 &j_3  }{ k_2 &k_3 }
 + \ddsvno{ j_2 & j_1  }{ k_2 &k_1 } \ddsvno{ j_3 }{ k_3 }
 -  \ddsvno{j_3 &j_2 & j_1 }{ k_3& k_2 &k_1 }\, .
\notag
\end{align}
As will be explained in appendix \ref{app.B}, the detailed structure of (\ref{beqv.new31})
harmonizes with an all-weight formula for $\overline{\alphaBR{\ldots}{\ldots}{\tau}}$ of depth three.
The right-hand side features $\beta_-$ of depth $0\leq \ell \leq 2$, for instance, depth-two integrals
in the first line and in the contributions 
$\nicedel_\phi
\beta_- \big[ \begin{smallmatrix}j_3 &j_2 &j_1 \\ k_3 &k_2 &k_1\end{smallmatrix} \big]\, \big|_{\rm depth \ 2}
- \nicedel_\phi
\beta_- \big[ \begin{smallmatrix} j_2 &j_1 \\  k_2 &k_1\end{smallmatrix} \big]
\beta_- \big[ \begin{smallmatrix}j_3   \\ k_3 \end{smallmatrix} \big]$ to the
fourth line. The last line of (\ref{beqv.new31}) in turn has depth zero and
again features at least one power of $\bar \tau$ in each term.

\subsubsection{Comparison with Brown's single-valued iterated Eisenstein integrals}
\label{sec:4.2.3}

According to section 8.1 of \cite{Brown:2017qwo2}, right-multiplication of
the series $J^{\rm eqv}$ of EIEIs with $(B^{\rm sv})^{-1}$ yields Brown's 
{\it single-valued iterated Eisenstein integrals},
\begin{align}
J^{\rm sv}\big( \{\epsilon_k\};\tau \big) &= J^{\rm eqv}\big( \{\epsilon_k\};\tau \big)
B^{\rm sv}\big( \{ \epsilon_k\};\tau\big)^{-1}   \label{beqv.new41} \\
&=
J_{+}\big( \{ \epsilon_k\};\tau\big)
B^{\rm sv}\big( \{ \epsilon_k\};\tau\big)
 \phi^{\rm sv}\big( \widetilde{J_{-}}( \{ \epsilon_k\};\tau)\big)
B^{\rm sv}\big( \{ \epsilon_k\};\tau\big)^{-1} \, ,
\notag
\end{align}
that, using~\eqref{beqv.new00}, can also be written as
\begin{align}
J^{\rm sv}\big( \{\epsilon_k\};\tau \big) &=
J_{+}\big( \{ \epsilon_k\};\tau\big)  \mathbb{N}^{\rm sv}
 \widetilde{J_{-}}\big( \{ \epsilon_k\};\tau\big) ( \mathbb{N}^{\rm sv})^{-1}
 \,,
 \label{reminisc}
\end{align}
where $\mathbb{N}^{\rm sv} = B^{\rm sv}\big( \{ \epsilon_k\};\tau\big) \mathbb{M}^{\rm sv}$ 
depends on both the $\epsilon_k$ and $z_m$. The freedom to redefine
$B^{\rm sv}\rightarrow B^{\rm sv} a^{-1}$ and $\phi^{\rm sv} ( \widetilde{J_{-}} ) \rightarrow a
\phi^{\rm sv} ( \widetilde{J_{-}} ) a^{-1}$ discussed below
(\ref{beqv.new00}) amounts to the left-multiplication 
$\mathbb{M}^{\rm sv}\rightarrow a \mathbb{M}^{\rm sv}$
and therefore drops out from the product $\mathbb{N}^{\rm sv} = B^{\rm sv}\mathbb{M}^{\rm sv}$. 
That is why $J^{\rm sv}$ is canonically defined and does 
admit redefinitions analogous to
$J^{\rm eqv}\rightarrow J^{\rm eqv} a^{-1}$ \cite{Brown:2017qwo2}.
 
 The form (\ref{reminisc}) is reminiscent of known constructions of single-valued functions and periods (e.g. \cite{Brown:2013gia,Brown:2004ugm,Broedel:2016kls, DelDuca:2016lad}), which are made up of a combination of holomorphic and anti-holomorphic parts where the anti-holomorphic parts are transformed using for example -- in the case of generating functions -- a conjugation.

The expansion of $J^{\rm sv}$ is to be contrasted with the generating series of the $\beta^{\rm sv}$ 
in the physics literature
\begin{align}
\sum_P \beta^{\rm sv}[P;\tau]  \epsilon[P]  &=  J^{\rm eqv}\big( \{\epsilon_k\};\tau \big)
D^{\rm sv}\big( \{ \epsilon_k\};\tau\big)^{-1} 
 \label{beqv.new42}
\\
&=
J_{+}\big( \{ \epsilon_k\};\tau\big)
B^{\rm sv}\big( \{ \epsilon_k\};\tau\big)
\phi^{\rm sv}\big( \widetilde{J_{-}}( \{ \epsilon_k\};\tau)\big)
D^{\rm sv}\big( \{ \epsilon_k\};\tau\big)^{-1} \, , \notag
 \end{align}
which follows from (\ref{MGFtoBR12}) and (\ref{beqv.new21}) and features a right-multiplicative
inverse $D^{\rm sv}\big( \{ \epsilon_k\};\tau\big)^{-1} $ instead of
the $B^{\rm sv}\big( \{ \epsilon_k\};\tau\big)^{-1} $ in
(\ref{beqv.new41}). Thus it follows that 
\begin{equation}
 J^{\rm sv}\big( \{\epsilon_k\};\tau \big) = \left(\sum_P \beta^{\rm sv}[P;\tau]  \epsilon[P] \right)D^{\rm sv}\big( \{ \epsilon_k\};\tau\big)
 B^{\rm sv}\big( \{ \epsilon_k\};\tau\big)^{-1}\,.
\end{equation}
The depth-one components $\bbsvno{j}{k} =  \ddsvno{j}{k} $ 
still happen to agree, and $\bsvBRno{j}{k}$ coincide with Brown's single-valued
iterated Eisenstein integrals at depth one. At depth $\ell \geq 2$, however,
the $\bbsvpure$ generically depart from $\ddsvpure$ by
the polynomial dependence on $\bar \tau$ in (\ref{MGFtoBR64}). Hence, by the mismatch
between (\ref{beqv.new41}) and (\ref{beqv.new42}), the 
$\beta^{\rm sv} \big[ \begin{smallmatrix} j_1 &\ldots &j_\ell \\
k_1 &\ldots &k_\ell \end{smallmatrix} \big]$ at $\ell \geq 2$
differ from Brown's single-valued iterated Eisenstein integrals by terms
involving at least one non-trivial MZV and one power of $\bar \tau$ in the
numerator. One may view the $\beta^{\rm sv}$ as $T$-invariantized versions 
of Brown's single-valued iterated Eisenstein integrals since, in contrast to the inverse
$B^{\rm sv}$ in (\ref{beqv.new41}), the series
$D^{\rm sv}$ in (\ref{beqv.new42}) is invariant under $\tau \rightarrow \tau{+}1$.

\subsection{Relation to Brown's equivariant double iterated integrals}
\label{sec:4.3}

So far, we have only matched the subspace of the modular forms $\beta^{\rm eqv}$
with Brown's work, where iterated integrals involving holomorphic cusp forms
are projected out by the accompanying $\epsilon_k$. In this section, we go beyond this
subspace and connect the cusp-form contributions $\beta^{\rm sv}_{\Delta}$ to
$\beta^{\rm eqv}$ of depth two in (\ref{MGFtoBR31}) and (\ref{bequv.12})
with Brown's equivariant double iterated integrals of \cite{Brown:2017qwo}.

Following the normalization conventions in the reference, Brown's double integrals
are constructed from holomorphic $(1,0)$-forms\footnote{The expressions
(\ref{deftheta}) for $\underline{\rm G}_k[X,Y;\tau]$
and (\ref{beqv.new51}) below for $M_{k}[X,Y;\tau]$ are denoted by 
$\underline{E}_k$ and ${\cal E}_{k-2}$ in \cite{Brown:2017qwo}.}
\beq
\underline{\rm G}_k[X,Y;\tau] = \frac{ (k{-}1)! }{2 (2\pi i)^{k-1} } \, (X{-}\tau Y)^{k-2} \, {\rm G}_k(\tau)\, \dd \tau
\label{deftheta}
\eeq
involving commutative bookkeeping variables $X,Y$. These one-forms 
are in fact, multiples of the generating functions (\ref{omegavsXY}) of the earlier forms 
$ \ompm{j}{k}{\tau,\tau_1}$ in (\ref{MGFtoBR09}).
Accordingly, iterated integrals of $\underline{\rm G}_{k_1}[X_1,Y_1;\tau_1]
\underline{\rm G}_{k_2}[X_2,Y_2;\tau_2]\ldots$ 
generate the constituents $\beta_{\pm}$ of MGFs in (\ref{MGFtoBR13})
upon expansion in the combinations $(X_i{-}\tau Y_i)^{j_i}(X_i{-}\bar \tau Y_i)^{k_i-j_i-2}$ \cite{Brown:2017qwo}. 

At depth one, the real-analytic combinations 
\beq
M_{k}[X,Y;\tau]  = -\frac{1}{2} \int_\tau^{i \infty} \underline{\rm G}_k[X,Y;\tau_1] 
- \frac{1}{2} \int_{\bar \tau}^{-i \infty} \overline{ \underline{\rm G}_k[X,Y;\tau_1]} 
+ \frac{(k{-}2)!  \zeta_{k-1}}{2(2\pi i)^{k-2}}  \, Y^{k-2}
 \label{beqv.new51}
\eeq
with $k\geq 4$ as well as $\overline{X}=X$ and $\overline{Y}=Y$
 generate non-holomorphic Eisenstein series (\ref{moreex.01}) via
\begin{align}
M_{k}[X,Y;\tau]&= -  \frac{1}{4} (k{-}1)! 
\sum_{j=0}^{k-2}
 \frac{ {k-2 \choose j}  }{(-4y)^{j}}
 \beqv{j}{k}{\tau}  
(X{-}\tau Y)^{j} (X{-}\bar \tau Y)^{k-2-j}  \, ,
  \label{beqv.new52}
  \end{align}
where the additive constant $\sim Y^{k-2}$ in (\ref{beqv.new51})
(not to be confused with $y = \pi \Im \tau$) introduces the odd zeta value into 
the modular forms $\beta^{\rm eqv}$ of depth
one in (\ref{MGFtoBR24}) \cite{Brown:2017qwo}.
The object (\ref{beqv.new52}) is referred to as an equivariant Eisenstein
integral since the action on the bookkeeping variables $(X,Y)$  leads to 
the $SL(2,\mathbb Z)$ invariance
\beq
M_{k}\bigg[aX{+}bY,cX{+}dY;\frac{a\tau{+}b}{c\tau{+}d}\bigg]  = M_{k}[X,Y;\tau] \, .
  \label{beqv.new50}
 \eeq

\subsubsection{The non-modular primitives at depth two}
\label{sec:4.3.1}

At depth two, Brown's construction \cite{Brown:2017qwo} of equivariant double iterated integrals
starts from closed one-forms $\sim \underline{\rm G}_{k_1}[X_1,Y_1;\tau] M_{k_2}[X_2,Y_2;\tau]
+M_{k_1}[X_1,Y_1;\tau] \overline{ \underline{\rm G}_{k_2}[X_2,Y_2;\tau] }$ and considers their 
real-analytic primitive
\begin{align}
&K_{k_1,k_2}[X_1,Y_1,X_2,Y_2;\tau]= \frac{1}{4}\int_{\tau}^{i\infty}  \underline{\rm G}_{k_1}[X_1,Y_1;\tau_1]  \int_{\tau_1}^{i\infty} \underline{\rm G}_{k_2}[X_2,Y_2;\tau_2] 
\label{beqv.new53} \\
&\quad 
+ \frac{1}{4}\int_{\tau}^{i\infty}  \underline{\rm G}_{k_1}[X_1,Y_1;\tau_1]  
 \int_{\bar \tau}^{-i\infty} \overline{ \underline{\rm G}_{k_2}[X_2,Y_2;\tau_2] } 
 + \frac{1}{4}\int_{\bar \tau}^{-i\infty} \overline{ \underline{\rm G}_{k_2}[X_2,Y_2;\tau_2] }   
\int_{\bar \tau_2}^{-i\infty} \overline{ \underline{\rm G}_{k_1}[X_1,Y_1;\tau_1] } 
  \notag\\
  &\quad 
  - \frac{(k_2{-}2)!  \zeta_{k_2-1} }{4(2\pi i)^{k_2-2}}\,Y_2^{k_2-2}  \int_{\tau}^{i\infty}  \underline{\rm G}_{k_1}[X_1,Y_1;\tau_1]  
  - \frac{(k_1{-}2)! \zeta_{k_1-1}}{4(2\pi i)^{k_1-2}} \, Y_1^{k_1-2}  \int_{\bar \tau}^{-i\infty} \overline{ \underline{\rm G}_{k_2}[X_2,Y_2;\tau_2] }  \, ,   \notag
\end{align}
where we have rewritten the integrals of the reference in a manifestly homotopy-invariant
way. In order to avoid cluttering, 
we shall no longer spell out the dependence of $K_{k_1,k_2}$
and related objects on the commutative bookkeeping variables $X_i,Y_i$.
By decomposing the integration kernels $X_i{-}\tau_iY_i$ and $X_i{-}\bar \tau_i Y_i$
as in (\ref{omegavsXY}), one can rewrite (\ref{beqv.new53}) as
\begin{align}
K_{k_1,k_2}& =
\frac{(k_1{-}1)!(k_2{-}1)!}{16}\sum_{j_1=0}^{k_1-2}\sum_{j_2=0}^{k_2-2}
 \frac{ {k_1{-}2 \choose j_1}{k_2{-}2 \choose j_2} }{ (-4y)^{j_1+j_2}}
\Big(
 \bplusno{j_2 &j_1}{k_2 &k_1} - \bplusno{j_1}{k_1} \bminusno{j_2}{k_2}  + \bminusno{ j_1 &j_2}{k_1 &k_2} \Big) \notag\\
&\quad \quad \quad \times    (X_1{-}\tau Y_1)^{j_1}(X_1{-} \bar{\tau} Y_1)^{k_1-2-j_1}
 (X_2{-}\tau Y_2)^{j_2}(X_2{-} \bar{\tau} Y_2)^{k_2-2-j_2}  
\label{eq:depthtwoKeval}  \\
&\quad - \frac{(k_2{-}2)!(k_1{-}1)! \zeta_{k_2-1}}{8 (2\pi i)^{k_2-2}}\sum_{j=0}^{k_1-2}{k_1{-}2 \choose j}\frac{(-1)^j}{(4y)^j}\bplusno{j}{k_1}  (X_1{-}\tau Y_1)^j(X_1{-}\bar{\tau}Y_1)^{k_1-2-j}  Y_2^{k_2-2} \notag\\
&\quad - \frac{(k_1{-}2)!(k_2{-}1)! \zeta_{k_1-1}}{8 (2\pi i)^{k_1-2}}\sum_{j=0}^{k_2-2}{k_2{-}2 \choose j}\frac{(-1)^j}{(4y)^j}\bminusno{j}{k_2} Y_1^{k_1-2} (X_2{-}\tau Y_2)^j(X_2{-}\bar{\tau}Y_2)^{k_2-2-j} \, ,\notag 
\end{align}
where the round bracket in the first line features
$\beta^{\rm sv} \big[ \begin{smallmatrix} j_2 &j_1 \\
k_2 &k_1 \end{smallmatrix} \big]$
up to the lower-depth contributions from $\overline{\alphaBR{ \ldots }{ \ldots }{\tau} }$
in (\ref{MGFtoBR10}). One can anticipate from the variety of lower-depth
terms in the constructions (\ref{MGFtoBR31}) or (\ref{beqv.new06}) of modular forms
$\beta^{\rm eqv} \big[ \begin{smallmatrix} j_1 &j_2 \\
k_1 &k_2 \end{smallmatrix} \big]$ that (\ref{eq:depthtwoKeval}) is not yet 
equivariant in the sense of (\ref{beqv.new50}).

\subsubsection{The modular completion at depth two}
\label{sec:4.3.2}

In the same way as the additive constant $\zeta_{k-1} Y^{k-2}$
in (\ref{beqv.new51}) ensures equivariance of $M_{k}$
at depth one, there is a systematic modular completion $M_{k_1,k_2}$ of $K_{k_1,k_2}$
in (\ref{beqv.new53}) and (\ref{eq:depthtwoKeval}): On top of a polynomial
$c^{\gamma}_{k_1,k_2} $ in $X_i,Y_i$ independent on $\tau$,
the additional complexity at depth two generically requires the addition of
holomorphic and antiholomorphic depth-one integrals \cite{Brown:2017qwo},
\beq
M_{k_1,k_2} = K_{k_1,k_2} - c^{\gamma}_{k_1,k_2} - \frac{1}{2}
\bigg\{ 
\int_{\tau}^{i\infty}  \underbar{f}^{\, 0}_{k_1,k_2}(\tau_1)
+\int_{\bar \tau}^{-i\infty} \!   \overline{\underbar{g}^{\, 0}_{k_1,k_2}(\tau_1)}
+\int_{\bar \tau}^{-i\infty} \!   \overline{\underbar{g}^{\, \rm E}_{k_1,k_2}(\tau_1)} \bigg\}\, .
\label{KtoMd2}
\eeq
As detailed in the reference, $\underbar{f}^{\, 0}_{k_1,k_2}(\tau_1)$ is an equivariant $(1,0)$-form
in $\tau_1$ while $\overline{\underbar{g}^{\, 0}_{k_1,k_2}(\tau_1)}$ and 
$\overline{\underbar{g}^{\, {\rm E}}_{k_1,k_2}(\tau_1)}$  
are equivariant $(0,1)$-forms in $\tau_1$ composed of\footnote{In~\cite{Brown:2017qwo}, the antiholomorphic forms  $\overline{\underbar{g}^{\, 0}_{k_1,k_2}(\tau_1)}$ and $\overline{\underbar{g}^{\, {\rm E}}_{k_1,k_2}(\tau_1)}$ are combined into a single object, but we find it useful to separate their contributions. We have also introduced a superscript $^0$ for the cuspidal parts compared to the notation in~\cite{Brown:2017qwo}.}
\begin{itemize}
\item a holomorphic cusp form $\Delta_{2s}(\tau_1)$ in the case of $\underbar{f}\,{}^0_{k_1,k_2}(\tau_1)$
\item an antiholomorphic cusp form  $\overline{\Delta_{2s}(\tau_1)}$ in the case of  $ \overline{\underbar{g}\,{}^0_{k_1,k_2}(\tau_1)}$ and an antiholomorphic Eisenstein 
series $\overline{{\rm G}_{k}(\tau_1)}$ in the case of
$ \overline{\underbar{g}\,{}^{\rm E}_{k_1,k_2}(\tau_1)}$
\item a polynomial in the bookkeeping variables $X_i,Y_i$
as well as $\tau_1$ in the case of $\underbar{f}^{\, 0}_{k_1,k_2}(\tau_1)$ 
or $\bar \tau_1$ in the case of $\overline{\underbar{g}^{\,0}_{k_1,k_2}(\tau_1)}$ or $ \overline{\underbar{g}^{\,\rm E}_{k_1,k_2}(\tau_1)}$
\end{itemize}
All of $c^{\gamma}_{k_1,k_2}$, $\underbar{f}^{\, 0}_{k_1,k_2}(\tau_1)$, 
$\overline{\underbar{g}^{\, 0}_{k_1,k_2}(\tau_1)}$ and $\overline{\underbar{g}^{\, \rm E}_{k_1,k_2}(\tau_1)}$
are understood to have homogeneity degrees $k_1{-}2$ in $X_1,Y_1$ and $k_2{-}2$ in $X_2,Y_2$ 
compatible with $K_{k_1,k_2}$ in (\ref{beqv.new53}). As detailed in section 9.2 of \cite{Brown:2017qwo},
the existence of modular completions $M_{k_1,k_2}$ in (\ref{KtoMd2}) for any $k_1,k_2 \in 2\mathbb N{+}2$
follows from the Eichler--Shimura theorem \cite{Eichler:1957, Shimura:1959}. At depth one, 
in turn, the contribution $\zeta_{k-1} Y^{k-2}$ to (\ref{beqv.new51}) already suffices to compensate 
the $SL(2,\mathbb Z)$-transformation (or ``cocycle'') of the $ \underline{\rm G}_k[\ldots]$- 
and $\overline{  \underline{\rm G}_k[\ldots]} $ integrals: The cocycle of the real-analytic combination 
of Eisenstein integrals in (\ref{beqv.new51}) is a coboundary \cite{Brown:mmv}.

The modular completion $K_{k_1,k_2} \rightarrow M_{k_1,k_2}$ in (\ref{KtoMd2}) assembles 
exactly the kinds of constituents needed to convert the $\beta_{\pm}$ in (\ref{eq:depthtwoKeval})
to the modular forms $\beta^{\rm eqv} \big[ \begin{smallmatrix} j_1 &j_2 \\
k_1 &k_2 \end{smallmatrix} \big]$:
\begin{itemize}
\item The $\mathbb Q[(2\pi i)^{-1}]$-linear combinations of MZVs in the 
coefficients of $c^{\gamma}_{k_1,k_2}$ generate the depth-zero terms
$\bbsvpure\big[ \begin{smallmatrix} j_1 &j_2 \\
k_1 &k_2 \end{smallmatrix} \big]$ in (\ref{beqv.new06}).
\item The Eisenstein part $\overline{\underbar{g}^{\, \rm E}_{k_1,k_2}(\tau_1)}$ 
yields the contribution $\nicedel_\phi\beta_- \big[ \begin{smallmatrix}
j_2 &j_1\\k_2&k_1\end{smallmatrix} \big]$ to (\ref{beqv.new06}) from
the change of alphabet $\phi^{\rm sv}$ in section \ref{sec:4.1.2}. We therefore find 
non-zero $\overline{\underbar{g}^{\, \rm E}_{k_1,k_2}(\tau_1)}$
for any pair $k_1,k_2 \in 2\mathbb N{+}2$ subject to $k_1\neq k_2$.
\item Finally, $\underbar{f}^{\, 0}_{k_1,k_2}(\tau_1)$ and 
$\overline{\underbar{g}^{\, 0}_{k_1,k_2}(\tau_1)}$ combine to the contributions 
$\beta_{\Delta}^{\rm sv}  \big[ \begin{smallmatrix} j_2 &j_1\\k_2&k_1\end{smallmatrix} \big]$ 
to (\ref{beqv.new06}) and vanish for $k_1{+}k_2< 14$. More precisely, 
the holomorphic and antiholomorphic integrals in
(\ref{deltavar}) are captured by $\underbar{f}^{\, 0}_{k_1,k_2}(\tau_1)$ and 
$\overline{\underbar{g}^{\, 0}_{k_1,k_2}(\tau_1)}$, respectively.
\end{itemize}
On these grounds, it is not surprising that the transition from
$K_{k_1,k_2}$ to $M_{k_1,k_2}$ promotes the depth-two terms in the
first line of (\ref{eq:depthtwoKeval}) to $\beta^{\rm eqv} \big[ \begin{smallmatrix} j_1 &j_2 \\
k_1 &k_2 \end{smallmatrix} \big]$,
\begin{align}
M_{k_1,k_2} &= \frac{1}{16} (k_1{-}1)! (k_2{-}1)!
\sum_{j_1=0}^{k_1-2} \sum_{j_2=0}^{k_2-2} 
 \frac{ {k_1-2 \choose j_1} {k_2-2 \choose j_2}    }{(-4y)^{j_1+j_2}}
 \beqvno{j_2 & j_1}{k_2 & k_1}
 \label{bequv.21}   \\
&\quad \times
(X_1{-}\tau Y_1)^{j_1} (X_1{-}\bar \tau Y_1)^{k_1-2-j_1}  
(X_2{-}\tau Y_2)^{j_2}  (X_2{-}\bar \tau Y_2)^{k_2-2-j_2} \, . \notag
\end{align}
Moreover, since the relation between $\bbsvpure$ and $\ccsvpure$ in (\ref{MGFtoBR64})
implements the change of alphabet between $(X_i{-}\tau Y_i),(X_i{-}\bar \tau Y_i)$ and 
$X_i,Y_i$, we can express the constant $c_{k_1,k_2}^\gamma$ in (\ref{KtoMd2}) as
\begin{align}
c_{k_1,k_2}^\gamma  &= - \frac{1}{16} (k_1{-}1)!(k_2{-}1)!
\sum_{j_1=0}^{k_1-2} \sum_{j_2=0}^{k_2-2} { k_1{-}2 \choose j_1 } { k_2{-}2 \choose j_2 }  \ccsv{j_2  &j_1 }{ k_2 &k_1}\notag \\
&\quad \times 
\bigg( \frac{ i Y_1 }{2\pi } \bigg)^{j_1}  
X_1^{k_1-2-j_1} 
 \bigg( \frac{ i Y_2 }{2\pi } \bigg)^{j_2}
  X_2^{k_2-2-j_2} \, .
 \label{bequv.25}
\end{align}
Finally, the Eisenstein part $\overline{\underbar{g}^{\, \rm E}_{k_1,k_2}(\tau)}$ 
can be given in the closed form
\beq
\overline{\underbar{g}^{\, \rm E}_{k_1,k_2}(\tau)} =  \left\{ \begin{array}{rl}
\displaystyle 
- \frac{ \zeta_{k_1-1}  (k_1{-}2)! (k_2{-}k_1{+}1) (k_2{-}k_1{+}2)\, {\rm B}_{k_2}}{
2 (2\pi i)^{k_1-2}  k_2(k_2{-}1)\, {\rm B}_{k_2-k_1+2}} \vspace{0.2cm} \ \ \ \ \
& \\
\hspace{0.5cm}
\times (X_1 Y_2{-}X_2 Y_1)^{k_1-2}
\overline{ \underline{{\rm G}}_{k_2-k_1+2}[X_2,Y_2;\tau]} &: k_1<k_2\, , \\ \\
\displaystyle 
\phantom{-} \frac{ \zeta_{k_2-1}  (k_2{-}2)! (k_1{-}k_2{+}1) (k_1{-}k_2{+}2) \, {\rm B}_{k_1}}{
2 (2\pi i)^{k_2-2}  k_1(k_1{-}1)\, {\rm B}_{k_1-k_2+2}} \vspace{0.2cm}\ \ \ \ \
& \\
\hspace{0.5cm}
\times
(X_1 Y_2{-}X_2 Y_1)^{k_2-2}\, 
\overline{ \underline{{\rm G}}_{k_1-k_2+2}[X_1,Y_1;\tau]} &: k_2<k_1 \, ,
\\ \\
0 \ \ \ \ \ \ \ \ \ \ \ \ \ \ \ \ \ \ \ \ \ \ \ \ \ \ \  &: \, k_1 = k_2\, ,
\end{array} \right.
  \label{beqv.new56}
\eeq
see (\ref{deftheta}) for the definition of $ \underline{\rm G}_k[X,Y;\tau]$.
As can be anticipated from the powers of the modular invariant $X_1 Y_2{-}X_2 Y_1$, the Eisenstein
integrals $\overline{ \underline{\rm G}_{k_2-k_1+2}[\ldots]}$ only contribute to a 
single $SL(2,\mathbb R)$ multiplet of MGFs that can be extracted from $M_{k_1,k_2}$ 
through the projectors $\delta^{k}$ in section 7 of \cite{Brown:2017qwo}. 
The $SL(2,\mathbb R)$ representation theory of MGFs of depths two and three will be 
discussed in \cite{depth3paper}. We have checked (\ref{bequv.21}), (\ref{bequv.25}) 
and (\ref{beqv.new56}) for all depth-two cases up to and including $k_1{+}k_2 = 28$.

\subsubsection{On uniqueness of $M_{k_1,k_2}$ and choices in $J^{\rm eqv}$}
\label{sec:unique}

As noted in section 9.2 of \cite{Brown:2017qwo}, the $SL(2,\mathbb R)$-singlet component of
$M_{k_1,k_2}$ may be shifted by a constant $c\in \mathbb C$ without altering the modular
properties. Such singlets only occur in the case of $k_1\!=\!k_2\!=\!k$ and are then proportional 
to $(X_1Y_2{-}X_2 Y_1)^{k-2}$. Since $k$ is even, the singlet will be symmetric under 
exchange of $X_1,Y_1 \leftrightarrow X_2,Y_2$ and therefore involve combinations
$\beta^{\rm eqv}  \big[ \begin{smallmatrix} j_1 &j_2\\k&k\end{smallmatrix} \big]
+ \beta^{\rm eqv}  \big[ \begin{smallmatrix} j_2 &j_1\\k&k\end{smallmatrix} \big]$ in (\ref{bequv.21}).
These combinations conspire to the shuffles $\beta^{\rm eqv}  \big[ \begin{smallmatrix} j_1 \\k\end{smallmatrix} \big] \beta^{\rm eqv}  \big[ \begin{smallmatrix} j_2 \\k\end{smallmatrix} \big]$
which would no longer be the case when adding $c(X_1Y_2{-}X_2 Y_1)^{k-2}$ to $M_{k,k}$
and thereby redefining $\beta^{\rm eqv}  \big[ \begin{smallmatrix} j_1 &j_2\\k&k\end{smallmatrix} \big]
+ \beta^{\rm eqv}  \big[ \begin{smallmatrix} j_2 &j_1\\k&k\end{smallmatrix} \big]$.
Hence, the ambiguity of adding a constant to the $SL(2,\mathbb R)$-singlet components at
depth two is fixed by imposing shuffle relations between the components $\beta^{\rm eqv}$
of $M_{k}$ and $M_{k,k}$.

Using this imposition, the modular completion given by equation \eqref{KtoMd2} now uniquely determines the $c^{\gamma}_{k_1,k_2}$, $\underbar{f}^{\, 0}_{k_1,k_2}(\tau_1)$,
$\overline{\underbar{g}^{\, 0}_{k_1,k_2}(\tau_1)}$ and $\overline{\underbar{g}^{\, \rm E}_{k_1,k_2}(\tau_1)}$. We note, however, that in this equation we have chosen the Eisenstein addition to appear in the form of $\overline{\underbar{g}^{\, \rm E}_{k_1,k_2}(\tau_1)}$. We can, in fact, modify $M_{k_1,k_2}$ by a suitable multiple of $M_k$ to remove $\overline{\underbar{g}^{\, \rm E}_{k_1,k_2}(\tau_1)}$ whilst retaining modular completion, but this will cause a function $\underbar{f}^{\, \rm E}_{k_1,k_2}(\tau_1)$ composed of a holomorphic Eisenstein series to appear instead. The equation defining $c^{\gamma}_{k_1,k_2}$ would also be affected by this modification.

The uniqueness argument for $M_{k_1,k_2}$ also explains why the redefinitions $B^{\rm sv}\rightarrow B^{\rm sv} a^{-1}$
and $\phi^{\rm sv} ( \widetilde{J_{-}} ) \rightarrow a
\phi^{\rm sv} ( \widetilde{J_{-}} ) a^{-1}$ 
by some series $a$ in $\epsilon_k^{(j)}$ (with MZV coefficients)
mentioned below (\ref{beqv.new00}) can be ruled out below depth three:
Only an $SL(2,\mathbb R)$ singlet of modular forms $\beta^{\rm eqv}$ admits 
a redefinition of modular invariants by a constant from the series $a$ without 
spoiling the differential equations (\ref{MGFtoBR33}) or (\ref{MGFtoBR61}). The absence
of singlets at depth one rules out any components with a single factor of $ \epsilon_k^{(j)}$ 
in $a$. Since the series $a$ is imposed to be group like \cite{Brown:2017qwo2} (one would otherwise
give up shuffle properties), the only viable depth-two contributions are commutators
$ [ \epsilon_{k_1}^{(j_1)} , \epsilon_{k_2}^{(j_2)} ]$. However, the latter do not contribute to the
$SL(2,\mathbb R)$-singlet combination of $\beta^{\rm eqv}$ entering $M_{k,k}$
with $(X_1Y_2{-}Y_1 X_2)^{k-2}$ which rules out depth-two components in $a$.

At depth three, in turn, generic $(k_1,k_2,k_3)$
admit $SL(2,\mathbb R)$-singlet combinations of modular forms
$\beta^{\rm eqv} [ \begin{smallmatrix} j_1 &j_2 &j_3
 \\ k_1&k_2 &k_3 \end{smallmatrix} ]$ that are independent under shuffle relations.
That is why contributions $ [ \epsilon_{k_1}^{(j_1)} , [ \epsilon_{k_2}^{(j_2)} , \epsilon_{k_3}^{(j_3)} ]]$ to $a$ compatible with the shuffle properties of $\beta^{\rm eqv}$
are conceivable such as the examples $ \zeta_7 [\epsilon^{(j_1)}_4,[\epsilon^{(j_2)}_4,\epsilon^{(j_3)}_6]]$ and $ \zeta_3 \zeta_5 [\epsilon^{(j_1')}_6,[\epsilon^{(j_2')}_6,\epsilon^{(j_3')}_4]] $ in the discussion below (\ref{beqv.new00}). This freedom of modifying depth-three contributions
to $J^{\rm eqv}$ translates into depth-four modifications 
$\phi^{\rm sv} ( \widetilde{J_{-}} ) \rightarrow a
\phi^{\rm sv} ( \widetilde{J_{-}} ) a^{-1}$ and can be
fixed by imposing the change of alphabet $\phi^{\rm sv}$ to take the form (\ref{brown.17}).
Since we have not yet constructed the relevant $\beta^{\rm eqv}$ at depth four, 
we can only determine the preferred 
$\ccsvpure [ \begin{smallmatrix} j_1 &j_2&j_3 \\ 
4 &4 &6 \end{smallmatrix} ]$ and $\ccsvpure [ \begin{smallmatrix} j_1 &j_2&j_3 \\ 
4 &6 &6 \end{smallmatrix} ]$ up to one free parameter each (see the ancillary files
and the undetermined $c_{446} \in \mathbb Q$ in (\ref{moreex.11})) 
and relegate their fixing via (\ref{brown.17}) to the future. We stress that these free 
parameters merely concern canonical choices in shifting modular invariants by constants
and do not reflect any shortcomings in the construction of modular forms at depth three or beyond.

\subsubsection{Examples}
\label{sec:4.3.3}

The general discussion of Brown's equivariant double iterated integrals 
$M_{k_1,k_2}$ calls for examples of the building blocks
 $c^{\gamma}_{k_1,k_2}$, $\underbar{f}^{\, 0}_{k_1,k_2}(\tau)$, $\overline{\underbar{g}^{\, 0}_{k_1,k_2}(\tau)}$ and $\overline{\underbar{g}^{\, \rm E}_{k_1,k_2}(\tau)}$
of the modular completion in (\ref{KtoMd2}).

Based on the examples of single-valued MZVs
$ \ccsvpure\big[ \begin{smallmatrix} j_1 &j_2 \\k_1 &k_2 \end{smallmatrix} \big]$
of weight $(j_1{+}j_2{+}2)$ in
(\ref{bequv.5}) and (\ref{bequv.564}), the constants in (\ref{bequv.25}) specialize to
\begin{align}
c^\gamma_{4,4} &= -\frac{ i \zeta_3 }{960 \pi} X_1 X_2 (X_1 Y_2 {-} X_2 Y_1)   - \frac{ 5i \zeta_5}{ 192 \pi^3} Y_1 Y_2  (X_1 Y_2 {-} X_2 Y_1) 
- \frac{  \zeta_3^2 }{32\pi^4} Y_1^2 Y_2^2\, , \notag \\
c^\gamma_{6,4} &= \frac{ i \zeta_3}{10080 \pi}
 X_1^3 X_2(2 X_1 Y_2 {-} X_2 Y_1)  
  - \frac{ i \zeta_5 }{960 \pi^3} X_1 Y_1(3 X_2^2 Y_1^2 {-} 3 X_1 X_2 Y_1 Y_2 {+} X_1^2 Y_2^2)  \label{bequv.26} \\
&\quad
+ \frac{ \zeta_3^2 }{112 \pi^4}  Y_1^2 (X_1 Y_2 {-} X_2 Y_1)^2
+\frac{7i \zeta_7 }{128 \pi^5} Y_1^3 Y_2 (X_1 Y_2{ -} X_2 Y_1)
+\frac{ 3 \zeta_3 \zeta_5 }{32\pi^6} Y_1^4 Y_2^2
\notag
\end{align}
in case of double integrals over ${\rm G}_4 {\rm G}_4$ and ${\rm G}_4 {\rm G}_6$.
Examples at higher weights can be generated from the list of all
$ \ccsvpure\big[ \begin{smallmatrix} j_1 &j_2 \\k_1 &k_2 \end{smallmatrix} \big]$ 
at $k_1{+}k_2 \leq 28$ given in the ancillary files. We reiterate that the
$\ccsvpure$ at depth two only involve odd zeta values and bilinears thereof,
whereas irreducible single-valued MZVs beyond depth one are relegated to the modular
completion of triple Eisenstein integrals and higher depth \cite{Saad:2020mzv}, 
see section \ref{sec:4.1.4}. 

The closed formula (\ref{beqv.new56}) for the Eisenstein part 
 $\overline{\underbar{g}^{\,\rm E}_{k_1,k_2}(\tau)}$ specializes as follows 
in the simplest non-vanishing examples
\begin{align}
\overline{\underbar{g}^{\,\rm E}_{4,6}(\tau)} &= -  \frac{ \zeta_3 }{14 \pi^2} (X_1 Y_2{-}X_2 Y_1)^2
\, \overline{ \underline{\rm G}_4[X_2,Y_2;\tau]} \, ,\notag \\
\overline{\underbar{g}^{\,\rm E}_{4,8}(\tau)} &= -  \frac{3 \zeta_3 }{16 \pi^2} (X_1 Y_2{-}X_2 Y_1)^2
\, \overline{ \underline{\rm G}_6[X_2,Y_2;\tau]} \, ,  \label{beqv.new57}  \\
\overline{\underbar{g}^{\,\rm E}_{6,8}(\tau)} &= -  \frac{9 \zeta_5 }{56 \pi^4} (X_1 Y_2{-}X_2 Y_1)^4
\, \overline{\underline{\rm G}_4[X_2,Y_2;\tau]} \, .\notag
\end{align}
Finally, the cusp-form contributions $\underbar{f}^{\, 0}_{k_1,k_2}(\tau), 
 \overline{\underbar{g}^{\, 0}_{k_1,k_2}(\tau)}$ to (\ref{KtoMd2}) always arise in pairs
by the structure of the $\beta^{\rm sv} \big[\begin{smallmatrix} j \\ \Delta^{\pm}_{k} \end{smallmatrix}  ;\tau \big] $ in (\ref{bequv.12}), (\ref{deltavar}) and are most conveniently
expressed in terms of
\beq
\newuscdelta{k}{X_1 &X_2 }{ Y_1 &Y_2}{r_1 &r_2}{\tau}
= i \pi \, (k{-}1)! \, (X_1{-}\tau Y_1)^{r_1}
(X_2{-}\tau Y_2)^{r_2}\, \Delta_k(\tau) \, \dd \tau\, ,
\ \ \ \ r_1{+}r_2 = k{-}2 \, .
\eeq
This equivariant $(1,0)$-form may be viewed as the cuspidal analogue
of $\underline{\rm G}_k[X,Y;\tau]$ in (\ref{deftheta}), where $(X,Y)$ is split into two pairs $(X_i,Y_i)$
with partial homogeneity degrees $r_i$. The simplest non-vanishing examples of
$\underbar{f}^{\, 0}_{k_1,k_2}(\tau)$ and $\overline{\underbar{g}^{\, 0}_{k_1,k_2}(\tau)}$ are
\begin{align}
\big( \underbar{f}^{\, 0}_{4,10}(\tau),\,  \overline{\underbar{g}^{\, 0}_{4,10}(\tau)} \big) &=  
-\frac{1}{2700 \cdot 11!}\frac{ \Lambda(\Delta_{12},12)}{  \Lambda(\Delta_{12}, 10)}\Big(
 \newuscdelta{12}{X_1 &X_2 }{ Y_1 &Y_2}{2 &8}{\tau}
, \, {-}\overline{ \newuscdelta{12}{X_1 &X_2 }{ Y_1 &Y_2}{2 &8}{\tau}} \Big)\, , \notag \\
\big( \underbar{f}^{\, 0}_{6,8}(\tau),\,  \overline{\underbar{g}^{\, 0}_{6,8}(\tau)} \big) &=  
\frac{1}{4200 \cdot 11!}\frac{ \Lambda(\Delta_{12},12)}{  \Lambda(\Delta_{12}, 10)}\Big(
 \newuscdelta{12}{X_1 &X_2 }{ Y_1 &Y_2}{4 &6}{\tau}, \, {-}\overline{ \newuscdelta{12}{X_1 &X_2 }{ Y_1 &Y_2}{4 &6}{\tau}} \Big) \, , \label{fgex.01} \\
\big( \underbar{f}^{\, 0}_{4,12}(\tau),\,  \overline{\underbar{g}^{\, 0}_{4,12}(\tau)} \big) &= -\frac{1}{1382\cdot 11!} \frac{ \Lambda(\Delta_{12},13)}{  \Lambda(\Delta_{12}, 11)} 
\frac{(X_1 Y_2{-}X_2 Y_1)}{ (i\pi)}
 \Big(
  \newuscdelta{12}{X_1 &X_2 }{ Y_1 &Y_2}{1 &9}{\tau}, \, \overline{\newuscdelta{12}{X_1 &X_2 }{ Y_1 &Y_2}{1 &9}{\tau}} \Big) \, ,
\notag\\
\big( \underbar{f}^{\, 0}_{6,10}(\tau),\,  \overline{\underbar{g}^{\, 0}_{6,10}(\tau)} \big) &= 
\frac{5}{8292\cdot 11!} \frac{ \Lambda(\Delta_{12},13)}{  \Lambda(\Delta_{12}, 11)}
\frac{    (X_1 Y_2{-}X_2 Y_1)}{(i\pi)}
 \Big(
 \newuscdelta{12}{X_1 &X_2 }{ Y_1 &Y_2}{3 &7}{\tau} 
 , \, \overline{ \newuscdelta{12}{X_1 &X_2 }{ Y_1 &Y_2}{3 &7}{\tau}} \Big)\, , \notag 
\\
\big( \underbar{f}^{\, 0}_{8,8}(\tau),\,  \overline{\underbar{g}^{\, 0}_{8,8}(\tau)} \big) &= - \frac{3}{5528\cdot 11!}\frac{ \Lambda(\Delta_{12},13)}{  \Lambda(\Delta_{12}, 11)} 
\frac{ (X_1 Y_2{-}X_2 Y_1) }{(i \pi)}
 \Big(  \newuscdelta{12}{X_1 &X_2 }{ Y_1 &Y_2}{5 &5}{\tau}
, \, \overline{ \newuscdelta{12}{X_1 &X_2 }{ Y_1 &Y_2}{5 &5}{\tau} } \Big) \, .
\notag
\end{align}
Starting from $k_1{+}k_2=18$, generic $\underbar{f}^{\, 0}_{k_1,k_2}(\tau)$ and 
$\overline{\underbar{g}^{\, 0}_{k_1,k_2}(\tau)}$ comprise several cusp forms,
\begin{align}
\big( \underbar{f}^{\, 0}_{8,10}(\tau),\,  \overline{\underbar{g}^{\, 0}_{8,10}(\tau)} \big) &= 
-\frac{1 }{7840 {\cdot}15!}
 \frac{ \Lambda(\Delta_{16}, 16) }{ \Lambda(\Delta_{16}, 14) }
 \Big( 
 \newuscdelta{16}{X_1 &X_2 }{ Y_1 &Y_2}{6 &8}{\tau}
 , \, {-}\overline{ \newuscdelta{16}{X_1 &X_2 }{ Y_1 &Y_2}{6 &8}{\tau}} \Big)  \label{fgex.02}\\
 &\quad
 - \frac{1}{1560{\cdot} 11!}\frac{ \Lambda(\Delta_{12}, 14) }{ \Lambda(\Delta_{12}, 10) }  \frac{ (X_1Y_2{-}X_2Y_1)^2 }{ (i\pi)^2}
 \Big(
  \newuscdelta{12}{X_1 &X_2 }{ Y_1 &Y_2}{4 &6}{\tau}
  , \, {-}\overline{\newuscdelta{12}{X_1 &X_2 }{ Y_1 &Y_2}{4 &6}{\tau} }\Big) \, ,
 \notag
\end{align}
and the relative sign between $\underline{\Delta}_{k}$ and $\overline{\underline{\Delta}_{k}}$
in $M_{k_1,k_2}$ is $+1$ ($-1$) if $\frac{1}{2}(k_1{+}k_2{+}k)$ is even (odd).
Similar to the Eisenstein case in (\ref{beqv.new56}), the cuspidal contributions
with $k_1\leftrightarrow k_2$ interchanged can be obtained from
 \begin{align}
 \underbar{f}^{\, 0}_{k_1,k_2}[X_1,Y_1,X_2,Y_2;\tau] &= -
  \underbar{f}^{\, 0}_{k_2,k_1}[X_2,Y_2,X_1,Y_1;\tau] \, , \notag \\
  \overline{   \underbar{g}^{\, 0}_{k_1,k_2}[X_1,Y_1,X_2,Y_2;\tau] } &= -
 \overline{ \underbar{g}^{\, 0}_{k_2,k_1}[X_2,Y_2,X_1,Y_1;\tau] }\, .
  \label{cuspsymm}
 \end{align}
The coefficients of $\underline{\Delta}_{k}$ and $\overline{\underline{\Delta}_{k}}$
are always $\mathbb Q[ (X_1Y_2{-}X_2Y_1)/(i\pi)]$-linear combinations of
$\xi^{\Delta_{k}}_{k_1,k_2}$ in (\ref{bequv.12}) and (\ref{xiexpl}),
with $\frac{1}{2}(k_1{+}k_2{-}k)-1$ powers of $(X_1Y_2{-}X_2Y_1)/(i\pi)$.

The integrals over $\underbar{f}^{\, 0}_{k_1,k_2}(\tau)$ and 
$\overline{\underbar{g}^{\, 0}_{k_1,k_2}(\tau)}$
exemplified in this section generate the cusp-form contributions
(\ref{bequv.12}) to depth-two $\beta^{\rm eqv}$ in (\ref{bequv.21}). In contrast to the derivation-valued
generating series $J^{\rm eqv}$ of modular forms $\beta^{\rm eqv}$ in (\ref{select.1}), the equivariant
double integrals $M_{k_1,k_2}$ in (\ref{KtoMd2}) retain the integrals of holomorphic
cusp forms. One can equivalently generalize $J^{\rm eqv}$ by 
re-interpreting the $\epsilon_{k}^{(j)}$ at $k\geq4$ and $0\leq j\leq k{-}2$ as generators of a free 
algebra rather than brackets of Tsunogai's derivations and thereby preventing the dropouts 
of modular forms due to relations among commutators of $\epsilon_{k}^{(j)}$. This viewpoint is
taken in intermediate steps to motivate (\ref{MGFtoBR42}) as a generating series of {\it all} 
the $\overline{\alpha_{\rm hard}[\begin{smallmatrix}j_1 &j_2 \\ k_1 &k_2 \end{smallmatrix};\tau]}$, 
without any dropouts. Based on the combinations $M_{k_1,k_2,k_3}$ of $\beta^{\rm eqv}$ 
at depth three in upcoming work \cite{depth3paper}, the same logic applies to the generating series 
(\ref{appb.08}) of $\overline{\alpha_{\rm hard}[\begin{smallmatrix}j_1 &j_2 &j_3 \\ 
k_1 &k_2 &k_3 \end{smallmatrix};\tau]}$.

\section{Conclusions and further directions}
\label{sec:5}

\vspace{-0.3cm}
\noindent
In this work, we have established a dictionary between Brown's non-holomorphic modular
forms obtained from equivariant iterated Eisenstein integrals (EIEIs) and the modular graph 
forms (MGFs) in the low-energy expansion of one-loop closed-string amplitudes.
In spite of rapid progress in representing MGFs via iterated Eisenstein integrals
and their complex conjugates \cite{Broedel:2018izr, Gerken:2020yii, Gerken:2020xfv, Dorigoni:2021jfr}, it was a long-standing problem to pinpoint
their connection to Brown's EIEIs \cite{Brown:mmv, Brown:2017qwo, Brown:2017qwo2}. Based on 
new structural results on the building blocks of MGFs with detailed studies at depths two and 
three, see section \ref{sec.3}, we spell out their explicit relations with the key quantities 
of Brown's construction in section \ref{sec:4}.

One central ingredient of Brown's EIEIs is a change of alphabet $\phi^{\rm sv}$ for 
Tsunogai's derivations $\epsilon_{k}$ that appear as non-commutative bookkeeping variables 
in the integration kernels.
We extract an all-order expression for $\phi^{\rm sv}(\epsilon_{k})$ as an infinite series of
derivations with single-valued MZVs in their coefficients from the implicit characterization
of $\phi^{\rm sv}$ in \cite{Brown:mmv, Brown:2017qwo2}. 
This series-representation of $\phi^{\rm sv}(\epsilon_{k})$ involves nested
commutators of $\epsilon_k$ with additional derivations $z_3,z_5,\ldots$ studied in the
mathematics literature. While the depth-two parts of $[z_m, \epsilon_k]$
are known from Hain and Matsumoto \cite{hain_matsumoto_2020}, we provide 
the higher-depth completions for a variety of such commutators beyond the
examples in \cite{Pollack}.

Our new results on the derivation algebra, as well as a variety of examples for
other quantities of interest in this work, can be found in ancillary files of the arXiv submission.
A longer follow-up paper \cite{depth3paper} will elaborate on intermediate steps and
extensions of our results, including
\begin{itemize}
\item depth-three analogues of the solutions ${\rm F}^{\pm(s)}_{m,k}$ to 
Laplace eigenvalue equations at depth two \cite{Dorigoni:2021jfr} that were used in
the derivation of some of our results
\item double integrals mixing holomorphic Eisenstein series and cusp
forms that were studied from a Poincar\'e-series perspective in 
\cite{Diamantis:2020} and enter our modular integrals at depth three
\item depth-three analogues of Brown's equivariant double iterated integrals (\ref{KtoMd2})
as well as their cocycles and $SL(2,\mathbb R)$ multiplet structure
\end{itemize}
Apart from building a bridge between the mathematics and physics literature
on non-holomorphic modular forms, this work stimulates numerous research lines
for the future:

First, Brown's construction of purely $\tau$-dependent non-holomorphic modular forms
calls for an extension to so-called elliptic modular graph forms
\cite{DHoker:2018mys, Dhoker:2020gdz} which additionally depend on marked points $z$
on a torus. Elliptic modular graph forms extend Zagier's single-valued elliptic polylogarithms
\cite{Ramakrish} to higher depth and were recently translated into $z$-dependent analogues of the
real-analytic iterated integrals $\beta^{\rm sv}$ for conventional modular graph forms 
\cite{new:eMGF}. The dictionary between $\beta^{\rm sv}$ and EIEIs established in this work 
should guide the construction of similar generating functions 
of elliptic modular graph forms that manifest their (anti-)meromorphic iterated-integral constituents.

Second, it would be interesting to revisit the proposal \cite{Gerken:2020xfv} of MGFs 
being single-valued versions of elliptic MZVs in the light of this work: By relating 
MGFs with Brown's single-valued iterated Eisenstein integrals, our results pave the 
way to compare the proposal in the reference with the general frameworks of single-valued
integration \cite{Schnetz:2013hqa, Brown:2018omk} and single-valued periods 
\cite{Brown:2013gia, FrancisLecture}. This kind of follow-up study aims to extract
information on loop amplitudes of closed strings from single-valued open-string data.

Third, the variety of perspectives obtained on MGFs at genus one is
expected to inspire the study of higher-genus modular graph forms 
\cite{DHoker:2017pvk, DHoker:2018mys} and tensors \cite{DHoker:2020uid}.
The quest for the algebraic \cite{DHoker:2020tcq, DHoker:2020uid} and 
differential \cite{DHoker:2014oxd, Basu:2018bde, Basu:2021xdt} relations of higher-genus 
modular graph forms might greatly benefit from an organizing principle based on iterated
primitives of meromorphic modular tensors. The description of genus-one MGFs 
via iterated Eisenstein integrals in this work is hoped to find an echo
at higher genus.

\section*{Acknowledgements}

We are grateful to Johannes Broedel, Francis Brown, Eric D'Hoker, Nikolaos Diamantis, 
Herbert Gangl, Jan Gerken, Carlos Mafra, Erik Panzer, Aaron Pollack and Federico Zerbini 
for combinations of inspiring discussions, collaboration on related topics and helpful comments 
on a draft version of this work. Moreover, we are indebted to Francis Brown for sharing important insights, to 
Herbert Gangl for providing valuable context from the mathematical literature and to
Carlos Mafra for his crucial suggestions on basis manipulations in free Lie algebras.

The authors would like to thank the Isaac Newton Institute of Mathematical Sciences for support 
and hospitality during the programme ``New connections in number theory and physics'' when 
work on this paper was undertaken. This work was supported by EPSRC Grant Number EP/R014604/1. This research was supported by the Munich Institute for Astro-, Particle and BioPhysics (MIAPbP), which is funded by the Deutsche Forschungsgemeinschaft (DFG, German Research Foundation) under Germany's Excellence Strategy -- EXC-2094 -- 390783311. OS is grateful
to the organizers of the workshop ``Elliptic integrals in fundamental physics'' for the opportunity
to present this work and the participants for many interesting comments. OS furthermore thanks
the Mainz Institute for Theoretical Physics (MITP) of the
Cluster of Excellence PRISMA$^+$ (Project ID 39083149), for its hospitality and its
partial support during the completion of this work. 

The research of MD was supported by the IMPRS for Mathematical and Physical Aspects of Gravitation, Cosmology and Quantum Field Theory. JD is supported by the Engineering and Physical Sciences Research Council (Grant No.\ EP/W52251X/1). MH, OS and BV are supported by the European Research Council under
ERC-STG-804286 UNISCAMP, and BV is furthermore supported by the Knut and Alice Wallenberg Foundation under grant KAW2018.0162.

\appendix

\section{Explicit formula for $\overline{ \alpha_{\rm hard}}$ at depth two}
\label{app.A}

In this appendix, we reformulate the all-order solution for the $\overline{\alpha_{\rm hard}}$ at depth two in
(\ref{closedform}) as an explicit formula instead of a generating-series identity (\ref{MGFtoBR42}).
By inserting the commutators (\ref{MGFtoBR44}) into the generating series and treating
all the $\epsilon_{k_1}^{(j_1)} \epsilon_{k_2}^{(j_2)}$ as independent,
one can solve for
\begin{align}
\overline{\alphahard{ j_1 &j_2  }{ k_1 &k_2 }{\tau} }   &= \theta(k_1>k_2) \theta(k_2{-}2\leq j_1{+}j_2\leq k_1{-}2) \overline{ \rhopart{ j_1 &j_2  }{ k_1 &k_2 }{\tau}  }
\label{ahard.01} \\
&\quad 
 - \theta(k_2>k_1) \theta(k_1{-}2\leq j_1{+}j_2\leq k_2{-}2) \overline{ \rhopart{ j_2 &j_1  }{ k_2 &k_1 }{\tau}  }  \, ,
 \notag
 \end{align}
where the $\theta(\ldots)$ are taken to be 1 if the inequalities in the bracket hold and zero
otherwise. Moreover, we have introduced the shorthand
\begin{align}
&\overline{ \rhopart{ j_1 &j_2  }{ k_1 &k_2 }{\tau}  } = \frac{ 2 \zeta_{k_2-1} (k_1{-}k_2{+}1) (k_1{-}k_2{+}2)! (k_2{-}j_2{-}2)!
(k_1{-}j_1{-}2)! \, {\rm B}_{k_1}}{(k_2{-}1) k_1! (k_1{-}1)! \, {\rm B}_{k_1-k_2+2}}  \label{ahard.02} \\
&\quad \! \times  (j_1{+}j_2{-}k_2{+}2)!  \overline{ {\cal E}_0(k_1{-}k_2{+}2,0^{j_1+j_2-k_2+2};\tau) }
\! \! \sum_{\ell = {\rm max} (0,j_1-k_1+k_2)}^{k_2-j_2-2} \!  \frac{ (-1)^\ell (k_1{-}k_2{+}\ell)! }{\ell!
(k_2{-}j_2{-}2{-}\ell)! (\ell{-}j_1{+}k_1{-}k_2)!} \notag 
\end{align}
for the antiholomorphic $T$-invariants $\overline{{\cal E}_0}$ defined in (\ref{MGFtoBR41}).

\section{ Antiholomorphic $T$-invariants $\overline{ \alpha}$ at depth three}
\label{app.B}

This appendix provides a conjectural all-order expression for the 
$\overline{\alpha[\begin{smallmatrix}j_1 &j_2&j_3 \\ k_1 &k_2 &k_3 \end{smallmatrix};\tau]}$ 
at depth three that resembles (\ref{closedform}) for 
$\overline{\alpha[\begin{smallmatrix}j_1 &j_2 \\ k_1 &k_2 \end{smallmatrix};\tau]}$
at depth two. The subsequent depth-three results were crucial to guide us towards
a matching of the two generating series (\ref{MGFtoBR63}) and (\ref{earlierconst}) 
of modular forms $\beta^{\rm eqv}$. 
It will be convenient to represent the $T$-invariants (\ref{MGFtoBR41})
entering the $\overline{\alpha[\begin{smallmatrix}j_1 &j_2 \\ k_1 &k_2 \end{smallmatrix};\tau]}$ via
\beq
\ezero{j_1 }{ k_1}{\tau} = j_1! \overline{ {\cal E}_0(k_1,0^{j_1};\tau) }\,.
\label{appb.01}
\eeq
Moreover, the expressions in this appendix involve their generalization
\begin{align}
\ezero{j_1 &j_2 }{ k_1 &k_2 }{\tau} &= j_2! \sum_{r=0}^{j_2} \frac{ (j_1{+}r)! }{r!} \,
 \overline{ {\cal E}_0(k_2,0^{j_2-r},k_1,0^{j_1+r};\tau) }
\label{appb.02} \\
&\quad+ \frac{(j_1{+}j_2{+}1)!\, {\rm B}_{k_1} }{k_1! (j_1{+}1)} \,
 \overline{ {\cal E}_0(k_2,0^{j_1+j_2+1};\tau) }
 - \frac{(j_1{+}j_2{+}1)! \, {\rm B}_{k_2} }{k_2! (j_2{+}1)} \,
 \overline{ {\cal E}_0(k_1,0^{j_1+j_2+1};\tau) }\, ,
 \notag
\end{align}
where the first line of the right-hand side features double integrals of the 
$T$-invariant kernels ${\rm G}^0_k(\tau) = {\rm G}_k(\tau) - 2\zeta_k ={ O}(q)$
\cite{Broedel:2015hia}:
\begin{align}
{\cal E}_0(k_1,0^{p_1} ,k_2,0^{p_2} ;\tau) &= \frac{ (2\pi i)^{p_1+p_2-k_1-k_2+2} }{p_1! \, p_2!}
 \int_{ \tau}^{i\infty} \dd \tau_2 \, (\tau{-} \tau_2)^{p_2} {\rm G}^0_{k_2}(\tau_2) 
  \int_{ \tau_2}^{i\infty} \dd \tau_1 \, (\tau_2{-} \tau_1)^{p_1} {\rm G}^0_{k_1}(\tau_1)  \notag \\
  &= \frac{4}{(k_1{-}1)! (k_2{-}1)!} \sum_{m_1,n_1,m_2,n_2 = 1}^{\infty}
  \frac{ m_1^{k_1-1} m_2^{k_2-1} q^{m_1n_1+m_2 n_2} }{(m_1 n_1)^{p_1+1} (m_1n_1{+}m_2 n_2)^{p_2+1} }
  \, . 
\label{appb.03} 
\end{align}
As will be detailed in section \ref{app.B.just} below, the antiholomorphic
$T$-invariants in (\ref{appb.01}) and (\ref{appb.02}) are related via
shuffle relations 
\beq
\ezero{j_1 }{ k_1}{\tau} \,  \ezero{j_2 }{ k_2}{\tau} = 
\ezero{j_1 &j_2 }{ k_1 &k_2 }{\tau}  + \ezero{j_2 &j_1 }{ k_2 &k_1 }{\tau} 
\label{appb.04} 
\eeq
and can be rewritten in terms of Brown's iterated Eisenstein integrals over kernels
$\bar \tau^j \overline{ {\rm G}_k(\tau)}$ with $0\leq j \leq k{-}2$ and powers of
$2\pi i\bar \tau$ as coefficients. More specifically, the results of section \ref{app.B.just}
imply the following alternative formula for $\overline{ {\cal E}_0\big[\begin{smallmatrix} j_1 &j_2 \\ k_1 &k_2 \end{smallmatrix} ;\tau \big] }$ which generalizes (\ref{forlater})
for $\overline{ {\cal E}_0\big[\begin{smallmatrix} j_1 \\ k_1 \end{smallmatrix} ;\tau \big] }$
\begin{align}
\ezero{j_1 &j_2 }{ k_1 &k_2 }{\tau} &= 
\sum_{\ell_1=0}^{k_1-2-j_1} \sum_{\ell_2=0}^{k_2-2-j_2}
 \frac{\binom{k_1-2-j_1}{\ell_1}\binom{k_2-2-j_2}{\ell_2}}{(-4y)^{\ell_1+\ell_2}} \bminus{ j_2+\ell_2 &j_1+\ell_1}{k_2 &k_1}{\tau}  \label{appb.05}  \\
 &\quad 
  -  \frac{   {\rm B}_{k_2} \, (-2\pi i \bar \tau)^{j_2+1} }{(j_2{+}1)  k_2!}
 \sum_{\ell_1=0}^{k_1-2-j_1} 
 \frac{\binom{k_1-2-j_1}{\ell_1} }{(-4y)^{\ell_1}} \bminus{ j_1+\ell_1}{ k_1}{\tau} 
 + \frac{ {\rm B}_{k_1} \, {\rm B}_{k_2} \, (-2\pi i \bar \tau)^{j_1+j_2+2} }{(j_2{+}1) (j_1{+}j_2{+}2) k_1! k_2!} \, .
\notag
\end{align}
Similar to (\ref{closedform}), we shall split the desired $T$-invariants into two parts
\beq
\overline{\alphaBR{ j_1 &j_2 &j_3 }{ k_1 &k_2 &k_3 }{\tau} } = 
\overline{\alphaeasy{ j_1 &j_2 &j_3 }{ k_1 &k_2 &k_3 }{\tau} }  + 
\overline{\alphahard{ j_1 &j_2 &j_3  }{ k_1 &k_2 &k_3 }{\tau} }\, , 
\label{appb.06}
\eeq
and discuss the construction of $\overline{\alpha_{\rm easy}}$ and $\overline{\alpha_{\rm hard}}$
separately in the next two subsections. We have checked that all the
$\overline{\alphaBR{\ldots}{\ldots}{\tau}}$ with $k_1{+}k_2{+}k_3\leq20$ and
$0\leq j_i \leq k_i{-}2$ obtained from a depth-three generalization 
of ${\rm F}^{\pm (s)}_{m,k}$ \cite{depth3paper} are reproduced by
(\ref{appb.06}) with the $\overline{\alpha_{\rm easy}}$ and $\overline{\alpha_{\rm hard}}$ 
below. Our results are consistent with the necessary condition
\beq
\overline{\alphaBR{ j_1 &j_2 &j_3 }{ k_1 &k_2 &k_3 }{\tau} }
+ \overline{\alphaBR{ j_2 &j_1 &j_3 }{ k_2 &k_1 &k_3 }{\tau} } 
+ \overline{\alphaBR{ j_2 &j_3 &j_1 }{ k_2 &k_3 &k_1 }{\tau} } =0
\label{appb.90}
\eeq
for the shuffle relations (\ref{shff.02}) of the $\beta^{\rm sv}$ at depth three.

\subsection{$\overline{\alpha_{\rm easy}}$ at depth three}
\label{app.B.1}

The first part of (\ref{appb.06}) dubbed $\overline{\alpha_{\rm easy}}$ is determined by
the $\ccsvpure$ at depth $\leq 2$ multiplied by the above $\overline{{\cal E}_0}$
as well as the $\overline{ \alpha_{\rm hard}\big[\begin{smallmatrix} j_1 &j_2 \\ k_1 &k_2
 \end{smallmatrix} ;\tau \big] }$ of appendix \ref{app.A},
\begin{align}
\overline{\alphaeasy{ j_1 &j_2 &j_3 }{ k_1 &k_2 &k_3 }{\tau} } &= 
\ezero{j_1 &j_2 }{ k_1 &k_2 }{\tau} \ccsv{j_3 }{k_3 } 
+ \ezero{j_3 &j_2 }{ k_3 &k_2 }{\tau} \ccsv{j_1 }{k_1 }
- \ezero{j_1  }{ k_1   }{\tau} \, \ezero{j_3 }{ k_3 }{\tau} \ccsv{j_2 }{k_2 }
\notag \\
&\quad +    \ezero{j_1  }{ k_1   }{\tau}  \ccsv{j_2 &j_3}{k_2 &k_3}
+    \ezero{j_3 }{ k_3 }{\tau}  \ccsv{j_2 &j_1}{k_2 &k_1}
-\ezero{j_2 }{ k_2 }{\tau}  \ccsv{j_1}{k_1}   \ccsv{j_3}{k_3} 
\notag \\
&\quad+ \ccsv{j_3}{k_3} \overline{\alphahard{ j_1 &j_2   }{ k_1 &k_2  }{\tau} }
-\ccsv{j_1}{k_1}  \overline{\alphahard{ j_2 &j_3   }{ k_2 &k_3 }{\tau} }\, .
\label{appb.07}
\end{align}
At leading depth of the respective $\beta_-$ in (\ref{forlater}) and (\ref{appb.05}), the conversion 
of (\ref{appb.07}) to $\overline{ \kappa[\begin{smallmatrix}\ldots \\ \ldots\end{smallmatrix};\tau]}$ via (\ref{MGFtoBR14}) reproduces the first 
three lines of (\ref{beqv.new31}). In particular, the contributions of $\overline{\alpha_{\rm hard}}$ to 
(\ref{appb.07}) line up with the terms 
$ \nicedel_\phi \beta_- \big[\begin{smallmatrix}
 j_2 &j_1 \\ k_2 &k_1  \end{smallmatrix} \big] d^{\rm sv} \big[\begin{smallmatrix}   j_3  \\ k_3 \end{smallmatrix} \big] -  d^{\rm sv} \big[\begin{smallmatrix}  j_1  \\ k_1 \end{smallmatrix} \big] \nicedel_\phi \beta_- \big[\begin{smallmatrix}  j_3 &j_2 \\ k_3 &k_2 \end{smallmatrix} \big]$ in
the third line of (\ref{beqv.new31}). Note that the first two lines of (\ref{appb.07}) intuitively
generalize the rewriting
\beq
\overline{\alphaeasy{ j_1 &j_2 }{ k_1 &k_2  }{\tau} } = 
\ezero{j_1  }{ k_1  }{\tau} \ccsv{j_2 }{k_2 } 
- \ezero{j_2 }{  k_2 }{\tau} \ccsv{j_1 }{k_1 }
\label{rewrd2}
\eeq
of the $\overline{\alpha_{\rm easy}}$ at depth two in (\ref{closedform}).

\subsection{$\overline{\alpha_{\rm hard}}$ at depth three}
\label{app.B.2}

Similar to (\ref{MGFtoBR42}), the $\overline{\alpha_{\rm hard}}$ at depth three are most 
conveniently encoded in generating series (see the ancillary files for the
$[z_{2m+1},\epsilon_k]$ on the right-hand side),
\begin{align}
&\sum_{k_1,k_2,k_3=4}^\infty \sum_{j_1=0}^{k_1-2} \sum_{j_2=0}^{k_2-2}  \sum_{j_3=0}^{k_3-2}
\frac{ (k_1{-}1)(k_2{-}1) (k_3{-}1) (-1)^{j_1+j_2+j_3} }{(k_1{-}j_1{-}2)!
(k_2{-}j_2{-}2)! (k_3{-}j_3{-}2)!}   \notag \\
&\quad \quad \quad \quad \times
\overline{\alphahard{ j_1 &j_2 &j_3  }{ k_1 &k_2 &k_3 }{\tau} }  
\epsilon_{k_3}^{(k_3-j_3-2)} \epsilon_{k_2}^{(k_2-j_2-2)}
\epsilon_{k_1}^{(k_1-j_1-2)}  \notag\\
&\ \ =
\sum_{m=1}^\infty 2 \zeta_{2m+1} \sum_{k=4}^{\infty} \sum_{j=0}^{k-2}
\frac{ (k{-}1) (-1)^j }{(k{-}j{-}2)!}  \, \ezero{j }{ k }{\tau}\, 
[z_{2m+1}, \epsilon_k^{(k-j-2)}] \, \big|_{\rm depth\ 3} \notag \\
&\quad \quad +  \sum_{m_1,m_2=1}^\infty 2 \zeta_{2m_1+1} \zeta_{2m_2+1} 
  \sum_{k=4}^{\infty} \sum_{j=0}^{k-2}\frac{ (k{-}1) (-1)^j }{(k{-}j{-}2)!} \label{appb.08} \\
&\quad \quad \quad \quad \times \ezero{j }{ k }{\tau}\, 
\big[ z_{2m_1+1},[z_{2m_2+1}, \epsilon_k^{(k-j-2)}] \big] \, \big|_{\rm depth\ 3}
\notag \\
&\quad \quad +
\sum_{m=1}^\infty 2 \zeta_{2m+1} \sum_{k_1,k_2=4}^{\infty} 
\sum_{j_1=0}^{k_1-2} \sum_{j_2=0}^{k_2-2}
\frac{ (k_1{-}1)(k_2{-}1) (-1)^{j_1+j_2} }{(k_1{-}j_1{-}2)! (k_2{-}j_2{-}2)!} 
\notag \\
 &\quad \quad \quad \quad \times
 \ezero{j_1 &j_2}{ k_1 &k_2 }{\tau}\,
  \big[ [z_{2m+1}, \epsilon_{k_2}^{(k_2-j_2-2)}]
  \, \big|_{\rm depth\ 2}, \epsilon_{k_1}^{(k_1-j_1-2)} \big] \, .\notag 
 \end{align}
The $\overline{ {\cal E}_0\big[\begin{smallmatrix} j \\ k\end{smallmatrix}  ;\tau\big] }$
on the right-hand side contribute $\beta_-$ of depth one via (\ref{forlater})
which match the $\nicedel_\phi \beta_-
\big[\begin{smallmatrix} j_3 &j_2 &j_1 \\ k_3 &k_2 &k_1\end{smallmatrix}  \big] 
 \, \big|_{\rm depth \ 1}$ in the fourth line of (\ref{beqv.new31}) (again after 
 conversion (\ref{MGFtoBR14}) to $\overline{ \kappa[\begin{smallmatrix}\ldots\\ \ldots \end{smallmatrix};\tau]}$).
The nested brackets in the last two lines of (\ref{appb.08}) reproduce the
contributions $\nicedel_\phi \beta_- \big[\begin{smallmatrix} j_3 &j_2 &j_1  \\ k_3 &k_2 &k_1 
\end{smallmatrix}  \big]  \, \big|_{\rm depth \ 2}$
$- \nicedel_\phi \beta_- \big[\begin{smallmatrix} j_2 &j_1 \\ k_2 &k_1 \end{smallmatrix}   \big]  \beta_- \big[\begin{smallmatrix} j_3 \\ k_3 \end{smallmatrix}  \big] $ in
the fourth line of (\ref{beqv.new31}) when the $\overline{ {\cal E}_0\big[\begin{smallmatrix}
 j_1 &j_2 \\ k_1 &k_2\end{smallmatrix} ;\tau \big] }$ are restricted to the $\beta_-$ of
depth two in (\ref{appb.05}).
Finally, the last line of (\ref{beqv.new31}) has depth zero as well as at least one power
of $\bar \tau$ in each term and should account for the contributions to (\ref{forlater})
and (\ref{appb.05}) without any factor of $\beta_-$.

Similar to the comments before section \ref{sec:cusp}, the coefficients of 
$\epsilon_{k_1}^{(j_1)}\epsilon_{k_2}^{(j_2)}\epsilon_{k_3}^{(j_3)}$ in the 
generating-series identity (\ref{appb.08}) are understood to be equated before
employing any commutator relation in Tsunogai's derivation algebra.
The $\overline{ \alpha_{\rm hard}\big[\begin{smallmatrix} j_1 &j_2 &j_3 \\ k_1 &k_2 &k_3
 \end{smallmatrix} ;\tau \big] }$ are defined individually from (\ref{appb.08})
when the commutators among $z_m$ and $\epsilon_k$ are evaluated through
(\ref{MGFtoBR44}) and the expressions in the ancillary files.

\subsection{${\cal E}_0$ at depth two from iterated integrals}
\label{app.B.just}

We conclude this appendix by elaborating on the origin of the antiholomorphic
$T$-invariants at depth two in (\ref{appb.02}) from iterated integrals over kernels
${\rm G}_k^0(\tau) = {\rm G}_k(\tau) - 2 \zeta_k = { O}(q)$. The
first line of (\ref{appb.02}) can be traced back to
\begin{align}
&(2\pi i)^{2+j_1+j_2-k_1-k_2} 
 \int_{\bar \tau}^{-i\infty} \dd \bar \tau_1 \, (\bar \tau_1{-}\bar \tau)^{j_1} \overline{ {\rm G}^0_{k_1}(\tau_1)}
  \int_{\bar \tau_1}^{-i\infty} \dd \bar \tau_2 \, (\bar \tau_2{-}\bar \tau)^{j_2} \overline{ {\rm G}^0_{k_2}(\tau_2)} \notag \\
  &= (2\pi i)^{2+j_1+j_2-k_1-k_2}  \sum_{r=0}^{j_2} { j_2 \choose r}
 \int_{\bar \tau}^{-i\infty} \dd \bar \tau_1 \, (\bar \tau_1{-}\bar \tau)^{j_1+r} \overline{ {\rm G}^0_{k_1}(\tau_1)}
  \int_{\bar \tau_1}^{-i\infty} \dd \bar \tau_2 \, (\bar \tau_2{-}\bar \tau_1)^{j_2-r} \overline{ {\rm G}^0_{k_2}(\tau_2)}
\notag\\
&=   j_2! \sum_{r=0}^{j_2} \frac{ (j_1{+}r)! }{r!} \,
 \overline{ {\cal E}_0(k_2,0^{j_2-r},k_1,0^{j_1+r};\tau) } \, ,
\label{appb.91} 
\end{align}
where we have rewritten $(\bar \tau_2{-}\bar \tau)^{j_2} =  \sum_{r=0}^{j_2} { j_2 \choose r}
(\bar \tau_1{-}\bar \tau)^r (\bar \tau_2 {-}\bar \tau_1)^{j_2-r}$ and identified the
$\overline{ {\cal E}_0 }$ through their iterated-integral representation in (\ref{appb.03}).
By comparing with the integral representation (\ref{MGFtoBR41}) of $\overline{ {\cal E}_0\big[
\begin{smallmatrix} j \\ k \end{smallmatrix} ;\tau \big] }$, already the restriction of
$\overline{ {\cal E}_0\big[\begin{smallmatrix} j_1 &j_2 \\ k_1 &k_2 \end{smallmatrix} ;\tau \big] }$
to the double integrals (\ref{appb.91}) is easily seen to obey the shuffle relation (\ref{appb.04}).
The second line of (\ref{appb.02}) involving depth-one integrals over ${\rm G}_{k_i}^0$
in turn is antisymmetric under $(j_1,k_1) \leftrightarrow (j_2,k_2)$ and therefore drops
out from the symmetrized combination on the right-hand side of the shuffle relation
(\ref{appb.04}). Hence, the shuffle property of the $\overline{ {\cal E}_0[\ldots;\tau] }$
is a consequence of the integral representation (\ref{appb.91}).

We shall now state an alternative integral representation of the
$\overline{ {\cal E}_0\big[\begin{smallmatrix} j_1 &j_2 \\ k_1 &k_2 \end{smallmatrix} ;\tau \big] }$ 
which manifests their relation with Brown's kernels 
$\bar \tau^j \overline{ {\rm G}_k(\tau)}$ with $0\leq j \leq k{-}2$: Following
the tangential-basepoint regularization of \cite{Brown:mmv}, one can retrieve
(\ref{appb.02}) from
\begin{align}
\ezero{j_1 &j_2 }{ k_1 &k_2 }{\tau} &= (2\pi i)^{2+j_1+j_2-k_1-k_2} 
 \int_{\bar \tau}^{-i\infty} \dd \bar \tau_1 \, (\bar \tau_1{-}\bar \tau)^{j_1} \overline{ {\rm G}_{k_1}(\tau_1)}
  \int_{\bar \tau_1}^{-i\infty} \dd \bar \tau_2 \, (\bar \tau_2{-}\bar \tau)^{j_2} \overline{ {\rm G}_{k_2}(\tau_2)}
  \notag \\
  &\quad
  - \frac{ {\rm B}_{k_2} \, (-2\pi i \bar \tau)^{j_2+1} }{(j_2{+}1) k_2!} \, \ezero{j_1 }{ k_1  }{\tau} 
  - \frac{ {\rm B}_{k_1} \, {\rm B}_{k_2} \, (-2\pi i \bar \tau)^{j_1+j_2+2}}{(j_1{+}1) (j_1{+}j_2{+}2) k_1!  k_2! } \, .
\label{appb.92} 
\end{align}
Each term on the right-hand side is an (iterated) integral of kernels
$\bar \tau_i^{j_i} \overline{ {\rm G}_{k_i}(\tau_i)}$ with $0\leq j_i \leq k_i{-}2$ with
$\mathbb Q[2\pi i \bar \tau]$-coefficients: This is evidently the case for the first
line of (\ref{appb.92}) upon binomial expansion of both factors of 
$ (\bar \tau_i{-}\bar \tau)^{j_i}$, and the depth-one term
$\overline{ {\cal E}_0\big[\begin{smallmatrix} j_1 \\ k_1 \end{smallmatrix} ;\tau \big] }$
in the second line can be checked to have the same property via binomial expansion 
in (\ref{MGFtoBR41}). The alternative form (\ref{appb.05}) in terms of $\beta_-$ 
follows from rearranging the integral in the first line of (\ref{appb.92}) and
applying (\ref{forlater}) to its second line.

\providecommand{\href}[2]{#2}\begingroup\raggedright\endgroup


\begin{thebibliography}{10}

\bibitem{Schlotterer:2012ny}
O.~Schlotterer and S.~Stieberger, ``{Motivic Multiple Zeta Values and
  Superstring Amplitudes},''
  \href{http://dx.doi.org/10.1088/1751-8113/46/47/475401}{{\em J. Phys.} {\bf
  A46} (2013)  475401},
\href{http://arxiv.org/abs/1205.1516}{{\tt arXiv:1205.1516 [hep-th]}}.

\bibitem{Terasoma}
T.~Terasoma, ``{Selberg integrals and multiple zeta values},''
  \href{http://dx.doi.org/10.1023/A:1016377828316}{{\em Compositio Math.} {\bf
  133} (2002) no.~1, 1--24}.

\bibitem{Drummond:2013vz}
J.~M. Drummond and E.~Ragoucy, ``{Superstring amplitudes and the associator},''
  \href{http://dx.doi.org/10.1007/JHEP08(2013)135}{{\em JHEP} {\bf 08} (2013)
  135},
\href{http://arxiv.org/abs/1301.0794}{{\tt arXiv:1301.0794 [hep-th]}}.

\bibitem{Broedel:2013aza}
J.~Broedel, O.~Schlotterer, S.~Stieberger, and T.~Terasoma, ``{All order
  $\alpha^{\prime}$-expansion of superstring trees from the Drinfeld
  associator},'' \href{http://dx.doi.org/10.1103/PhysRevD.89.066014}{{\em Phys.
  Rev.} {\bf D89} (2014) no.~6, 066014},
\href{http://arxiv.org/abs/1304.7304}{{\tt arXiv:1304.7304 [hep-th]}}.

\bibitem{Kaderli:2019dny}
A.~Kaderli, ``{A note on the Drinfeld associator for genus-zero superstring
  amplitudes in twisted de Rham theory},''
  \href{http://dx.doi.org/10.1088/1751-8121/ab9462}{{\em J. Phys. A} {\bf 53}
  (2020) no.~41, 415401}, \href{http://arxiv.org/abs/1912.09406}{{\tt
  arXiv:1912.09406 [hep-th]}}.

\bibitem{Stieberger:2013wea}
S.~Stieberger, ``{Closed superstring amplitudes, single-valued multiple zeta
  values and the Deligne associator},''
  \href{http://dx.doi.org/10.1088/1751-8113/47/15/155401}{{\em J. Phys.} {\bf
  A47} (2014)  155401},
\href{http://arxiv.org/abs/1310.3259}{{\tt arXiv:1310.3259 [hep-th]}}.

\bibitem{Stieberger:2014hba}
S.~Stieberger and T.~R. Taylor, ``{Closed String Amplitudes as Single-Valued
  Open String Amplitudes},''
  \href{http://dx.doi.org/10.1016/j.nuclphysb.2014.02.005}{{\em Nucl. Phys.}
  {\bf B881} (2014)  269--287},
\href{http://arxiv.org/abs/1401.1218}{{\tt arXiv:1401.1218 [hep-th]}}.

\bibitem{Schlotterer:2018abc}
O.~Schlotterer and O.~Schnetz, ``{Closed strings as single-valued open strings:
  A genus-zero derivation},''
  \href{http://dx.doi.org/10.1088/1751-8121/aaea14}{{\em J. Phys.} {\bf A52}
  (2019) no.~4, 045401},
\href{http://arxiv.org/abs/1808.00713}{{\tt arXiv:1808.00713 [hep-th]}}.

\bibitem{Vanhove:2018elu}
P.~Vanhove and F.~Zerbini, ``{Single-valued hyperlogarithms, correlation
  functions and closed string amplitudes},''{\em Adv. Theor. Math. Phys.} {\bf
  26} (12, 2022)  , \href{http://arxiv.org/abs/1812.03018}{{\tt
  arXiv:1812.03018 [hep-th]}}.

\bibitem{Brown:2019wna}
F.~Brown and C.~Dupont, ``{Single-valued integration and superstring amplitudes
  in genus zero},'' \href{http://dx.doi.org/10.1007/s00220-021-03969-4}{{\em
  Commun. Math. Phys.} {\bf 382} (2021) no.~2, 815--874},
  \href{http://arxiv.org/abs/1910.01107}{{\tt arXiv:1910.01107 [math.NT]}}.

\bibitem{BrownLev}
F.~Brown and A.~Levin, ``{Multiple elliptic polylogarithms},''
  \href{http://arxiv.org/abs/1110.6917}{{\tt arXiv:1110.6917 [math]}}.

\bibitem{Enriquez:Emzv}
B.~Enriquez, ``Analogues elliptiques des nombres multiz\'etas,''
  \href{http://dx.doi.org/10.24033/bsmf.2718}{{\em Bull. Soc. Math. France}
  {\bf 144} (2016) no.~3, 395--427}, \href{http://arxiv.org/abs/1301.3042}{{\tt
  1301.3042}}.

\bibitem{Broedel:2014vla}
J.~Broedel, C.~R. Mafra, N.~Matthes, and O.~Schlotterer, ``{Elliptic multiple
  zeta values and one-loop superstring amplitudes},''
  \href{http://dx.doi.org/10.1007/JHEP07(2015)112}{{\em JHEP} {\bf 07} (2015)
  112},
\href{http://arxiv.org/abs/1412.5535}{{\tt arXiv:1412.5535 [hep-th]}}.

\bibitem{Broedel:2017jdo}
J.~Broedel, N.~Matthes, G.~Richter, and O.~Schlotterer, ``{Twisted elliptic
  multiple zeta values and non-planar one-loop open-string amplitudes},''
  \href{http://dx.doi.org/10.1088/1751-8121/aac601}{{\em J. Phys.} {\bf A51}
  (2018) no.~28, 285401},
\href{http://arxiv.org/abs/1704.03449}{{\tt arXiv:1704.03449 [hep-th]}}.

\bibitem{Green:1999pv}
M.~B. Green and P.~Vanhove, ``{The Low-energy expansion of the one loop type II
  superstring amplitude},''
  \href{http://dx.doi.org/10.1103/PhysRevD.61.104011}{{\em Phys.Rev.} {\bf D61}
  (2000)  104011},
\href{http://arxiv.org/abs/hep-th/9910056}{{\tt arXiv:hep-th/9910056
  [hep-th]}}.

\bibitem{Green:2008uj}
M.~B. Green, J.~G. Russo, and P.~Vanhove, ``{Low energy expansion of the
  four-particle genus-one amplitude in type II superstring theory},''
  \href{http://dx.doi.org/10.1088/1126-6708/2008/02/020}{{\em JHEP} {\bf 02}
  (2008)  020},
\href{http://arxiv.org/abs/0801.0322}{{\tt arXiv:0801.0322 [hep-th]}}.

\bibitem{DHoker:2015gmr}
E.~D'Hoker, M.~B. Green, and P.~Vanhove, ``{On the modular structure of the
  genus-one Type II superstring low energy expansion},''
  \href{http://dx.doi.org/10.1007/JHEP08(2015)041}{{\em JHEP} {\bf 08} (2015)
  041},
\href{http://arxiv.org/abs/1502.06698}{{\tt arXiv:1502.06698 [hep-th]}}.

\bibitem{Gerken:2018jrq}
J.~E. Gerken, A.~Kleinschmidt, and O.~Schlotterer, ``{Heterotic-string
  amplitudes at one loop: modular graph forms and relations to open strings},''
  \href{http://dx.doi.org/10.1007/JHEP01(2019)052}{{\em JHEP} {\bf 01} (2019)
  052},
\href{http://arxiv.org/abs/1811.02548}{{\tt arXiv:1811.02548 [hep-th]}}.

\bibitem{DHoker:2015wxz}
E.~D'Hoker, M.~B. Green, {\"O}.~G{\"u}rdogan, and P.~Vanhove, ``Modular graph
  functions,'' \href{http://dx.doi.org/10.4310/CNTP.2017.v11.n1.a4}{{\em
  Commun. Num. Theor. Phys.} {\bf 11} (2017)  165--218},
\href{http://arxiv.org/abs/1512.06779}{{\tt arXiv:1512.06779 [hep-th]}}.

\bibitem{DHoker:2016mwo}
E.~D'Hoker and M.~B. Green, ``Identities between modular graph forms,''
  \href{http://dx.doi.org/10.1016/j.jnt.2017.11.015}{{\em J. Number Theory}
  {\bf 189} (2018)  25--80},
\href{http://arxiv.org/abs/1603.00839}{{\tt arXiv:1603.00839 [hep-th]}}.

\bibitem{Gerken:review}
J.~E. Gerken, ``{Modular Graph Forms and Scattering Amplitudes in String
  Theory},'' \href{http://arxiv.org/abs/2011.08647}{{\tt arXiv:2011.08647
  [hep-th]}}.

\bibitem{Berkovits:2022ivl}
N.~Berkovits, E.~D'Hoker, M.~B. Green, H.~Johansson, and O.~Schlotterer,
  ``{Snowmass White Paper: String Perturbation Theory},'' in {\em {2022
  Snowmass Summer Study}}.
\newblock 3, 2022.
\newblock \href{http://arxiv.org/abs/2203.09099}{{\tt arXiv:2203.09099
  [hep-th]}}.

\bibitem{Dorigoni:2022iem}
D.~Dorigoni, M.~B. Green, and C.~Wen, ``{The SAGEX Review on Scattering
  Amplitudes, Chapter 10: Modular covariance of type IIB string amplitudes and
  their $\mathcal{N}=4$ supersymmetric Yang-Mills duals},''
  \href{http://arxiv.org/abs/2203.13021}{{\tt arXiv:2203.13021 [hep-th]}}.

\bibitem{DHoker:2022dxx}
E.~D'Hoker and J.~Kaidi, ``{Lectures on modular forms and strings},''
  \href{http://arxiv.org/abs/2208.07242}{{\tt arXiv:2208.07242 [hep-th]}}.

\bibitem{DHoker:2015sve}
E.~D'Hoker, M.~B. Green, and P.~Vanhove, ``{Proof of a modular relation between
  1-, 2- and 3-loop Feynman diagrams on a torus},''
  \href{http://dx.doi.org/10.1016/j.jnt.2017.07.022}{{\em J.\ Number Theory}
  (2018)  381},
\href{http://arxiv.org/abs/1509.00363}{{\tt arXiv:1509.00363 [hep-th]}}.

\bibitem{DHoker:2016quv}
E.~D'Hoker and J.~Kaidi, ``{Hierarchy of Modular Graph Identities},''
  \href{http://dx.doi.org/10.1007/JHEP11(2016)051}{{\em JHEP} {\bf 11} (2016)
  051},
\href{http://arxiv.org/abs/1608.04393}{{\tt arXiv:1608.04393 [hep-th]}}.

\bibitem{Basu:2016kli}
A.~Basu, ``{Proving relations between modular graph functions},''
  \href{http://dx.doi.org/10.1088/0264-9381/33/23/235011}{{\em Class. Quant.
  Grav.} {\bf 33} (2016) no.~23, 235011},
\href{http://arxiv.org/abs/1606.07084}{{\tt arXiv:1606.07084 [hep-th]}}.

\bibitem{Gerken:2018zcy}
J.~E. Gerken and J.~Kaidi, ``{Holomorphic subgraph reduction of higher-point
  modular graph forms},'' \href{http://dx.doi.org/10.1007/JHEP01(2019)131}{{\em
  JHEP} {\bf 01} (2019)  131},
\href{http://arxiv.org/abs/1809.05122}{{\tt arXiv:1809.05122 [hep-th]}}.

\bibitem{Zerbini:2015rss}
F.~Zerbini, ``{Single-valued multiple zeta values in genus 1 superstring
  amplitudes},'' \href{http://dx.doi.org/10.4310/CNTP.2016.v10.n4.a2}{{\em
  Commun. Num. Theor. Phys.} {\bf 10} (2016)  703--737},
\href{http://arxiv.org/abs/1512.05689}{{\tt arXiv:1512.05689 [hep-th]}}.

\bibitem{DHoker:2017zhq}
E.~D'Hoker and W.~Duke, ``Fourier series of modular graph functions,''
  \href{http://dx.doi.org/10.1016/j.jnt.2018.04.012}{{\em J. Number Theory}
  {\bf 192} (2018)  1--36}, \href{http://arxiv.org/abs/1708.07998}{{\tt
  arXiv:1708.07998 [math.NT]}}.

\bibitem{Panzertalk}
E.~Panzer, ``{Talk ``Modular graph functions as iterated Eisenstein integrals''
  given at the workshop ``Elliptic Integrals in Mathematics and Physics''
  (Ascona, Switzerland)}.''
  \url{https://indico.cern.ch/event/700233/contributions/3112451/attachments/1712442/2761239/elliptic.pdf},
  2018.

\bibitem{DHoker:2019xef}
E.~D'Hoker and M.~B. Green, ``{Absence of irreducible multiple zeta-values in
  melon modular graph functions},''
  \href{http://dx.doi.org/10.4310/CNTP.2020.v14.n2.a2}{{\em Commun. Num. Theor.
  Phys.} {\bf 14} (2020) no.~2, 315--324},
\href{http://arxiv.org/abs/1904.06603}{{\tt arXiv:1904.06603 [hep-th]}}.

\bibitem{Zagier:2019eus}
D.~Zagier and F.~Zerbini, ``{Genus-zero and genus-one string amplitudes and
  special multiple zeta values},''
  \href{http://dx.doi.org/10.4310/CNTP.2020.v14.n2.a4}{{\em Commun. Num. Theor.
  Phys.} {\bf 14} (2020) no.~2, 413--452},
\href{http://arxiv.org/abs/1906.12339}{{\tt arXiv:1906.12339 [math.NT]}}.

\bibitem{Vanhove:2020qtt}
P.~Vanhove and F.~Zerbini, ``{Building blocks of closed and open string
  amplitudes},'' \href{http://dx.doi.org/10.22323/1.383.0022}{{\em PoS} {\bf
  MA2019} (2022)  022}, \href{http://arxiv.org/abs/2007.08981}{{\tt
  arXiv:2007.08981 [hep-th]}}.

\bibitem{Brown:mmv}
F.~Brown, ``{Multiple modular values and the relative completion of the
  fundamental group of ${\cal M}_{1,1}$},''
  \href{http://arxiv.org/abs/1407.5167}{{\tt arXiv:1407.5167 [math.NT]}}.

\bibitem{Brown:2017qwo}
F.~Brown, ``{A class of non-holomorphic modular forms I},''
  \href{http://dx.doi.org/10.1007/s40687-018-0130-8}{{\em Res. Math. Sci.} {\bf
  5} (2018)  5:7}, \href{http://arxiv.org/abs/1707.01230}{{\tt arXiv:1707.01230
  [math.NT]}}.

\bibitem{Brown:2017qwo2}
F.~Brown, ``{A class of non-holomorphic modular forms II : equivariant iterated
  Eisenstein integrals},'' \href{http://dx.doi.org/10.1017/fms.2020.24}{{\em
  Forum~of~Mathematics,~Sigma} {\bf 8} (2020)  1},
\href{http://arxiv.org/abs/1708.03354}{{\tt arXiv:1708.03354 [math.NT]}}.

\bibitem{Brown:2004ugm}
F.~C.~S. Brown, ``{Polylogarithmes multiples uniformes en une variable},''
  \href{http://dx.doi.org/10.1016/j.crma.2004.02.001}{{\em Compt. Rend. Math.}
  {\bf 338} (2004) no.~7, 527--532}.

\bibitem{Diamantis:2019}
N.~Diamantis and J.~Drewitt, ``{Period functions associated to real-analytic
  modular forms},'' \href{http://dx.doi.org/10.1007/s40687-020-00221-8}{{\em
  Res. Math. Sci.} {\bf 7} (2020)  21},
  \href{http://arxiv.org/abs/1907.02895}{{\tt arXiv:1907.02895 [math.NT]}}.

\bibitem{Drewitt:2021}
J.~Drewitt, ``Laplace-eigenvalue equations for length three modular iterated
  integrals,''
  \href{http://dx.doi.org/https://doi.org/10.1016/j.jnt.2021.11.005}{{\em
  Journal of Number Theory} {\bf 239} (2022)  78--112},
  \href{http://arxiv.org/abs/2104.09916}{{\tt arXiv:2104.09916 [math.NT]}}.

\bibitem{Gerken:2019cxz}
J.~E. Gerken, A.~Kleinschmidt, and O.~Schlotterer, ``{All-order differential
  equations for one-loop closed-string integrals and modular graph forms},''
  \href{http://dx.doi.org/10.1007/JHEP01(2020)064}{{\em JHEP} {\bf 01} (2020)
  064},
\href{http://arxiv.org/abs/1911.03476}{{\tt arXiv:1911.03476 [hep-th]}}.

\bibitem{Gerken:2020yii}
J.~E. Gerken, A.~Kleinschmidt, and O.~Schlotterer, ``{Generating series of all
  modular graph forms from iterated Eisenstein integrals},''
  \href{http://dx.doi.org/10.1007/JHEP07(2020)190}{{\em JHEP} {\bf 07} (2020)
  no.~07, 190}, \href{http://arxiv.org/abs/2004.05156}{{\tt arXiv:2004.05156
  [hep-th]}}.

\bibitem{Dorigoni:2021jfr}
D.~Dorigoni, A.~Kleinschmidt, and O.~Schlotterer, ``{Poincar\'e series for
  modular graph forms at depth two. Part I. Seeds and Laplace systems},''
  \href{http://dx.doi.org/10.1007/JHEP01(2022)133}{{\em JHEP} {\bf 01} (2022)
  133}, \href{http://arxiv.org/abs/2109.05017}{{\tt arXiv:2109.05017
  [hep-th]}}.

\bibitem{Dorigoni:2021ngn}
D.~Dorigoni, A.~Kleinschmidt, and O.~Schlotterer, ``{Poincar\'e series for
  modular graph forms at depth two. Part II. Iterated integrals of cusp
  forms},'' \href{http://dx.doi.org/10.1007/JHEP01(2022)134}{{\em JHEP} {\bf
  01} (2022)  134}, \href{http://arxiv.org/abs/2109.05018}{{\tt
  arXiv:2109.05018 [hep-th]}}.

\bibitem{depth3paper}
D.~Dorigoni, M.~Doroudiani, J.~Drewitt, M.~Hidding, A.~Kleinschmidt,
  N.~Matthes, O.~Schlotterer, and B.~Verbeek, {\em {to appear}}.

\bibitem{Ihara:1990}
Y.~Ihara, ``Braids, {G}alois groups, and some arithmetic functions,'' in {\em
  Proceedings of the {I}nternational {C}ongress of {M}athematicians, {V}ol.
  {I}, {II} ({K}yoto, 1990)}, pp.~99--120.
\newblock Math. Soc. Japan, Tokyo, 1991.

\bibitem{Deligne:1987}
P.~Deligne, \href{http://dx.doi.org/10.1007/978-1-4613-9649-9\_3}{``Le groupe
  fondamental de la droite projective moins trois points,''} in {\em Galois
  groups over {${\bf Q}$} ({B}erkeley, {CA}, 1987)}, vol.~16 of {\em Math. Sci.
  Res. Inst. Publ.}, pp.~79--297.
\newblock Springer, New York, 1989.

\bibitem{IharaTakao:alt}
Y.~Ihara, ``Some arithmetic aspects of Galois actions in the pro-p fundamental group of $P1 \setminus \{0, 1, \infty\}$,'' in Arithmetic Fundamental Groups and Noncommutative Algebra (Berkeley, CA, 1999), Proceedings of Symposia in Pure Mathematics, Vol.\ 70, pp.\ 247--273 (American Mathematical Society, 2002).

\bibitem{Tsunogai}
H.~Tsunogai, ``On some derivations of {L}ie algebras related to {G}alois
  representations,'' \href{http://dx.doi.org/10.2977/prims/1195164794}{{\em
  Publ. Res. Inst. Math. Sci.} {\bf 31} (1995) no.~1, 113--134}.

\bibitem{Gonchtalk}
A.~B. Goncharov, ``Multiple $\zeta$-values, {G}alois groups, and geometry of
  modular varieties,'' in {\em European Congress of Mathematics},
  C.~Casacuberta, R.~M. Mir{\'o}-Roig, J.~Verdera, and S.~Xamb{\'o}-Descamps,
  eds., pp.~361--392.
\newblock Birkh{\"a}user Basel, Basel, 2001.

\bibitem{GKZ:2006}
H.~Gangl, M.~Kaneko, and D.~Zagier,
  \href{http://dx.doi.org/10.1142/9789812774415\_0004}{``Double zeta values and
  modular forms,''} in {\em Automorphic forms and zeta functions}, pp.~71--106.
\newblock World Sci. Publ., Hackensack, NJ, 2006.

\bibitem{Schneps:2006}
L.~Schneps, ``On the {P}oisson bracket on the free {L}ie algebra in two
  generators,'' {\em J. Lie Theory} {\bf 16} (2006) no.~1, 19--37.

\bibitem{Pollack}
A.~Pollack, ``{Relations between derivations arising from modular forms}.''
  \url{https://dukespace.lib.duke.edu/dspace/handle/10161/1281}, 2009.
\newblock Undergraduate thesis, Duke University.

\bibitem{BaumardSchneps:2015}
S.~Baumard and L.~Schneps, ``On the derivation representation of the
  fundamental {L}ie algebra of mixed elliptic motives,''
  \href{http://dx.doi.org/10.1007/s40316-015-0040-8}{{\em Ann. Math. Qu\'{e}.}
  {\bf 41} (2017) no.~1, 43--62}.

\bibitem{hain_matsumoto_2020}
R.~Hain and M.~Matsumoto, ``Universal mixed elliptic motives,''
  \href{http://dx.doi.org/10.1017/S1474748018000130}{{\em Journal of the
  Institute of Mathematics of Jussieu} {\bf 19} (2020) no.~3, 663--766},
  \href{http://arxiv.org/abs/1512.03975}{{\tt arXiv:1512.03975 [math.AG]}}.

\bibitem{Brown:Anatomy}
F.~Brown, ``{Talk ``Anatomy of the motivic Lie algebra'' given at the program
  ``Grothendieck-Teichm\"uller Groups, Deformation and Operads'' (Newton
  Institute, Cambridge, UK)}.'' \url{https://sms.cam.ac.uk/media/1459610},
  2013.

\bibitem{Brown:depth3}
F.~Brown, ``Zeta elements in depth 3 and the fundamental {L}ie algebra of the
  infinitesimal {T}ate curve,''
  \href{http://dx.doi.org/10.1017/fms.2016.29}{{\em Forum Math. Sigma} {\bf 5}
  (2017)  Paper No. e1, 56},
  \href{http://arxiv.org/abs/1504.04737}{{\tt arXiv:1504.04737 [math.NT]}}.

\bibitem{Gerken:2020aju}
J.~E. Gerken, ``{Basis Decompositions and a Mathematica Package for Modular
  Graph Forms},'' \href{http://dx.doi.org/10.1088/1751-8121/abbdf2}{{\em J.
  Phys. A} {\bf 54} (2021) no.~19, 195401},
  \href{http://arxiv.org/abs/2007.05476}{{\tt arXiv:2007.05476 [hep-th]}}.

\bibitem{Ganglzagier}
D.~Zagier and H.~Gangl, ``Classical and elliptic polylogarithms and special
  values of {$L$}-series,'' in {\em The arithmetic and geometry of algebraic
  cycles ({B}anff, {AB}, 1998)}, vol.~548 of {\em NATO Sci. Ser. C Math. Phys.
  Sci.}, pp.~561--615.
\newblock Kluwer Acad. Publ., Dordrecht, 2000.

\bibitem{Nilsnewarticle}
N.~Matthes, ``On the algebraic structure of iterated integrals of quasimodular
  forms,'' \href{http://dx.doi.org/10.2140/ant.2017.11.2113}{{\em Algebra \&
  Number Theory} {\bf 11-9} (2017)  2113--2130},
  \href{http://arxiv.org/abs/1708.04561}{{\tt arXiv:1708.04561}}.

\bibitem{Brown2019}
F.~Brown, ``{From the Deligne-Ihara conjecture to multiple modular values},''
  \href{http://arxiv.org/abs/1904.00179}{{\tt arXiv:1904.00179 [math.AG]}}.

\bibitem{LNT}
J.-G. Luque, J.-C. Novelli, and J.-Y. Thibon, ``{Period polynomials and Ihara
  brackets},'' {\em J. Lie Theory} {\bf 17} (2007)  229--239,
  \href{http://arxiv.org/abs/math/0606301}{{\tt arXiv:math/0606301
  [math.CO,math.NT]}}.

\bibitem{Schnetz:2013hqa}
O.~Schnetz, ``{Graphical functions and single-valued multiple
  polylogarithms},'' \href{http://dx.doi.org/10.4310/CNTP.2014.v8.n4.a1}{{\em
  Commun. Num. Theor. Phys.} {\bf 08} (2014)  589--675},
\href{http://arxiv.org/abs/1302.6445}{{\tt arXiv:1302.6445 [math.NT]}}.

\bibitem{Brown:2013gia}
F.~Brown, ``{Single-valued Motivic Periods and Multiple Zeta Values},''
  \href{http://dx.doi.org/10.1017/fms.2014.18}{{\em SIGMA} {\bf 2} (2014)
  e25},
\href{http://arxiv.org/abs/1309.5309}{{\tt arXiv:1309.5309 [math.NT]}}.

\bibitem{Broedel:2018izr}
J.~Broedel, O.~Schlotterer, and F.~Zerbini, ``{From elliptic multiple zeta
  values to modular graph functions: open and closed strings at one loop},''
  \href{http://dx.doi.org/10.1007/JHEP01(2019)155}{{\em JHEP} {\bf 01} (2019)
  155},
\href{http://arxiv.org/abs/1803.00527}{{\tt arXiv:1803.00527 [hep-th]}}.

\bibitem{Broedel:2015hia}
J.~Broedel, N.~Matthes, and O.~Schlotterer, ``{Relations between elliptic
  multiple zeta values and a special derivation algebra},''
  \href{http://dx.doi.org/10.1088/1751-8113/49/15/155203}{{\em J. Phys.} {\bf
  A49} (2016) no.~15, 155203},
\href{http://arxiv.org/abs/1507.02254}{{\tt arXiv:1507.02254 [hep-th]}}.

\bibitem{Gerken:2020xfv}
J.~E. Gerken, A.~Kleinschmidt, C.~R. Mafra, O.~Schlotterer, and B.~Verbeek,
  ``{Towards closed strings as single-valued open strings at genus one},''
  \href{http://dx.doi.org/10.1088/1751-8121/abe58b}{{\em J. Phys. A} {\bf 55}
  (2022) no.~2, 025401}, \href{http://arxiv.org/abs/2010.10558}{{\tt
  arXiv:2010.10558 [hep-th]}}.

\bibitem{Brown:2011ik}
F.~Brown, ``On the decomposition of motivic multiple zeta values,'' in {\em
  Galois-{T}eichm\"uller theory and arithmetic geometry}, vol.~63 of {\em Adv.
  Stud. Pure Math.}, pp.~31--58.
\newblock Math. Soc. Japan, Tokyo, 2012.
\newblock
\href{http://arxiv.org/abs/1102.1310}{{\tt arXiv:1102.1310 [math.NT]}}.
\newblock

\bibitem{Goncharov:2005sla}
A.~Goncharov, ``{Galois symmetries of fundamental groupoids and noncommutative
  geometry},'' \href{http://dx.doi.org/10.1215/S0012-7094-04-12822-2}{{\em Duke
  Math.J.} {\bf 128} (2005)  209},
\href{http://arxiv.org/abs/math/0208144}{{\tt arXiv:math/0208144 [math.AG]}}.

\bibitem{Brown:2011mot}
F.~Brown, ``{Mixed Tate motives over $\mathbb Z$},''
  \href{http://dx.doi.org/10.4007/annals.2012.175.2.10}{{\em Ann.\ Math.} {\bf
  175} (2012)  949}, \href{http://arxiv.org/abs/1102.1312}{{\tt arXiv:1102.1312
  [math.AG]}}.

\bibitem{Blumlein:2009cf}
J.~Bl{\"u}mlein, D.~J. Broadhurst, and J.~A.~M. Vermaseren, ``{The Multiple
  Zeta Value Data Mine},''
  \href{http://dx.doi.org/10.1016/j.cpc.2009.11.007}{{\em Comput. Phys.
  Commun.} {\bf 181} (2010)  582--625},
\href{http://arxiv.org/abs/0907.2557}{{\tt arXiv:0907.2557 [math-ph]}}.

\bibitem{DHoker:2019txf}
E.~D'Hoker and J.~Kaidi, ``{Modular graph functions and odd cuspidal functions.
  Fourier and Poincar{\'e} series},''
  \href{http://dx.doi.org/10.1007/JHEP04(2019)136}{{\em JHEP} {\bf 04} (2019)
  136},
\href{http://arxiv.org/abs/1902.04180}{{\tt arXiv:1902.04180 [hep-th]}}.

\bibitem{Saad:2020mzv}
A.~Saad, ``{Multiple zeta values and iterated Eisenstein integrals},''
  \href{http://arxiv.org/abs/2009.09885}{{\tt arXiv:2009.09885 [math.NT]}}.



\bibitem{Broedel:2016kls}
J.~Broedel, M.~Sprenger, and A.~Torres~Orjuela, ``{Towards single-valued
  polylogarithms in two variables for the seven-point remainder function in
  multi-Regge-kinematics},''
  \href{http://dx.doi.org/10.1016/j.nuclphysb.2016.12.016}{{\em Nucl. Phys. B}
  {\bf 915} (2017)  394--413}, \href{http://arxiv.org/abs/1606.08411}{{\tt
  arXiv:1606.08411 [hep-th]}}.

\bibitem{DelDuca:2016lad}
V.~Del~Duca, S.~Druc, J.~Drummond, C.~Duhr, F.~Dulat, R.~Marzucca,
  G.~Papathanasiou, and B.~Verbeek, ``{Multi-Regge kinematics and the moduli
  space of Riemann spheres with marked points},''
  \href{http://dx.doi.org/10.1007/JHEP08(2016)152}{{\em JHEP} {\bf 08} (2016)
  152}, \href{http://arxiv.org/abs/1606.08807}{{\tt arXiv:1606.08807
  [hep-th]}}.

\bibitem{Eichler:1957}
M.~Eichler, ``Eine {V}erallgemeinerung der {A}belschen {I}ntegrale,''
  \href{http://dx.doi.org/10.1007/BF01258863}{{\em Math. Z.} {\bf 67} (1957)
  267--298}.

\bibitem{Shimura:1959}
G.~Shimura, ``Sur les int\'{e}grales attach\'{e}es aux formes automorphes,''
  \href{http://dx.doi.org/10.2969/jmsj/01140291}{{\em J. Math. Soc. Japan} {\bf
  11} (1959)  291--311}.

\bibitem{Diamantis:2020}
N.~Diamantis, ``Modular iterated integrals associated with cusp forms,''
  \href{http://dx.doi.org/10.1515/forum-2021-0224}{{\em Forum Mathematicum}
  {\bf 34} (2022)  157--174}, \href{http://arxiv.org/abs/2009.07128}{{\tt
  arXiv:2009.07128 [math.NT]}}.

\bibitem{DHoker:2018mys}
E.~D'Hoker, M.~B. Green, and B.~Pioline, ``Asymptotics of the {$D^8{\cal R}^4$}
  genus-two string invariant,''
  \href{http://dx.doi.org/10.4310/CNTP.2019.v13.n2.a3}{{\em Commun. Num. Theor.
  Phys.} {\bf 13} (2019) no.~2, 351--462},
\href{http://arxiv.org/abs/1806.02691}{{\tt arXiv:1806.02691 [hep-th]}}.

\bibitem{Dhoker:2020gdz}
E.~D'Hoker, A.~Kleinschmidt, and O.~Schlotterer, ``{Elliptic modular graph
  forms. Part I. Identities and generating series},''
  \href{http://dx.doi.org/10.1007/JHEP03(2021)151}{{\em JHEP} {\bf 03} (2021)
  151}, \href{http://arxiv.org/abs/2012.09198}{{\tt arXiv:2012.09198
  [hep-th]}}.

\bibitem{Ramakrish}
D.~Zagier, ``The {B}loch-{W}igner-{R}amakrishnan polylogarithm function,''
  \href{http://dx.doi.org/10.1007/BF01453591}{{\em Math. Ann.} {\bf 286} (1990)
   613}.

\bibitem{new:eMGF}
M.~Hidding, O.~Schlotterer, and B.~Verbeek, ``{Elliptic modular graph forms II:
  Iterated integrals},'' \href{http://arxiv.org/abs/2208.11116}{{\tt
  arXiv:2208.11116 [hep-th]}}.

\bibitem{Brown:2018omk}
F.~Brown and C.~Dupont, ``{Single-valued integration and double copy},''
  \href{http://dx.doi.org/10.1515/crelle-2020-0042}{{\em J. Reine Angew. Math.}
  {\bf 2021} (2021) no.~775, 145--196},
  \href{http://arxiv.org/abs/1810.07682}{{\tt arXiv:1810.07682 [math.NT]}}.

\bibitem{FrancisLecture}
F.~Brown, ``{Notes on motivic periods},''
  \href{http://dx.doi.org/10.4310/CNTP.2017.v11.n3.a2}{{\em Communications in
  Number Theory and Physics} {\bf 11} (2015) no.~3, 557--655},
\href{http://arxiv.org/abs/1512.06410}{{\tt arXiv:1512.06410 [math.NT]}}.

\bibitem{DHoker:2017pvk}
E.~D'Hoker, M.~B. Green, and B.~Pioline, ``{Higher genus modular graph
  functions, string invariants, and their exact asymptotics},''
  \href{http://dx.doi.org/10.1007/s00220-018-3244-3}{{\em Commun. Math. Phys.}
  {\bf 366} (2019) no.~3, 927--979},
\href{http://arxiv.org/abs/1712.06135}{{\tt arXiv:1712.06135 [hep-th]}}.

\bibitem{DHoker:2020uid}
E.~D'Hoker and O.~Schlotterer, ``{Identities among higher genus modular graph
  tensors},'' \href{http://dx.doi.org/10.4310/CNTP.2022.v16.n1.a2}{{\em Commun.
  Num. Theor. Phys.} {\bf 16} (2022) no.~1, 35--74},
  \href{http://arxiv.org/abs/2010.00924}{{\tt arXiv:2010.00924 [hep-th]}}.

\bibitem{DHoker:2020tcq}
E.~D'Hoker, C.~R. Mafra, B.~Pioline, and O.~Schlotterer, ``{Two-loop
  superstring five-point amplitudes. Part II. Low energy expansion and
  S-duality},'' \href{http://dx.doi.org/10.1007/JHEP02(2021)139}{{\em JHEP}
  {\bf 02} (2021)  139}, \href{http://arxiv.org/abs/2008.08687}{{\tt
  arXiv:2008.08687 [hep-th]}}.

\bibitem{DHoker:2014oxd}
E.~D'Hoker, M.~B. Green, B.~Pioline, and R.~Russo, ``{Matching the $D^{6}R^{4}$
  interaction at two-loops},''
  \href{http://dx.doi.org/10.1007/JHEP01(2015)031}{{\em JHEP} {\bf 01} (2015)
  031},
\href{http://arxiv.org/abs/1405.6226}{{\tt arXiv:1405.6226 [hep-th]}}.

\bibitem{Basu:2018bde}
A.~Basu, ``{Eigenvalue equation for genus two modular graphs},''
  \href{http://dx.doi.org/10.1007/JHEP02(2019)046}{{\em JHEP} {\bf 02} (2019)
  046},
\href{http://arxiv.org/abs/1812.00389}{{\tt arXiv:1812.00389 [hep-th]}}.

\bibitem{Basu:2021xdt}
A.~Basu, ``{Poisson equation for genus two string invariants: a conjecture},''
  \href{http://dx.doi.org/10.1007/JHEP04(2021)050}{{\em JHEP} {\bf 04} (2021)
  050}, \href{http://arxiv.org/abs/2101.04597}{{\tt arXiv:2101.04597
  [hep-th]}}.

\end{thebibliography}
\end{document}